\documentclass[fleqn,usenatbib]{mnras}

\usepackage{newtxtext,newtxmath}

\usepackage{xcolor}
\usepackage{graphicx}
\usepackage{subcaption}
\usepackage{adjustbox}
\usepackage[normalem]{ulem}

\usepackage[T1]{fontenc}

\DeclareRobustCommand{\VAN}[3]{#2}
\let\VANthebibliography\thebibliography
\def\thebibliography{\DeclareRobustCommand{\VAN}[3]{##3}\VANthebibliography}

\usepackage{graphicx}	
\usepackage{amsmath}	

\title[Light-curve analysis and shape models of NEAs]{Light-curve analysis and shape models of NEAs 7335, 7822, 154244 and 159402}
\author[Javier Rodríguez Rodríguez et al.]{Javier Rodríguez Rodríguez,$^{1}$\thanks{E-mail: rodriguezrjavier@uniovi.es}
Enrique Díez Alonso,$^{1,2}$\thanks{E-mail: diezenrique@uniovi.es}
Santiago Iglesias Álvarez,$^{1}$
\newauthor
Saúl Pérez Fernández,$^{1}$
Alejandro Buendia Roca,$^{1}$
Julia Fernández Díaz,$^{1}$
Javier Licandro,$^{4,5}$\thanks{E-mail: jlicandr@iac.es}
\newauthor
Miguel R. Alarcon,$^{4,5}$
Miquel Serra-Ricart,$^{4,5,7}$
Noemi Pinilla-Alonso,$^{6}$
\newauthor
and Francisco Javier de Cos Juez$^{1,3}$
\\ \\
$^{1}$Instituto Universitario de Ciencias y Tecnologías Espaciales de Asturias (ICTEA), University of Oviedo, C. Independencia 13, 33004 Oviedo, Spain\\
$^{2}$Departamento de Matemáticas, Facultad de Ciencias, Universidad de Oviedo, 33007 Oviedo, Spain\\
$^{3}$Departamento de Explotación y Prospección de Minas, Universidad de Oviedo, 33004 Oviedo, Spain\\
$^{4}$Instituto de Astrof\'{\i}sica de Canarias (IAC), C/V\'{\i}a L\'actea sn, 38205 La Laguna, Spain\\
$^{5}$Departamento de Astrof\'{\i}sica, Universidad de La Laguna, 38206 La Laguna, Tenerife, Spain\\
$^{6}$Florida Space Institute, University of Central Florida, Orlando, FL 32816, USA\\
$^{7}$Light Bridges S. L., Avda. Alcalde Ram\'irez Bethencourt 17, E-35004 Las Palmas de Gran Canaria, Canarias, Spain \\
}

\date{Accepted XXX. Received YYY; in original form ZZZ}

\pubyear{2024}

\begin{document}
\label{firstpage}
\pagerange{\pageref{firstpage}--\pageref{lastpage}}
\maketitle

\begin{abstract}

In an attempt to further characterise the near-Earth asteroid (NEA) population we present 38 new light-curves acquired between September 2020 and November 2023 for NEAs (7335) 1989 JA, (7822) 1991 CS, (154244) 2002 KL6 and (159402) 1999 AP10, obtained from observations taken at the Teide Observatory (Tenerife, Spain). With these new observations along with archival data, we computed their first shape models and spin solutions by applying the light curve inversion method. The obtained rotation periods are in good agreement with those reported in previous works, with improved uncertainties. Additionally, besides the constant period models for (7335) 1989 JA, (7822) 1991 CS and (159402) 1999 AP10, our results for (154244) 2002 KL6 suggest that it could be affected by a Yarkovsky–O’Keefe–Radzievskii–Paddack acceleration with a value of $\upsilon \simeq -7\times10^{-9}$ rad d$^{-2}$. This would be one of the first detections of this effect slowing down an asteroid.

\end{abstract}




\section{Introduction} \label{sec:intro}

The known near-Earth asteroid (NEA) population is growing at a pace of $\sim$ 200 new asteroids discovered every month, with a total of 34467 NEAs discovered as of 8 March of 2024, according to the Center for Near Earth Object Studies (CNEOS)\footnote{\url{https://cneos.jpl.nasa.gov/stats/totals.html}}, with surveys such as the Asteroid Terrestrial-impact Last Alert System (ATLAS; \citealt{2018PASP..130f4505T}), the Panoramic Survey Telescope and Rapid Response System (Pan-STARRS; \citealt{2018AAS...23110201C}), the Catalina Sky Survey (CSS: \citealt{2009ApJ...696..870D}) or the Lowell Observatory Near-Earth-object Search (LONEOS; \citealt{1995DPS....27.0110B, 1999DPS....31.1202K}) among others being the responsible of this number of discoveries\footnote{See \url{https://www.minorplanetcenter.net/iau/lists/MPDiscsNum.html} for a full list of objects discovered per survey.}. Since its number is high and constantly growing, most of them barely have estimations of their rotation period and diameter. Among this group lies another important subgroup know as Potentially Hazardous Asteroids (PHAs), which is even more interesting because its Minimum Orbit Intersection Distance (MOID) is less than 0.05 AU. This makes this subgroup dangerous to Earth because of a possible collision. In addition, from a future resource exploitation perspective, NEAs are extremely important due to their potential as resource sources. Their periodic close approaches to Earth make them ideal targets for extraction missions.

The characterization of the physical properties of the NEA population is one of the hot topics in asteroid research. Particularly important is the determination of their rotational properties (rotation period and pole) and shape. A widely used technique for determining the rotational properties and shapes of asteroids is light-curve inversion as we did in \cite{2024MNRAS.527.6814R}. 

In this work we present the light-curves and derive the rotational properties and shape for NEAs (7335) 1989 JA, (7822) 1991 CS and (159402) 1999 AP10. The observations were obtained in the framework of the Visible NEAs Observations Survey (ViNOS; \citealt{2023MNRAS.521.3784L}). Targets were selected because they have also observations done using radar techniques with the aim of providing complementary data. Among the asteroids studied in this work, 7335 and 7822 are PHAs, their MOID and absolute magnitude (H) are 0.022203 and 17.8, and 0.021713 and 17.292 respectively, values  are below the threshold for PHAs according to CNEOS\footnote{\label{neo_groups}\url{https://cneos.jpl.nasa.gov/about/neo_groups.html}} of MOID $\leq 0.05$ AU and H $\leq 22$.

For computing the shape models, we opted for the Convex Inversion Method as detailed in \cite{2001Icar..153...24K} and \cite{2001Icar..153...37K}. This method generates convex models with its spin parameters from a set of light-curves, which can be either dense, sparse or a suitable or appropriate set of both. Dense light-curves is data collected from high cadence observations, taken in spans of hours during a single night; this kind of data is typically acquired in the frame of follow-up programmes, as in ViNOS. The other kind, sparse light-curves, is data collected from observations taken during several nights, in a low cadence with a temporal span from months to years, typically extracted from sky patrol programmes such as ATLAS, the Near-Earth Object Wide-field Infrared Survey Explorer (NEOWISE; \citealt{2011ApJ...731...53M}) or Pan-STARRS. 

For this work, all data consisted of dense light-curves, as our new observations were conducted by the ViNOS project. Obtaining several light-curves of an asteroid at different epochs is crucial for the proper application of the inversion methods. At different epochs the viewing geometry, in particular the aspect angle (the angle between the observer's line of sight and the asteroid's rotation axis) varies, providing diverse perspectives of the asteroid's shape and surface features, which is reflected in significant differences in the light-curve shapes, especially in their amplitude. Single-epoch observations, or observations done at similar aspect angles can only provide limited and potentially misleading information about the object's shape and rotational characteristics. Unlike main belt asteroids, the viewing geometry of NEAs varies over a short period of time as they pass close to Earth, which in some cases allows for obtaining reasonable shape models with data collected during a single close approach. 


Interestingly, NEAs can be affected by the Yarkovsky (\citealt{yarkovsky1901density, 2006AREPS..34..157B, 2015aste.book..509V}) and the YORP \cite{yarkovsky1901density,1952AZh....29..162R,1969JGR....74.4379P,o1976tektites,2006AREPS..34..157B,2015aste.book..509V} effects. The Yarkovsky effect is caused by the thermal re-emission of the incident solar radiation over the asteroid´s surface, which is then re-emitted, changing the orbit's semi major axis, increasing it for prograde rotation asteroids and decreasing it for retrograde ones, while the YORP effect produce changes in the asteroid spin state due to the anisotropic character of the thermal re-emission. Determining asteroid rotational variations due to YORP provide information on its thermal properties. As of 18 April 2024 there are only 8 asteroids confirmed to be affected by YORP and another candidate: (6489) Golevka \citep{2003Sci...302.1739C}, (1862) Apollo \citep{2007Natur.446..420K}, (54509) 2000 PH5 \citep{2007Sci...316..272L, 2007Sci...316..274T}, (1620) Geographos \citep{2008A&A...489L..25D}, (25143) Itokawa \citep{2014A&A...562A..48L}, (1685) Toro, (3103) Eger and (161989) Cacus \citep{2018A&A...609A..86D}, while the candidate is (2100) Ra-Shalom \citep{2024MNRAS.527.6814R, 2024A&A...682A..93D}.

The paper is organized as follows:  Sec. \ref{sec:obs} presents the observations and data reduction, Sec. \ref{sec:meth} describes the methodology used to study the light-curves and derive information on the asteroid rotation period and shape, in Sec. \ref{sec:res} we present the results and discuss them for each asteroid and finally the conclusions are presented in Sec.\ref{sec:con}.

\section{Observations, data reductions and light-curves}\label{sec:obs}

Photometric observations were obtained using five telescopes located at Teide Observatory (TO, Tenerife, Canary Islands, Spain), the IAC80, the TTT1 and TTT2 (Two Meter Twin Telescope), and the TAR2 and TAR4 telescopes. The observational circumstances are shown in Table \ref{tab_obs_cir}. 
	
The Two-meter Twin Telescope facility (TTT) is located at the Teide Observatory (latitude: 28{\degr}~18'~01.8"~N; longitude: +16{\degr}~30'~39.2"~W; altitude:  2386.75~m), on the island of Tenerife (Canary Islands, Spain). Currently, it includes two 0.8m telescopes (TTT1 and TTT2) on altazimuth mounts. Each telescope has two Nasmyth ports with focal ratios of f/4.4 and f/6.8, respectively. The observations were made using QHY411M\footnote{\url{https://www.qhyccd.com/}}  CMOS cameras \citep{2023PASP..135e5001A} installed on the f/4.4 focus of each telescope. The QHY411M are equipped with scientific Complementary Metal–Oxide–Semiconductor (sCMOS) image sensors with 14K x 10K 3.76~$\mu$m~pixel$^{-1}$ pixels. This setup provide an effective FoV of 51.4$^{\prime}\times$38.3$^{\prime}$ (with an angular resolution of 0.22"~pixel$^{-1}$). Images were taken using the $Luminance$  filter, that covers the 0.4 to 0.7 $\mu$m wavelength range, and the exposure time was dynamically set between to ensure a signal-to-noise ration (S/N) higher than 50. Images were bias and flat-field corrected in the standard way. 

The IAC80 is a 82~cm telescope with an $f/D =$ 11.3 in the Cassegrain focus. It is equipped with the CAMELOT-2 camera, a back-illuminated e2v 4K x 4K pixels CCD of 15 \textmu m$^2$ pixels. This setupt provide a plate scale of 0.32 arcsec/pixel, and a field of view of 21.98 x 22.06 arcmin$^2$. Images were obtained using the Sloan $r$ filter with the telescope in sidereal tracking, so the individual exposure time of the images was selected such the asteroid trail were smaller than the typical FWHM of the IAC80 images ($\sim 1.0$ \arcsec). The images were bias and flat-field corrected in the standard way. 

TAR2 is a 46~cm $f/D =$ 2.8 in the prime focus robotic telescope. Until July 2022 TAR2 was equipped with a SBIG STL 11000 CCD camera since then replaced by a FLI-Kepler KL400 camera.  The SBIG STL 11000 has a front illuminated 4008 x 2672 píxeles CCD with 9$\mu$m pixel size. The FLI-Kepler KL400 camera has a back illuminated 2K x 2K pixels GPixel GSense400 CMOS with a pixel size of 11 $\mu m^2$. 

TAR4 is a 40cm MEADE 16" $f/D =$ 2.8 in the Cassegrain focus robotic telescope,  equipped with a FLI-Kepler KL400 camera. Images of both TAR telescopes were bias, dark and flat-field corrected in the standard way.

To obtain the light-curves we did aperture photometry of the final images using the Photometry Pipeline\footnote{\url{https://photometrypipeline.readthedocs.io/en/latest/}} (PP) software \citep{2017A&C....18...47M}, as we did in \cite{2023MNRAS.521.3784L}. The images obtained with the L-filter were calibrated to the $r$ SLOAN band using the Pan-STARRS catalogue while the other images were calibrated to the corresponding bands for the filters used.

\begin{table*}
\caption{Observational circumstances of the new light-curves acquired by ViNOS. The table includes the asteroids ID, telescope and filters used (r-sloan, V, Clear and Luminance), the date, the starting and end times (UT) of the observations, the phase angle ($\alpha$), the heliocentric ($r$) and geocentric ($\Delta$) distances and the phase angle bisector longitude (PABLon) and latitude (PABLat) of the asteroid at the time of observation.}
\label{tab_obs_cir} 
\resizebox{\textwidth}{!}{%
\begin{tabular}{lllrlllrrrrr}
\hline
 Asteroid          & Telescope   & Filter   &   Exp. Time (s) & Date        & UT (start)   & UT (end)    &   $\alpha (^\circ)$ &   $r$ (au) &   $\Delta$ (au) &   PABLon (deg) &   PABLat (deg) \\
\hline
 7335 (1989 JA)    & IAC80        & SR        &           135 & 2022-Apr-12 & 00:36:31.622 & 2:19:56.525 &               27.71 &    1.3251 &           0.3826 &       215.20  &        20.19 \\
 7335 (1989 JA)    & IAC80        & SR       &           135 & 2022-Apr-12 & 03:55:36.077 & 6:02:11.587 &               27.69 &     1.3241 &          0.3813 &       215.25   &        20.18  \\
 7335 (1989 JA)    & IAC80        & SR        &           255 & 2022-Apr-27 & 01:33:05.328 & 5:39:49.450 &               26.88 &     1.2187 &          0.2448 &       220.70  &        19.32 \\
 7335 (1989 JA)    & TAR4        & Clear        &           120 & 2022-May-01 & 00:14:15.619 & 5:44:19.795 &               26.89 &     1.1914 &          0.2114 &       222.13  &        18.81 \\
 7335 (1989 JA)    & TAR4        & Clear        &           120 & 2022-May-02 & 21:59:39.293 & 5:41:45.139 &               26.92 &     1.1784 &          0.1956 &       222.82  &        18.50 \\
 7335 (1989 JA)    & TAR4        & Clear        &           120 & 2022-May-03 & 21:18:39.830 & 5:34:46.099 &               26.94 &     1.1718 &          0.1876 &       223.17  &        18.31 \\
 7335 (1989 JA)    & TAR4        & Clear        &           120 & 2022-May-04 & 23:11:42.403 & 1:59:24.547 &               26.97 &     1.1645 &          0.1788 &       223.55  &        18.09 \\
 7335 (1989 JA)    & TAR4        & Clear        &           120 & 2022-May-05 & 02:47:45.082 & 3:41:45.859 &               26.97 &     1.1635 &          0.1776 &       223.60  &        18.06 \\
 7335 (1989 JA)    & TAR2        & Clear        &           90 & 2022-May-08 & 21:12:35.222 & 5:17:19.363 &               27.05 &     1.1385 &          0.1475 &       224.89  &        17.08 \\
 7335 (1989 JA)    & TAR2        & Clear        &           30 & 2022-May-11 & 22:09:10.138 & 0:52:10.013 &               27.05 &     1.1187 &          0.1238 &       225.85  &        15.97 \\
 7335 (1989 JA)    & TAR2        & Clear        &           20 & 2022-May-17 & 22:19:40.253 & 0:22:23.174 &               26.57 &     1.0810  &          0.0786 &       227.20  &        12.04 \\
 7335 (1989 JA)    & TAR2        & Clear        &           30 & 2022-May-20 & 20:45:46.800 & 1:04:18.797 &               26.45 &     1.0635 &          0.0579 &       227.21  &         8.10 \\
 7335 (1989 JA)    & TAR2        & Clear       &           90 & 2022-May-21 & 20:44:43.901 & 2:58:04.886 &               26.84 &     1.0576 &          0.0512 &       227.01   &         6.07 \\
 7335 (1989 JA)    & TAR2        & Clear       &           30 & 2022-May-21 & 01:04:48.778 & 4:09:20.419 &               26.52 &     1.0624 &          0.0567 &       227.16  &         7.76 \\
 7822 (1991 CS)    & TAR2        & Clear        &           23 & 2022-Feb-07 & 03:36:40.522 & 5:33:03.283 &               53.78 &     1.1465 &          0.3349 &       180.49  &        -5.80  \\
 7822 (1991 CS)    & TAR2        & Clear        &           116 & 2022-Feb-12 & 02:43:43.680 & 6:30:52.070 &               49.54 &     1.1592 &          0.3090  &       181.60  &        -0.04 \\
 7822 (1991 CS)    & TAR2        & Clear        &           23 & 2022-Feb-13 & 01:57:54.259 & 3:59:41.453 &               48.66 &     1.1617 &          0.3046 &       181.74  &         1.14 \\
 7822 (1991 CS)    & TAR2        & Clear        &           46 & 2022-Feb-14 & 01:57:56.592 & 6:58:00.278 &               47.75 &     1.1642 &          0.3005 &       181.86  &         2.38 \\
 7822 (1991 CS)    & IAC80        & SR        &           30 & 2022-Mar-03 & 05:01:19.632 & 6:27:18.490 &               37.53 &     1.2049 &          0.2893 &       179.72  &        23.55 \\
 7822 (1991 CS)    & TAR2        & Clear        &           50 & 2022-Mar-05 & 01:54:15.581 & 6:36:12.182 &               37.67 &     1.2091 &          0.2954 &       179.16  &        25.49 \\
 154244 (2002 KL6) & IAC80        & SR        &           31 & 2023-Jul-20 & 22:43:13.843 & 0:14:03.782 &               59.93 &     1.0573 &          0.0876 &       273.39  &        24.96 \\
 154244 (2002 KL6) & IAC80        & SR        &           25 & 2023-Jul-29 & 21:43:07.334 & 2:29:28.291 &               63.44 &     1.0440  &          0.0684 &       292.49  &        32.72  \\
 154244 (2002 KL6)  & TT2         & L        &           8 & 2023-Aug-04 & 01:03:42.854 & 3:35:06.086 &               63.27 &     1.0418 &          0.0637 &       308.85  &        34.79 \\
 154244 (2002 KL6)  & TT2         & L        &           8 & 2023-Aug-04 & 21:34:56.928 & 3:04:18.566 &               63.08 &     1.0418 &          0.0636 &       311.79  &        34.75 \\
 154244 (2002 KL6)  & TT1         & L        &           7 & 2023-Aug-17 & 04:31:54.134 & 5:01:05.030 &               56.84 &     1.0538 &          0.0793 &       344.59  &        25.06 \\
 154244 (2002 KL6) & IAC80        & SR        &           75 & 2023-Sep-23 & 23:05:06.000 & 1:17:28.493 &               18.34 &     1.2096 &          0.2197 &        12.39 &         5.19 \\
 154244 (2002 KL6) & IAC80        & SR        &           75 & 2023-Sep-24 & 01:18:43.920 & 3:23:41.971 &               18.24 &     1.2102 &          0.2202 &        12.42 &         5.17 \\
 154244 (2002 KL6) & IAC80        & SR        &           75 & 2023-Sep-24 & 03:24:57.485 & 5:50:03.235 &               18.14 &     1.2107 &          0.2206 &        12.45 &         5.16 \\
 154244 (2002 KL6) & IAC80        & SR        &           75 & 2023-Oct-04 & 23:30:00.202 & 5:28:11.078 &                7.35 &     1.2783 &          0.2809 &        15.96 &         3.10 \\
 154244 (2002 KL6) & IAC80        & SR        &           105 & 2023-Oct-11 & 21:06:13.507 & 4:27:37.008 &                2.57 &     1.3243 &          0.3264 &        17.99  &         2.12 \\
 154244 (2002 KL6) & IAC80        & SR        &           75 & 2023-Oct-18 & 23:33:10.282 & 4:11:46.090 &                5.82 &     1.3736 &          0.3800   &        20.04 &         1.32 \\
 154244 (2002 KL6) & IAC80        & SR        &           105 & 2023-Nov-08 & 19:36:10.656 & 2:52:10.330 &               18.16 &     1.5246 &          0.5793 &        26.22 &        -0.25 \\
 159402 (1999 AP10) & TAR2       & Clear     &           120 & 2020-Sep-20 & 20:00:51.034 & 2:11:02.285 &               15.05 &     1.1656 &          0.1683 &       348.03  &        -0.73 \\
 159402 (1999 AP10) & TAR2        & Clear     &           60  & 2020-Sep-21 & 19:56:52.483 & 2:10:23.750 &               15.88 &     1.1599 &          0.1633 &       348.51  &        -0.25 \\
 159402 (1999 AP10) & TAR2        & Clear     &           10  & 2009-Oct-10 & 02:41:03.840 & 5:38:15.965 &               42.78 &     1.0632 &          0.0905 &       356.74  &        15.11 \\
 159402 (1999 AP10) & TAR2        & Clear     &           10  & 2020-Oct-11 & 19:39:16.589 & 2:30:07.776 &               39.95 &     1.0646 &          0.0889 &         3.72 &        18.03  \\
 159402 (1999 AP10) & TAR2        & Clear     &           10  & 2020-Oct-11 & 23:50:27.773 & 2:19:14.880 &               40.27 &     1.0640  &          0.0886 &         3.93 &        18.29 \\
 159402 (1999 AP10) & TAR2        & Clear     &           10  & 2020-Oct-12 & 19:37:06.557 & 0:11:51.677 &               41.68 &     1.0611 &          0.0870  &         5.05 &        19.48 \\
 159402 (1999 AP10) & TAR2        & Clear     &           10  & 2020-Oct-12 & 20:00:31.594 & 2:06:11.635 &               41.72 &     1.0611 &          0.0870   &         5.08 &        19.51 \\
 159402 (1999 AP10) & TAR2        & Clear     &           10  & 2020-Oct-13 & 19:39:49.421 & 1:46:29.510 &               43.48 &     1.0577 &          0.0853 &         6.49 &        20.97 \\
 159402 (1999 AP10) & TAR2        & Clear     &           60  & 2020-Dec-15 & 00:45:10.627 & 2:32:36.499 &               44.17 &     1.1416 &          0.2392 &       104.52  &        24.22 \\
 159402 (1999 AP10) & TAR2        & Clear     &           60  & 2021-Jan-15 & 22:23:37.766 & 6:46:36.595 &               16.94 &     1.3513 &          0.3912 &       115.39  &        15.12 \\
 159402 (1999 AP10) & TAR2        & Clear     &           60  & 2021-Jan-21 & 19:53:00.672 & 1:23:18.758 &               15.50  &     1.3951 &          0.4335 &       116.88  &        13.70 \\
 159402 (1999 AP10) & TAR2        & Clear     &           120 & 2021-Jan-22 & 20:09:07.747 & 1:22:21.907 &               15.43 &     1.4026 &          0.4414 &       117.13  &        13.47 \\
\hline
\end{tabular}%
}
\end{table*}

\section{Light-curve inversion method}\label{sec:meth}

In this study, we follow the same methodology as outlined in \cite{2024MNRAS.527.6814R}, to derive pole solutions and morphology of the asteroids. Two codes were used for this purpose, the first is the code publicly available at the Database of Asteroid Models from Inversion Techniques (DAMIT; \citealt{2010A&A...513A..46D}) which generates models with constant rotation period ($P$) This code was previously used in studies by \cite{2011A&A...530A.134H, 2012A&A...547A..10D, 2018A&A...609A..86D, 2024A&A...682A..93D} among others. From now on, we will refer to this code as "No YORP" code. The second code is a modification of the former (No YORP code), allowing the rotation period to constantly change over time. This modification enables the code to take into account the YORP effect that may be affecting the asteroid. This code, previously used in studies such as \cite{2012A&A...547A..10D, 2018A&A...609A..86D, 2024A&A...682A..93D}, will be referred as "YORP code". In contrast with the No YORP code, the YORP code is not publicly available and was provided by Josef \v{D}urech.

Next, the key parameters for the model creation are discussed. One crucial parameter needed to generate accurate models that fit the data is the rotation period ($P$), representing the time taken for an asteroid to complete a spin around its rotation axis considering the background stars as a reference frame. The ecliptic coordinates to which the spin axis points to, are Lambda ($\lambda$) and Beta ($\beta$) (its ecliptic longitude and latitude respectively), being their ranges $0^\circ \leq \lambda \leq 360^\circ$ and $-90^\circ \leq \beta \leq 90^\circ$. It is possible to obtain a solution with a duality in $\lambda$, that is, the code could offer two solutions with 180$^{\circ}$ of separation in $\lambda$ values, while having almost identical value for $\beta$.
By computing the values of $\lambda$ and $\beta$ from the best fitting model, and the asteroid's inclination (i), longitude of ascending node ($\Omega$) and the argument of pericenter ($\omega$), obtained from the Horizons System \citep{1996DPS....28.2504G, 2001DPS....33.5813G} the obliquity ($\epsilon$) can be calculated. If $0^\circ \leq \epsilon \leq 90^\circ$, the asteroid would be a prograde rotator, whereas otherwise ($90^\circ < \epsilon \leq 180^\circ$) it would be retrograde.

The data used for computing the models was acquired from two sources: the new light-curves presented in Section \ref{sec:obs} and Table \ref{tab_obs_cir}, and the light-curves hosted on the Asteroid Lightcurve Data Exchange Format (ALCDEF; \citealt{2010DPS....42.3914S, 2011MPBu...38..172W, 2018DPS....5041703S}) database. Observational circumstances of the archival data from ALCDEF for asteroids 7822, 154244 and 159402 are detailed in Tables \ref{tab:7822 archive}, \ref{tab:154244 archive} and \ref{tab:159402 archive}, respectively.

The initial step in the model generation process is to determine a $P$ to adopt as initial value in the code. To accomplish this, we used the tool provided with both codes for this purpose. This tool identifies the best fitting period to the light-curves. The code, along other default parameters, requires the interval of periods to perform the search, a coefficient $p$ of the period step and the convexity regularization weight ($d$), used to maintain the dark facet area below 1\%. For each asteroid, an initial search was made around the synodic period of the light-curves to ensure the global minimum in terms of $\chi^{2}$. Once the global minimum is found, another search is performed around it to refine the $P$ that will be used as initial $P$ for the subsequent models.
It is worth noting that the $P$ values obtained for the asteroids in this study were in line with those already published and available on the ALCDEF database.

Upon defining the initial value for $P$ for the asteroid, the following procedure is followed independently of the code used (No YORP or YORP code). Both codes, as the period search tool, have several parameters that were left as default, only modifying the $\lambda, \beta, P, d$ value if the model presents a value $> 1\%$ in the dark facet area, and setting the YORP value to $\upsilon=1\times10^{-8}$ in the YORP code.

Initially, a medium resolution search is made across the entire sphere ($0^{\circ} < \lambda \leq 360^{\circ}, -90^{\circ} \leq \beta \leq 90^{\circ}$) with $5^{\circ}$ steps ($\sim$ 2700 poles) and the obtained value of $P$. This medium search yields an initial solution in terms of $\chi^{2}$. These initial poles are then reduced to the number of observations used for each asteroid, resulting in a solution in terms of $\chi_\mathrm{red}^{2}$ as follows:

\begin{equation}
    \chi_{red}^2=\frac{\chi^2}{\nu}
    \label{eqn:chi_red}
\end{equation}

In Equation \ref{eqn:chi_red}, $\chi_{red}^2$ is the reduced $\chi^2$ to the number of degrees of freedom $\nu$, which in this case is the number of measures used in the model for each asteroid minus the number of parameters ($\sim 100$) \citep{2011AJ....142..159V}. When reducing the $\chi^2$ to the data, the value of $\chi_{red}^2 \simeq 1$ means that that the model matches the data almost perfectly, but this is usually not possible to be achieved since the observations are not perfect and contain some uncertainties.

Subsequently, a fine search is performed, narrowing down the area searched to a $30^{\circ}$ x $30^{\circ}$ square centered on the lowest $\chi_\mathrm{red}^{2}$ solution, with $2^{\circ}$ steps ($\sim$ 250 poles) and the value for the initial period provided with that solution. Again, as in the previous search the poles are too in terms of $\chi^{2}$, so the same process of reduction is done obtaining a final solution in terms of $\chi_\mathrm{red}^{2}$.

As neither the No YORP nor the YORP code compute the uncertainties of the best-fitting solution, a bootstrapping approach is adopted to fix it. This approach involves creating 100 subsets of light-curves from the initial set, randomly removing 25\% of the measurements in it, since the data sets were large enough ($\sim$ 3000 measurements). For each of these 100 subsets, a fine search was applied around the best solution from the medium search in terms of $\chi_\mathrm{red}^{2}$, yielding 100 solutions. Then the mean (which is almost identical to the best solution from the fine search with the main set of measurements) and the standard deviation (3$\sigma$ level) were  calculated, with the latter adopted as the uncertainty of the solution.

In an attempt to further validate the obtained results for asteroids affected by YORP effect, the method proposed in \cite{2017AJ....153..270V} was followed. This method yields a value for $\upsilon$ and its uncertainty at 3$\sigma$ level. To apply this method, the YORP code is iterated, with all the values fixed on those obtained in the best fine solution, except the $\upsilon$ value, adopting the lowest value of $\upsilon$ in terms of $\chi_\mathrm{red}^{2}$ obtained by this method as the final solution (again the final value is almost identical to the one obtained in the best solution).

\section{Results \& discussion}\label{sec:res}

In the next subsections, the methods discussed in Section \ref{sec:meth} are applied for each asteroid individually and the results for each of them are discussed (see Table \ref{tab_results} for a summary of the results).

\begin{table*}
\caption{Results obtained in this work for each asteroid, with the type of model (linearly increasing period (L) and constant period (C)), rotation period, geocentric ecliptic coordinates of the spin pole ($\lambda, \beta$), obliquity ($\epsilon$) and YORP acceleration ($\upsilon$) if the model has linearly increasing period.}
\label{tab_results} 
\begin{tabular}{llrrrrr}
\hline
Asteroid                 & Model & Period [h]          & $\lambda [^\circ]$           & $\beta [^\circ]$            & $\epsilon [^\circ]$          & $\upsilon$ [rad d$^{-2}$] \\
\hline
7335 (1989JA)       & C     & 2.590432$\pm$0.000391  & 243$\pm$17          & -61$\pm$6        & 147$\pm$8           & - \\
7822 (1991CS)       & C     & 2.390157$\pm$0.000002  & 242$\pm$9           & -57$\pm$7        & 175$\pm$7           & - \\
154244 (2002KL6)    & C     & 4.610235$\pm$0.000001  & 153$\pm$21          & -90$\pm$4        & 177$\pm$3           & - \\
154244 (2002KL6)    & C     & 4.610235$\pm$0.000001  & 334$\pm$28          & -90$\pm$4        & 176$\pm$3           & - \\
154244 (2002KL6)    & L     & 4.610232$\pm$0.000001  & 152$\pm$15          & -90$\pm$2        & 177$\pm$2           & (-7.12$\pm$1.65)$\times10^{-9}$ \\
154244 (2002KL6)    & L     & 4.610232$\pm$0.000001  & 333$\pm$18          & -89$\pm$2        & 177$\pm$2           & (-7.14$\pm$1.93)$\times10^{-9}$ \\
159402 (1999AP10)   & C     & 7.921917$\pm$0.000005  & 49$\pm$2            & -60$\pm$3        & 155$\pm$2           & - \\
\hline
\end{tabular}
\end{table*}

\subsection{(7335) 1989 JA}\label{sec:7335}

This asteroid belongs to the Apollo group\footref{neo_groups}, the asteroids in this group have a semi-major axis (a) > 1.0 AU and a perihelion distance (q) < 1.017 AU. There are two diameter values reported using WISE data: 0.932$\pm$0.153 km  \citep{2011ApJ...741...90M} and 0.73$\pm$0.02 km \citep{2016AJ....152...63N}.
Radar observations using Arecibo and Goldstone between May 4 and June 9, 1999, were reported in \cite{2002P&SS...50..257M} and conclude that it has an effective diameter within a factor of two of 1 km and a rotation period less than half a day.
It is remarkable that this asteroid had also a close approach to Earth during May 2022, in which Goldstone Radar, besides confirming the estimated diameter by WISE (0.7 km), discovered a small satellite with a diameter between 100 and 200 meters with an orbital period of about 17 hours\footnote{\url{https://echo.jpl.nasa.gov/asteroids/1989JA/1989JA.2022.goldstone.planning.html}}. 

Asteroid 7335 has neither any light-curves published on the ALCDEF database nor any shape model published. Therefore, in this work, we present its first shape model determination, for which we used only the light-curves acquired by ViNOS. These 14 light-curves cover a temporal span of 39 days from 12 April 2022 to 21 May 2022. Other light-curves obtained during the same 2022 close approach were obtained and used to determine its rotation period: $P=2.58988 \pm 0.00005$ h (Pravec 2022web\footnote{\url{https://www.asu.cas.cz/~asteroid/07335_2022a_p1.png}}), $P=2.5900 \pm 0.0002$ h \citep{2022MPBu...49..289L}, $P=2.588 \pm 0.001$ h \citep{2022MPBu...49..342F}, $P=2.592 \pm 0.006$ h \citep{2023MPBu...50...16L} and $P=2.588 \pm 0.001$ h \citep{2023MPBu...50...43S}.

\begin{figure}
    \centering
    \includegraphics[width=\linewidth,keepaspectratio]{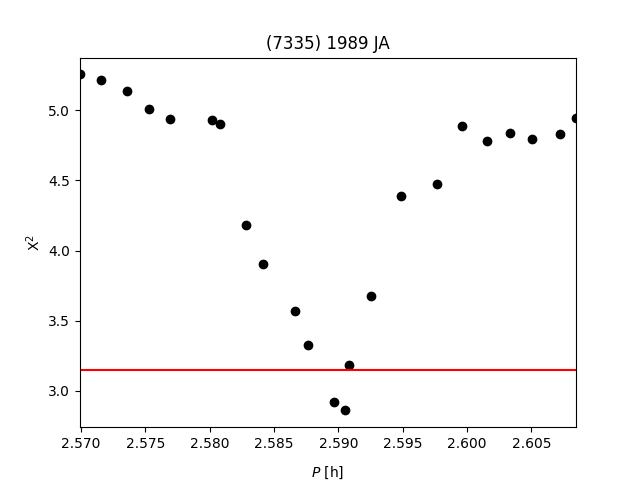}
    \caption{Period search tool output plot for (7335) 1989 JA. This search was performed in a interval from 2.57 h to 2.61 h, with a coefficient $p$ of 0.5. Each obtained period is represented as a black dot, with the red line representing a 10\% threshold from the lowest $\chi^{2}$ obtained. The presence of 2 values under this threshold (2.590536 h and 2.590704 h) is an indicator that more data will refine the period value.}
    \label{fig:7335_period}
\end{figure}

Taking into account the previously obtained periods, we performed several searches in their proximity, finding in most cases a minimum in $P=2.590536$ h, as shown in Figure \ref{fig:7335_period}. This value was adopted as the initial period to the No YORP code. Since the temporal span is so small, it is not a candidate to detect whether it is affected by YORP or not; nevertheless, the YORP code was run without success.

\begin{figure}
    \centering
    \includegraphics[width=\linewidth,keepaspectratio]{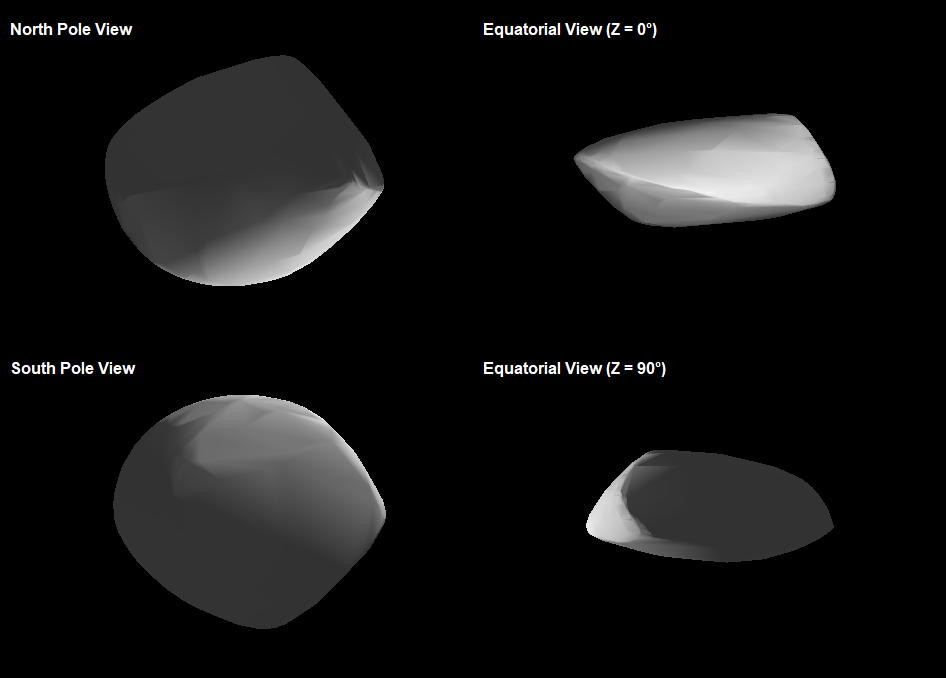}
    \caption{Shape model of (7335) 1989 JA obtained with a constant rotation period. Left top: North Pole View (Y axis = 0$^{\circ}$). Left bottom: South Pole View (Y axis = 180$^{\circ}$). Right top: Equatorial View with Z axis rotated 0$^{\circ}$. Right bottom: Equatorial View with Z axis rotated 90$^{\circ}$.}
    \label{fig:7335_shape_model}
\end{figure}

\begin{figure*}
    \centering
    \begin{subfigure}{0.49\textwidth}
        \centering
        \includegraphics[width=\textwidth]{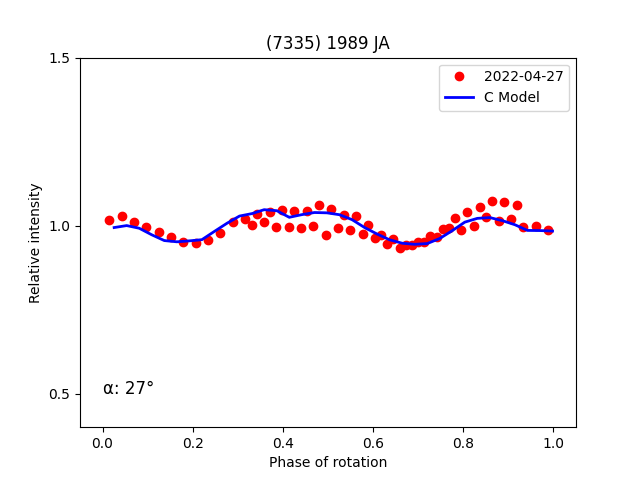}
    \end{subfigure}
    \begin{subfigure}{0.49\textwidth}
        \centering
        \includegraphics[width=\textwidth]{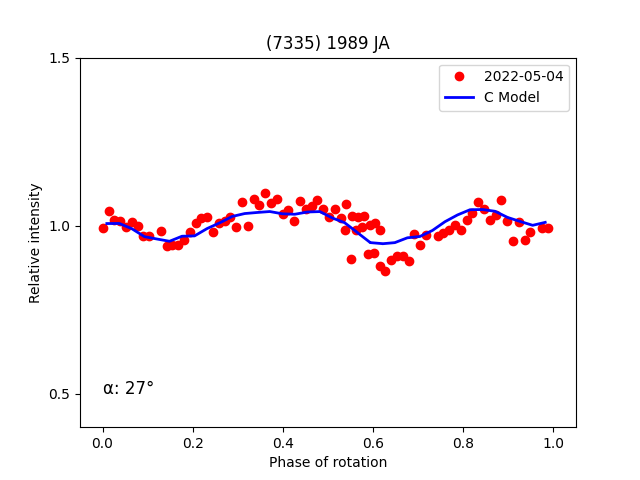}
    \end{subfigure}
    \begin{subfigure}{0.5\textwidth}
        \centering
        \includegraphics[width=\textwidth]{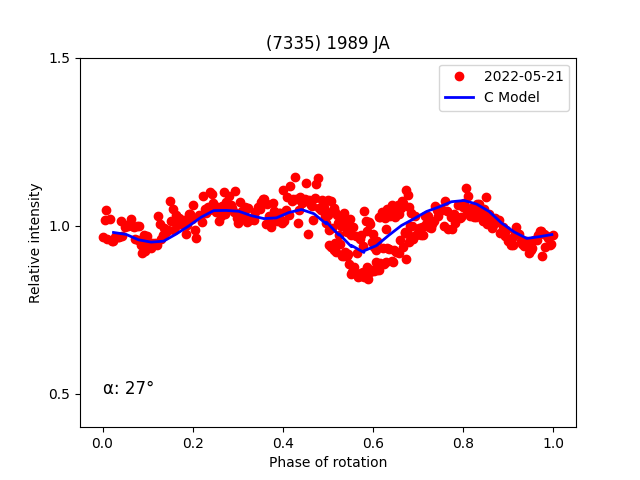}
    \end{subfigure}
    \caption{Fit between three of the new light-curves presented in this work from (7335) 1989 JA78 and the best-fitting constant period model (C Model). The data is plotted as red dots for each observation, while the model is plotted as a solid blue line. The geometry is described by its solar phase angle $\alpha$.}
    \label{fig:IAC_fit_7335_1}
\end{figure*}

With the initial period, we ran the code obtaining a medium solution with $P=2.590421$ h, $\lambda = 250^{\circ}$, $\beta = -60^{\circ}$,  $\chi_\mathrm{red}^{2}=1.05$ (see Figure \ref{fig:7335_plot} for a representation of the distribution of the best fitting solutions obtained in this medium search), and then a fine solution around that one of $P=2.590543$ h, $\lambda = 243^{\circ}$, $\beta = -61^{\circ}$,  $\chi_\mathrm{red}^{2}=1.04$ and $\epsilon \simeq 147^{\circ}$, which implies a retrograde rotation. The shape model presented in Figure \ref{fig:7335_shape_model} is the best fitting to the data (see Figure \ref{fig:IAC_fit_7335_1} and Figure \ref{fig:IAC_fit_7335_2} is a graphical representation of the fit between light-curves and shape model).

In order to obtain the uncertainties for the solution, 100 subsets from the main data set ($\sim$ 3300 measures) were created randomly removing 25\% of the measurements. After running the code in a fine search around the best medium solution the following result were obtained: $P=2.590432 \pm 0.000391$ h $\lambda = 243^{\circ} \pm 17^{\circ}$, $\beta = -61^{\circ} \pm 6^{\circ}$ and $\epsilon = 147^{\circ} \pm 8^{\circ}$. It is worth mentioning that more observations covering a wider range of viewing geometries are needed to compute a more consistent model.

\begin{figure*}
    \centering
    \begin{subfigure}{0.49\textwidth}
        \centering
        \includegraphics[width=\textwidth]{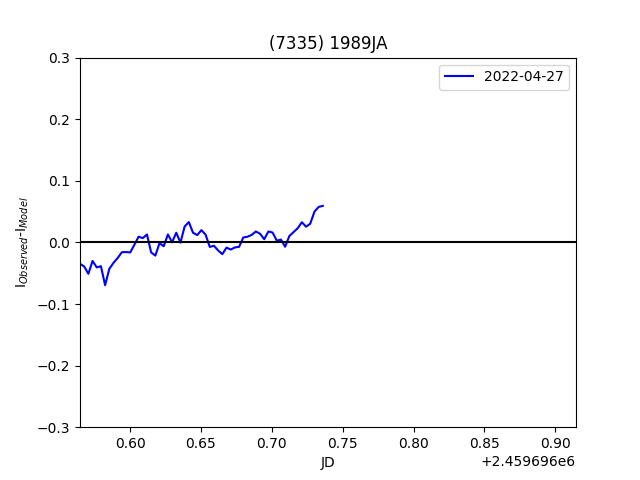}
    \end{subfigure}
    \begin{subfigure}{0.49\textwidth}
        \centering
        \includegraphics[width=\textwidth]{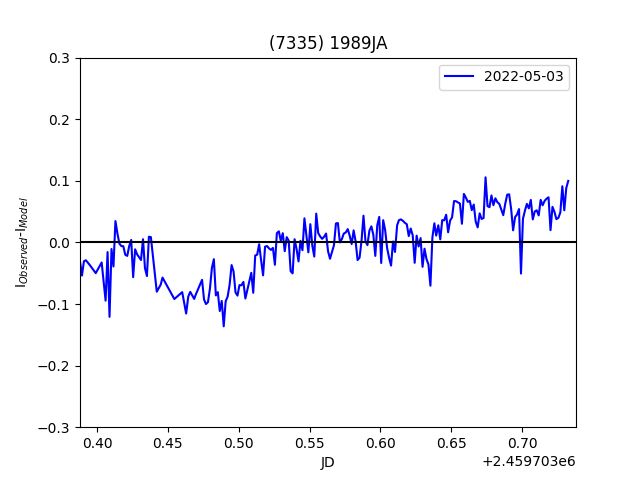}
    \end{subfigure}
    \begin{subfigure}{0.49\textwidth}
        \centering
        \includegraphics[width=\textwidth]{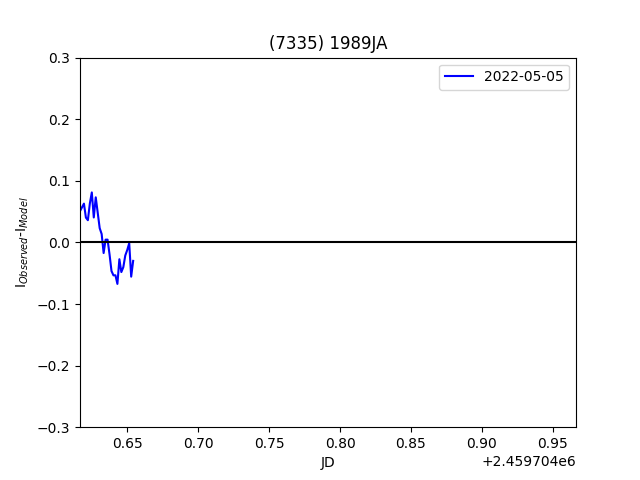}
    \end{subfigure}
    \begin{subfigure}{0.49\textwidth}
        \centering
        \includegraphics[width=\textwidth]{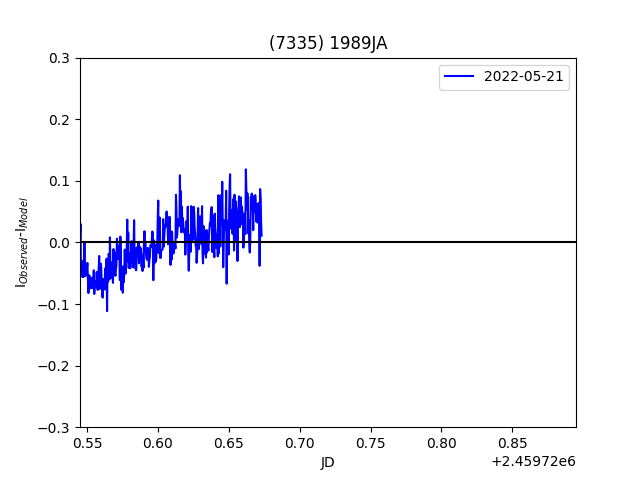}
    \end{subfigure}
    \caption{Plots of the intensity of the data minus the intensity of the model against the Julian date (JD) for the four nights with hints of falls in the intensity that could be attributed to mutual events of the asteroid and its satellite. A complete event is observed on May 3 2022, starting JD $\sim$ 2459703.44, ending $\sim$ 2459703.52, two partial events are recorded on May 5 (starting JD $\sim$ 2459705.02) and May 27 (ending $\sim$ 2459727.57 and a possible partial event is observed on April }
    \label{fig:7335_sat_det}
\end{figure*}

Finally we looked for mutual events between 7335 and its satellite. To do that we plotted the observed intensity minus the model intensity versus the Julian date (JD) of the observations. Results are shown in Figures \ref{fig:7335_sat_det} and \ref{fig:7335_sat_all}. 

We find four possible mutual events in the light-curves obtained on 27/4/2022, 3/5/2022, 5/5/2022 and 21/5/2022, where drops in the intensity are visible, only one (3/5/2022) is complete and lasted $\sim$ 2 hr, while the other three would be partially recorded. More data is needed to try to obtain reliable values for the orbital elements of the satellite and the relative sizes of both asteroids. The object will be observable again between September and November 2029 with medium-size telescopes.

\subsection{(7822) 1991 CS}\label{sec:7822}

In previous studies this asteroid has been classified as an S-type \citep{2002Icar..158..146B, 2014Icar..228..217T, 2019Icar..324...41B}. Dynamically is a member of the Apollo group. Radar observations of 7822 were obtained on August 1996 at Goldstone \cite{1999Icar..137..247B} concluded that the objects do not present a very elongated pole-on silhouette as the determine that "the hulls have a mean elongation and rms dispersion of 1.18 $\pm$ 0.02 and place a
lower bound on the maximum pole-on dimension of 1.3 km/cos($\delta$), where $\delta$ is the angle between the radar line-of-sight and the asteroid’s apparent equator". 
Other reported diameter are 0.83 km \citep{2010AJ....140..770T, 2011AJ....141...75H}; 1.602$\pm$0.012 km \citep{2011ApJ...741...90M} and 0.712$\pm$0.179 km \citep{2017AJ....154..168M}.

For this asteroid, there are 18 light-curves published on ALCDEF with a temporal span from 6 August 2015 to 23 August 2021, which were added to the new 6 presented in this work taken from 7 February 2022 to 5 March 2022. As for the previous object, there are five published periods: $P=2.391 \pm 0.001$ h \citep{2016MPBu...43...66W}, $P=2.389 \pm 0.001$ h \citep{2016MPBu...43..234K}, $P=2.392 \pm 0.002$ h \citep{2016MPBu...43..240W}, $P=2.3896 \pm 0.0005$ h (Pravec 2021web\footnote{\url{https://www.asu.cas.cz/~asteroid/1991cs_c1.png}}) and $P=2.388 \pm 0.002$ h \citep{2022MPBu...49...16W}, but not a shape model, so the one presented in this work is the first shape model for this asteroid.

\begin{figure}
    \centering
    \includegraphics[width=\linewidth,keepaspectratio]{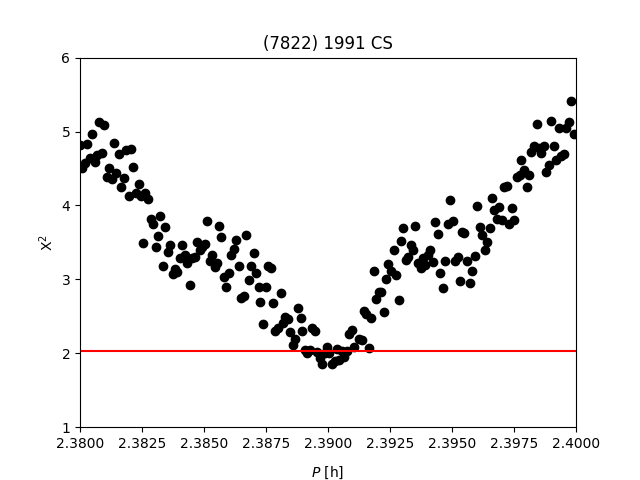}
    \caption{Period search tool output plot for (7822) 1991 CS. This search was made in a interval from 2.38 h to 2.40 h, with a coefficient $p$ of 1. Each obtained period is represented as a black dot, with the red line representing a 10\% threshold from the lowest $\chi^{2}$ obtained. The presence of several values under this threshold is an indicator that more future observations are needed to refine this value.}
    \label{fig:7822_period}
\end{figure}

With both archival data and the new light-curves (24 light-curves in total) the period search tool was run in an interval between 2.38 and 2.40 h, finding as the best fitting period $P=2.390159$ h. As shown in Figure \ref{fig:7822_period}, there are more periods that falls under the 10\% threshold from the lowest $\chi^{2}$, which implies that more data with different viewing geometries is needed from future observations to obtain a more precise value. It is worth noting that all the values under this threshold are among the accepted periods already published (2.388-2.391 h).

\begin{figure}
    \centering
    \includegraphics[width=\linewidth,keepaspectratio]{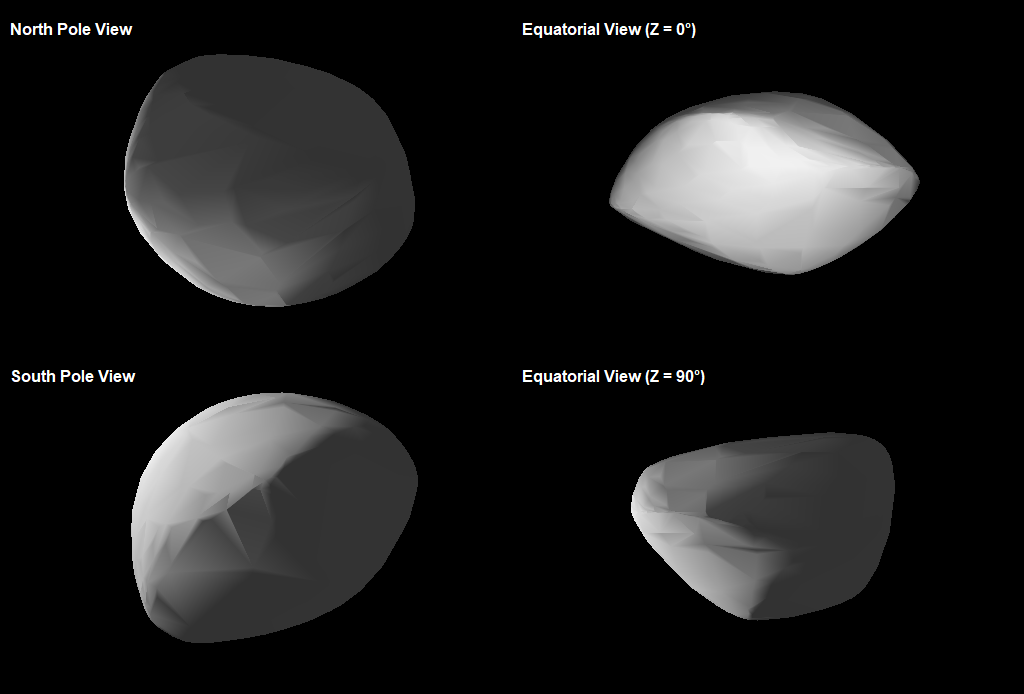}
    \caption{Shape model of (7822) 1991 CS obtained with a constant rotation period. Left top: North Pole View (Y axis = 0$^{\circ}$). Left bottom: South Pole View (Y axis = 180$^{\circ}$). Right top: Equatorial View with Z axis rotated 0$^{\circ}$. Right bottom: Equatorial View with Z axis rotated 90$^{\circ}$.}
    \label{fig:7822_shape_model}
\end{figure}

\begin{figure*}
    \centering
    \begin{subfigure}{0.49\textwidth}
        \centering
        \includegraphics[width=\textwidth]{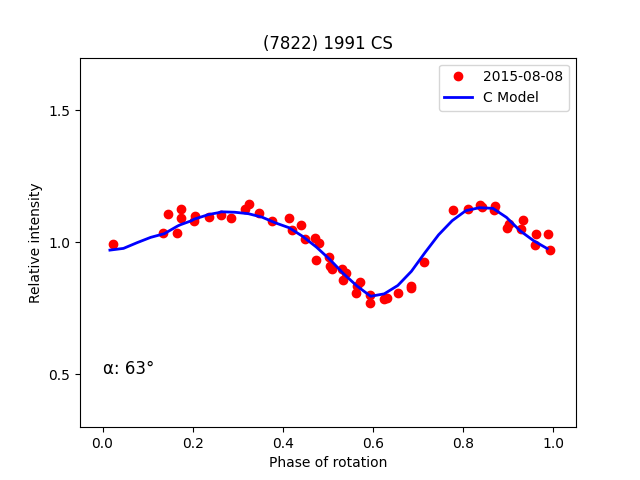}
    \end{subfigure}
    \begin{subfigure}{0.49\textwidth}
        \centering
        \includegraphics[width=\textwidth]{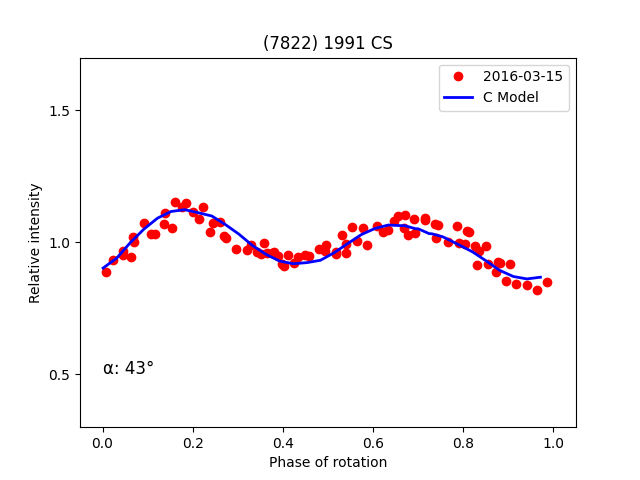}
    \end{subfigure}
    \begin{subfigure}{0.49\textwidth}
        \centering
        \includegraphics[width=\textwidth]{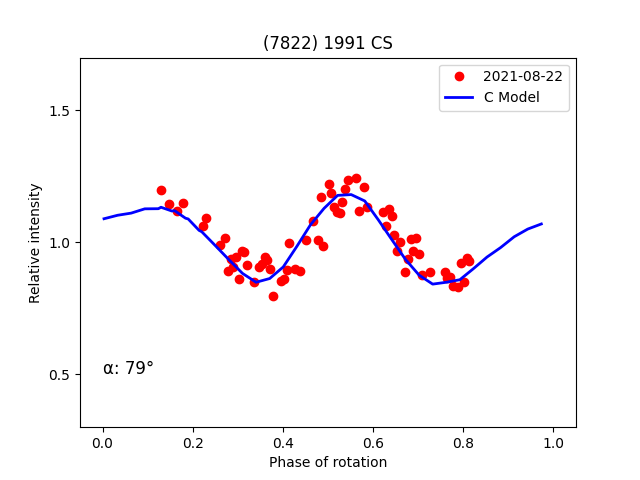}
    \end{subfigure}
    \begin{subfigure}{0.49\textwidth}
        \centering
        \includegraphics[width=\textwidth]{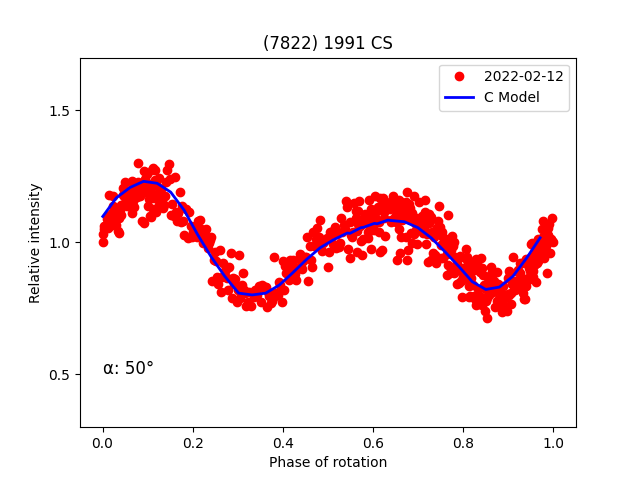}
    \end{subfigure}
    \caption{Fit between three of the ALCDEF light-curves and one (12 February 2022) of the new set presented in this work from (7822) 1991 CS and the best-fitting constant period model (C Model). The data is plotted as red dots for each observation, while the model is plotted as a solid blue line. The geometry is described by its solar phase angle $\alpha$.}
    \label{fig:fit 7822}
\end{figure*}

With this initial value of $P=2.390159$ h, the medium search was made with the No YORP code, obtaining the following initial solution: $P=2.390159$ h, $\lambda = 230^{\circ}$, $\beta = -60^{\circ}$, $\chi_\mathrm{red}^{2}=1.12$ (in Figure \ref{fig:7822_plot} a representation of the best fitting solutions is shown). With this initial solution, a fine search was performed, obtaining: $P=2.390157$ h, $\lambda = 240^{\circ}$, $\beta = -55^{\circ}$, $\epsilon \simeq 175^{\circ}$, which implies again retrograde rotation, and with $\chi_\mathrm{red}^{2}=1.10$ as the best fitting solution (see Figure \ref{fig:7822_shape_model} for a shape model of the best solution and Figures \ref{fig:fit 7822} and \ref{fig:IAC_fit_7822} for a fit between the best solution and the data). Is worth mentioning that since there are several periods under the threshold, models were created with the most significant of them as the initial period, but all of them resulted in worse $\chi_\mathrm{red}^{2}$ than the presented model. Although the temporal span is not sufficiently large (approximately 8 years), we still applied the YORP code to this asteroid. As expected, the YORP effect was not detected.

For the uncertainties the subsets were created from the main one ($\sim$ 3000 measures), removing randomly 25\% of the measurements, then run the code around the best fitting medium solution, obtaining: $P=2.390157 \pm 0.000002$ h $\lambda = 242^{\circ} \pm 9^{\circ}$, $\beta = -57^{\circ} \pm 7^{\circ}$ and $\epsilon = 175^{\circ} \pm 7^{\circ}$.

\subsection{(154244) 2002 KL6}\label{sec:154244}

This asteroid belongs to the Amor group\footref{neo_groups}, which has a > 1 AU and 1.017 < q < 1.3 AU. This asteroid has been reported as belonging to the taxonomic groups Q \citep{2014Icar..228..217T} or Sq \citep{2019Icar..324...41B}, and its diameter has been estimated to be within a factor of two of 1 km (between 0.5 and 2 km) by its optical albedo of 0.18 \footnote{\url{https://echo.jpl.nasa.gov/asteroids/2002KL6/2002KL6_planning.html}}. Radar observations of 154244 were obtained using Arecibo and Goldstone radiotelescopes during July 2016\footnote{\url{http://mel.epss.ucla.edu/radar/object/info.php?id=a0154244}}, but results are not published yet. 

Regarding 154244 rotation period, it was already studied in previous works, yielding the following results: $P=4.6063 \pm 0.0002$ h \citep{2010MPBu...37....9G}, $P=4.607 \pm 0.001$ h, $P=4.6081 \pm 0.0003$ h, $P=4.610 \pm 0.002$ h and $P=4.605 \pm 0.002$ h \citep{2014MPBu...41..286K}, $P=4.60869 \pm 0.00005$ h \citep{2016MPBu...43..343W}; $P=4.609 \pm 0.005$ h \citep{2017MPBu...44...22W}, $P=4.607 \pm 0.001$ h \citep{2017MPBu...44..206W}, $P=4.608 \pm 0.001$ h, $P=4.6052 \pm 0.0003$ h, $P=4.6060 \pm 0.0004$ h and $P=4.6096 \pm 0.0006$ h \citep{2019MPBu...46..458S}, $P=4.610 \pm 0.001$ h \citep{2023MPBu...50..304W}. It is also worth noting that \cite{2017MPBu...44..206W} published a shape model and a pole solution with $P=4.610233 \pm 0.000002$ h, $\lambda = 129^{\circ} \pm 10^{\circ}$, $\beta = -89^{\circ} \pm 10^{\circ}$.

\begin{figure}
    \centering
    \includegraphics[width=\linewidth,keepaspectratio]{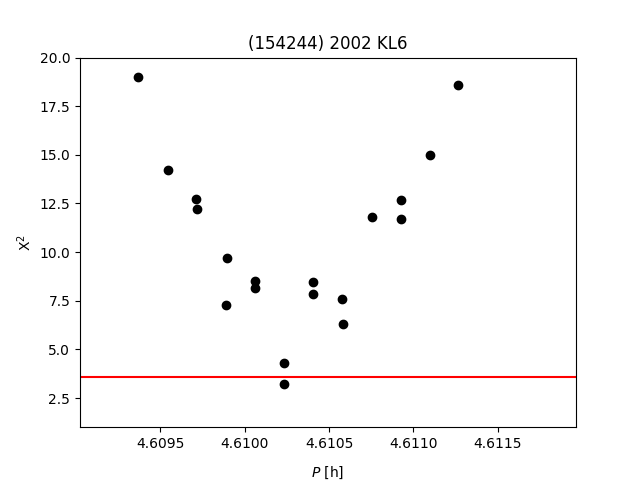}
    \caption{Period search tool output plot for (154244) 2002 KL6. This search was made in a interval from 4.609 h to 4.612 h, with a coefficient $p$ of 0.8. Each obtained period is represented as a black dot, with the red line representing a 10\% threshold from the lowest $\chi^{2}$ obtained. Only one period is under the threshold, which suggests that the period is well defined with the current data.}
    \label{fig:154244_period}
\end{figure}

In this work, the archival data available on ALCDEF (53 light-curves) covering a temporal span from 18 June 2009 to 5 August 2023, was used along with our 9 new light-curves (observations from 20 July 2023 to 11 October 2023). With this set of light-curves, a period search was made in a interval between 4.605 and 4.611 h, obtaining a best fitting period of $P=4.610235$ h as shown in Figure \ref{fig:154244_period}.

\begin{figure}
    \centering
    \includegraphics[width=\linewidth,keepaspectratio]{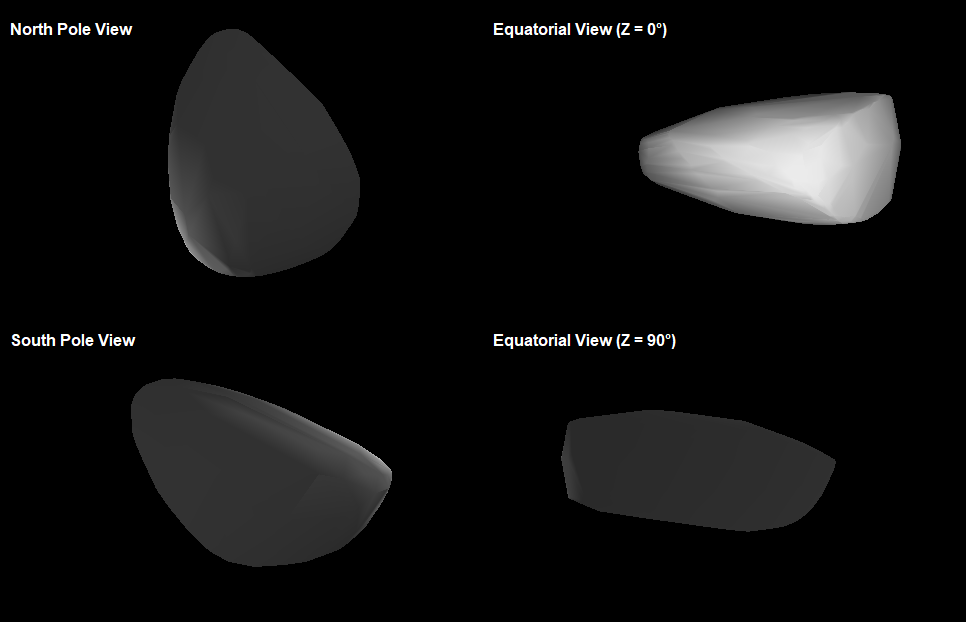}
    \caption{Shape model of (154244) 2002 KL6 obtained with a constant rotation period around ($\lambda = 149^{\circ}$, $\beta = -90^{\circ}$). Left top: North Pole View (Y axis = 0$^{\circ}$). Left bottom: South Pole View (Y axis = 180$^{\circ}$). Right top: Equatorial View with Z axis rotated 0$^{\circ}$. Right bottom: Equatorial View with Z axis rotated 90$^{\circ}$.}
    \label{fig:154244_150_ny_shape_model}
\end{figure}

\begin{figure}
    \centering
    \includegraphics[width=\linewidth,keepaspectratio]{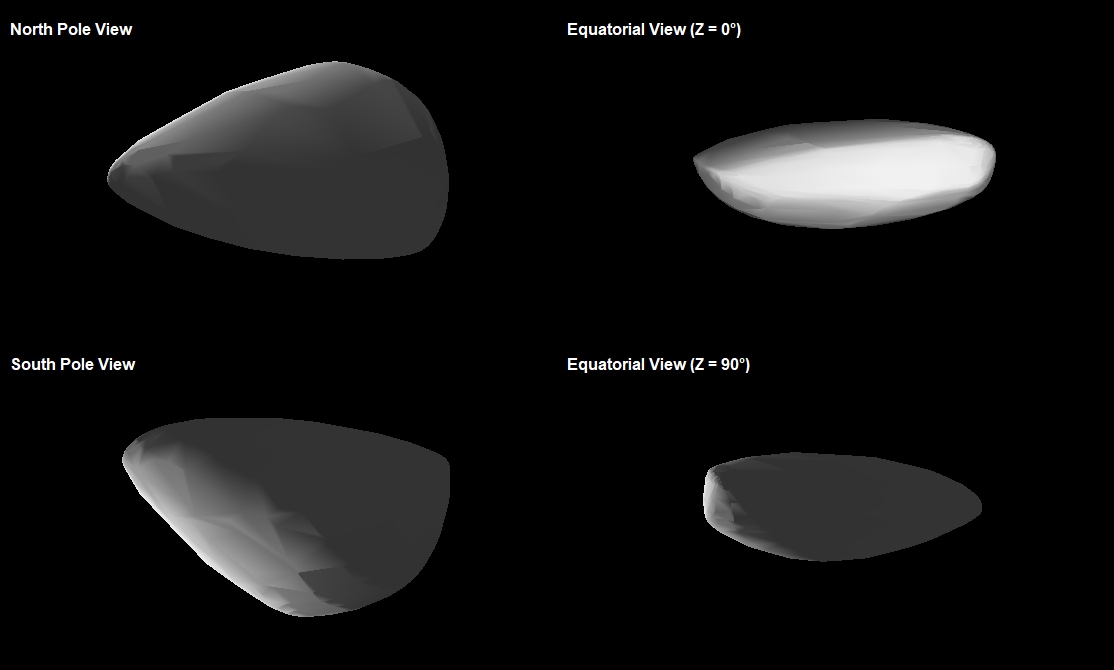}
    \caption{Shape model of (154244) 2002 KL6 obtained with a constant rotation period around ($\lambda = 326^{\circ}$, $\beta = -88^{\circ}$). Left top: North Pole View (Y axis = 0$^{\circ}$). Left bottom: South Pole View (Y axis = 180$^{\circ}$). Right top: Equatorial View with Z axis rotated 0$^{\circ}$. Right bottom: Equatorial View with Z axis rotated 90$^{\circ}$.}
    \label{fig:154244_330_ny_shape_model}
\end{figure}

\begin{figure*}
    \centering
    \begin{subfigure}{0.49\textwidth}
        \centering
        \includegraphics[width=\textwidth]{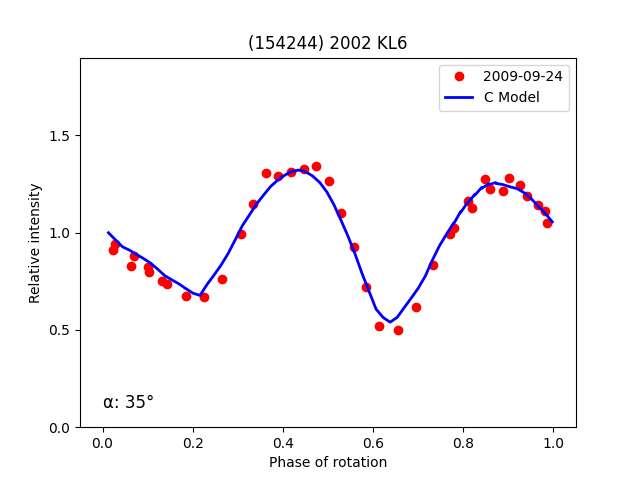}
    \end{subfigure}
    \begin{subfigure}{0.49\textwidth}
        \centering
        \includegraphics[width=\textwidth]{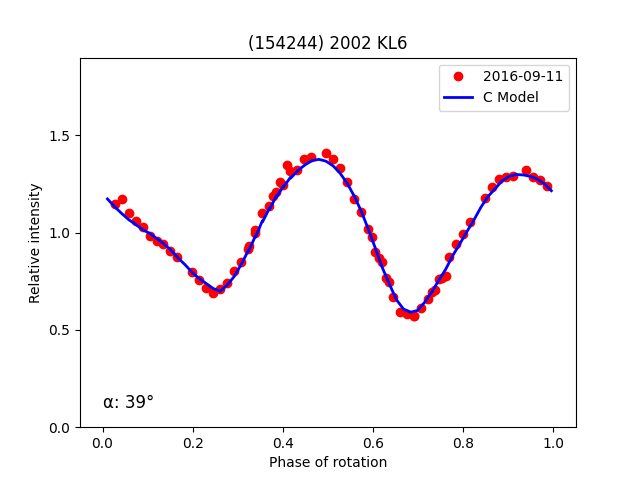}
    \end{subfigure}
    \begin{subfigure}{0.49\textwidth}
        \centering
        \includegraphics[width=\textwidth]{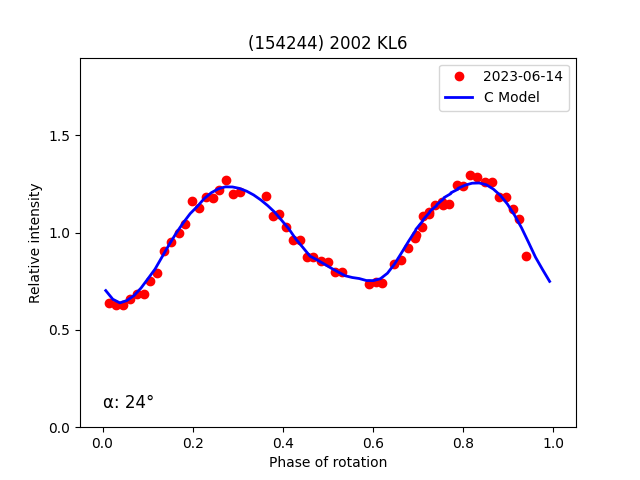}
    \end{subfigure}
    \begin{subfigure}{0.49\textwidth}
        \centering
        \includegraphics[width=\textwidth]{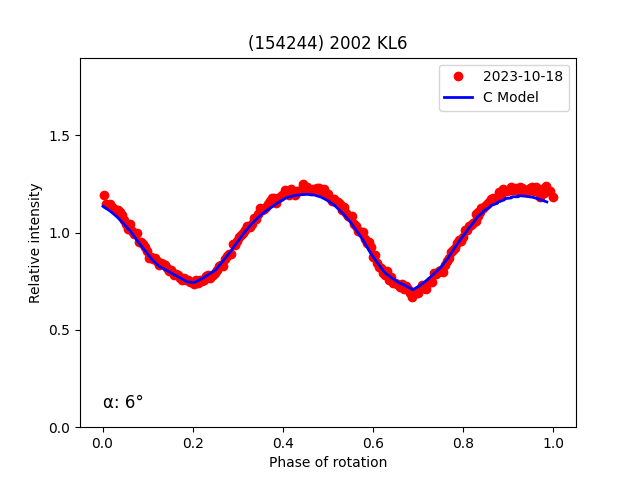}
    \end{subfigure}
    \caption{Fit between three of the ALCDEF light-curves and one (18 October 2023) of the new light-curves presented in this work from (154244) 2002 KL6 and the best-fitting constant period model (C Model) around ($\lambda = 149^{\circ}$, $\beta = -90^{\circ}$). The data is plotted as red dots for each observation, while the model is plotted as a solid blue line. The geometry is described by its solar phase angle $\alpha$.}
    \label{fig:fit_154244_ny_155}
\end{figure*}

\begin{figure*}
    \centering
    \begin{subfigure}{0.49\textwidth}
        \centering
        \includegraphics[width=\textwidth]{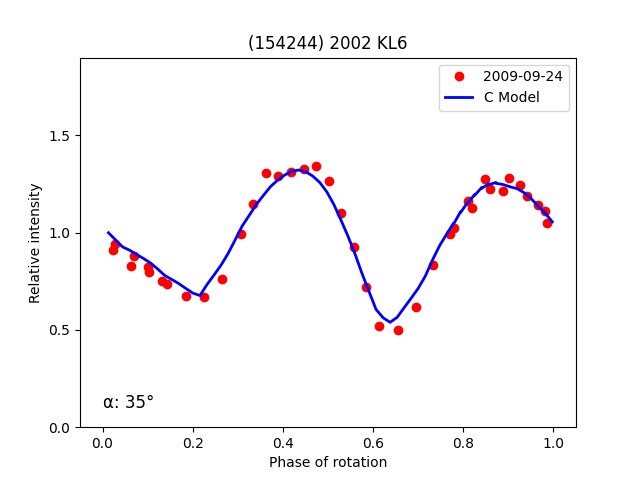}
    \end{subfigure}
    \begin{subfigure}{0.49\textwidth}
        \centering
        \includegraphics[width=\textwidth]{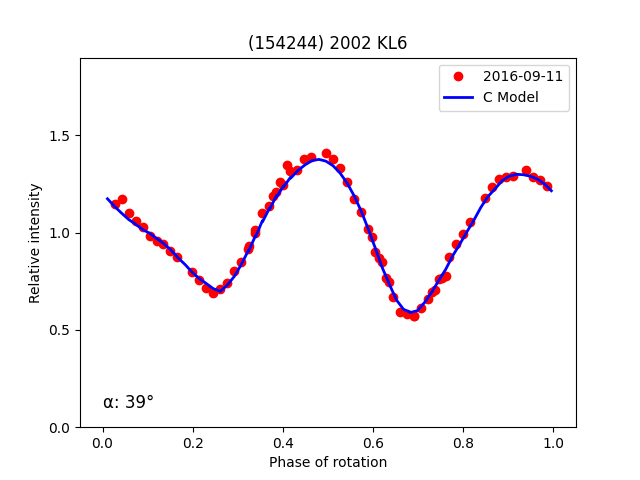}
    \end{subfigure}
    \begin{subfigure}{0.49\textwidth}
        \centering
        \includegraphics[width=\textwidth]{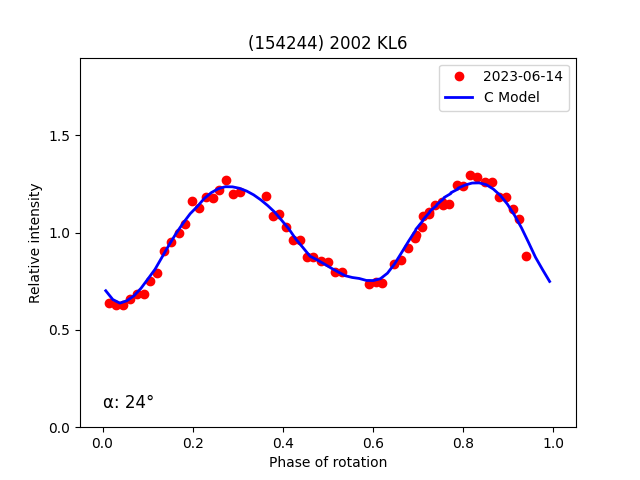}
    \end{subfigure}
    \begin{subfigure}{0.49\textwidth}
        \centering
        \includegraphics[width=\textwidth]{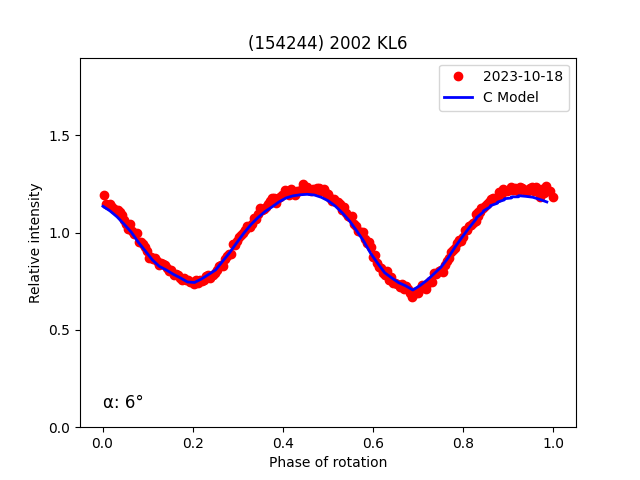}
    \end{subfigure}
    \caption{Fit between three of the ALCDEF light-curves and one (18 October 2023) of the new light-curves presented in this work from (154244) 2002 KL6 and the best-fitting constant period model (C Model) around ($\lambda = 326^{\circ}$, $\beta = -88^{\circ}$). The data is plotted as red dots for each observation, while the model is plotted as a solid blue line. The geometry is described by its solar phase angle $\alpha$.}
    \label{fig:fit_154244_ny_335}
\end{figure*}

With this initial period, the medium search was made with the No YORP code, obtaining two paired solutions: $P=4.610235$ h, $\lambda = 330^{\circ}$, $\beta = -90^{\circ}$,  $\chi_\mathrm{red}^{2}=1.16$ and $P=4.610235$ h, $\lambda = 150^{\circ}$, $\beta = -90^{\circ}$,  $\chi_\mathrm{red}^{2}=1.16$ (see Figure \ref{fig:15422_plot_ny} for a graphical representation of the medium pole search). These paired solutions are 180$^{\circ}$ away from each other in terms of $\lambda$, and the value of $\beta$ means that its uncertainty may be causing this change in $\lambda$ due to its position in the sphere, which could imply that it is the same solution. After these medium results, one fine search was made around each of them, obtaining the following results: $P=4.610235$ h, $\lambda = 150^{\circ}$, $\beta = -90^{\circ}$, $\epsilon \sim 178^{\circ}$, $\chi_\mathrm{red}^{2}=1.12$ and $P=4.610235$ h, $\lambda = 330^{\circ}$, $\beta = -89^{\circ}$, $\epsilon \sim 177^{\circ}$, $\chi_\mathrm{red}^{2}=1.12$ (see Figures \ref{fig:154244_150_ny_shape_model} and \ref{fig:154244_330_ny_shape_model} for shape models of both solutions and Figures \ref{fig:fit_154244_ny_155}, \ref{fig:IAC_fit_154244_ny_155} and \ref{fig:fit_154244_ny_335}, \ref{fig:IAC_fit_154244_ny_335} for the fit to the data respectively). In both solutions, $\epsilon$ implies a retrograde rotation.

To calculate the uncertainties, a random 25\% of the main data were removed ($\sim$ 5000 measurements) creating the 100 subsets to be iterated around the 2 obtained medium solutions, yielding the following results for each of them: $P=4.610235 \pm 0.000001$ h, $\lambda = 334^{\circ} \pm 28^{\circ}$, $\beta = -90^{\circ} \pm 4^{\circ}$ and $\epsilon = 176^{\circ} \pm 3^{\circ}$ (iteration around $\lambda=330^{\circ}$, $\beta=-90^{\circ}$, $P=4.610235$ h) and $P=4.610235 \pm 0.000001$ h, $\lambda = 153^{\circ} \pm 21^{\circ}$, $\beta = -90^{\circ} \pm 4^{\circ}$ and $\epsilon = 177^{\circ} \pm 3^{\circ}$ (iteration around $\lambda=150^{\circ}$, $\beta=-90^{\circ}$, $P=4.610235$ h), again both solutions imply retrograde rotation. One of the two obtained results is pretty close to the one already mentioned from \cite{2017MPBu...44..206W}, the difference in $\lambda$ may be related to our new data from 2023.

\begin{figure}
    \centering
    \includegraphics[width=\linewidth,keepaspectratio]{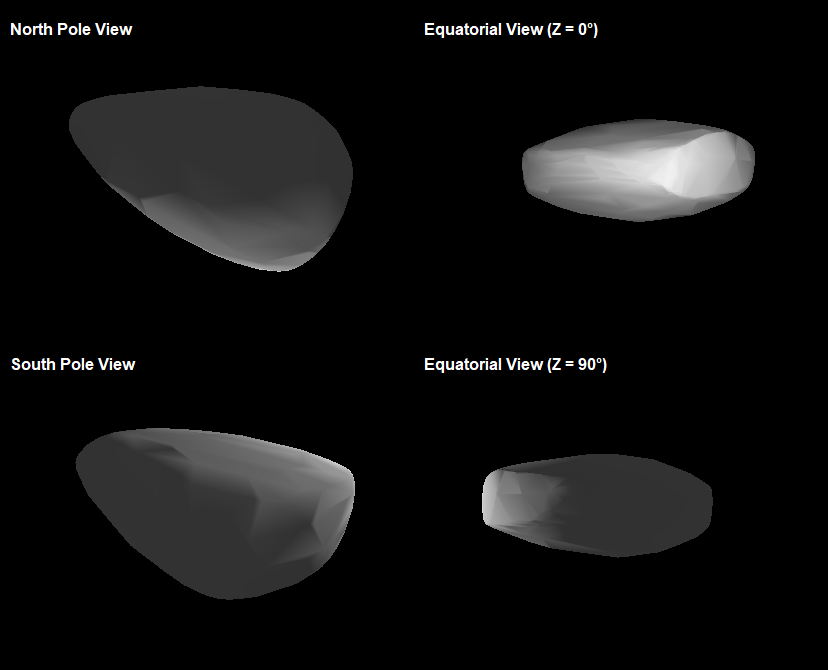}
    \caption{Shape model of (154244) 2002 KL6 obtained with a linearly increasing rotation period around ($\lambda = 151^{\circ}$, $\beta = -90^{\circ}$). Left top: North Pole View (Y axis = 0$^{\circ}$). Left bottom: South Pole View (Y axis = 180$^{\circ}$). Right top and bottom: Equatorial Views with Z axis rotated 0$^{\circ}$ and 90$^{\circ}$.}
    \label{fig:154244_155_y_shape_model}
\end{figure}

\begin{figure}
    \centering
    \includegraphics[width=\linewidth,keepaspectratio]{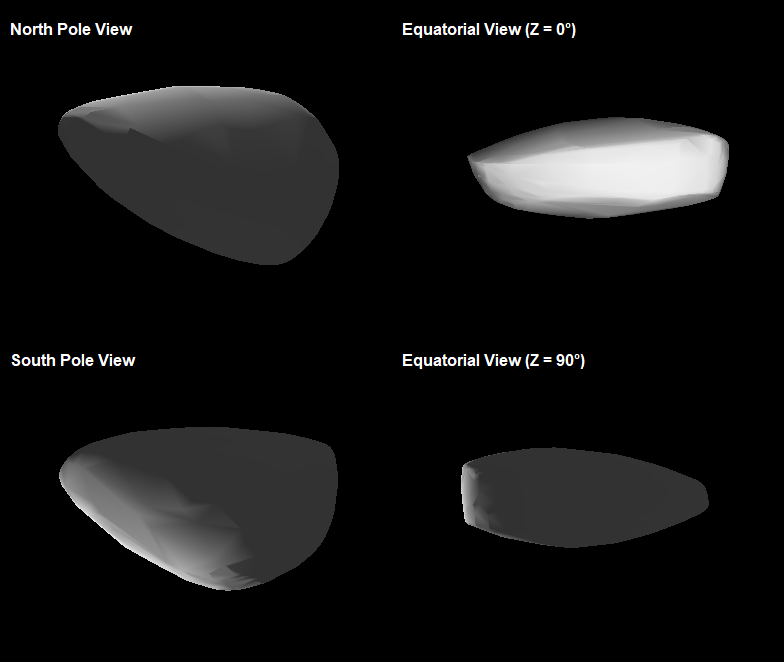}
    \caption{Shape model of (154244) 2002 KL6 obtained with a linearly increasing rotation period around ($\lambda = 330^{\circ}$, $\beta = -89^{\circ}$). Left top: North Pole View (Y axis = 0$^{\circ}$). Left bottom: South Pole View (Y axis = 180$^{\circ}$). Right top and bottom: Equatorial Views with Z axis rotated 0$^{\circ}$ and 90$^{\circ}$.}
    \label{fig:154244_335_y_shape_model}
\end{figure}

\begin{figure*}
    \centering
    \begin{subfigure}{0.49\textwidth}
        \centering
        \includegraphics[width=\textwidth]{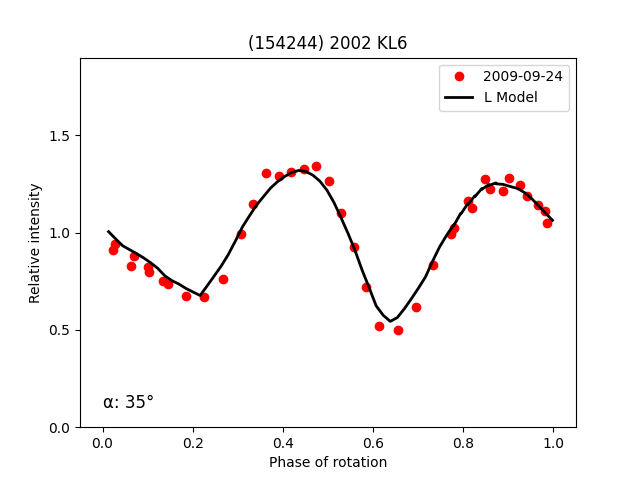}
    \end{subfigure}
    \begin{subfigure}{0.49\textwidth}
        \centering
        \includegraphics[width=\textwidth]{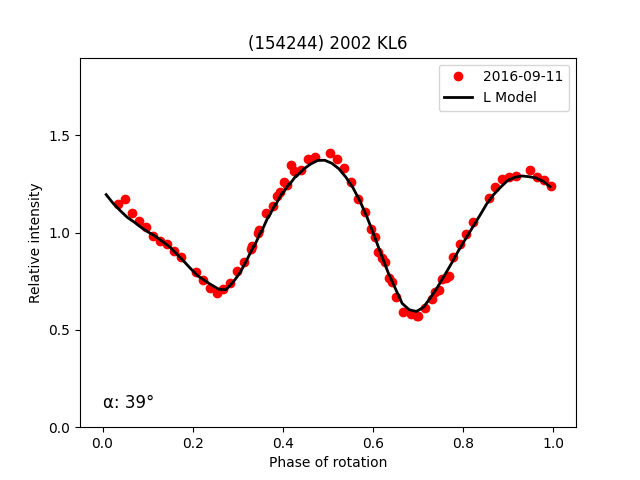}
    \end{subfigure}
    \begin{subfigure}{0.49\textwidth}
        \centering
        \includegraphics[width=\textwidth]{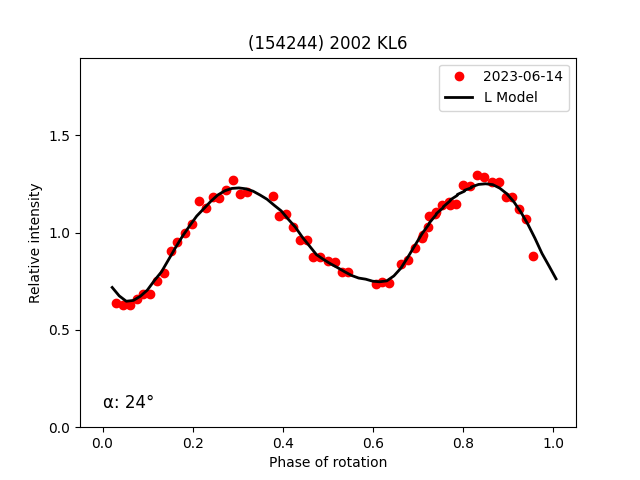}
    \end{subfigure}
    \begin{subfigure}{0.49\textwidth}
        \centering
        \includegraphics[width=\textwidth]{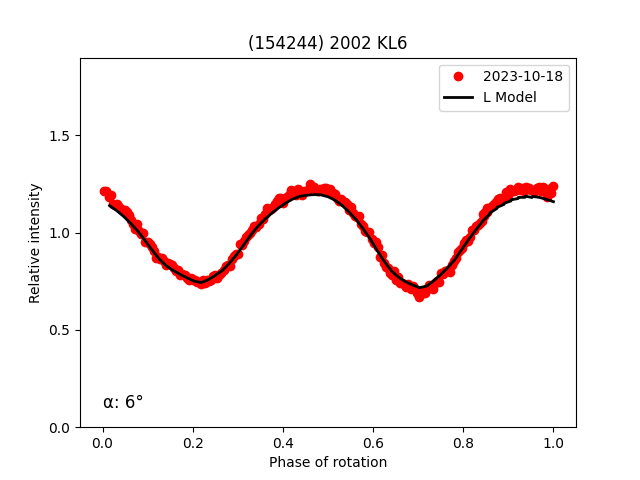}
    \end{subfigure}
    \caption{Fit between three of the ALCDEF light-curves and one (18 October 2023) of the new light-curves from (154244) 2002 KL6 presented in this work and the best-fitting linearly increasing period model (L Model) around ($\lambda = 151^{\circ}$, $\beta = -90^{\circ}$). The data is plotted as red dots for each observation, while the model is plotted as a solid black line. The geometry is described by its solar phase angle $\alpha$.}
    \label{fig:fit_154244_y_150}
\end{figure*}

\begin{figure*}
    \centering
    \begin{subfigure}{0.49\textwidth}
        \centering
        \includegraphics[width=\textwidth]{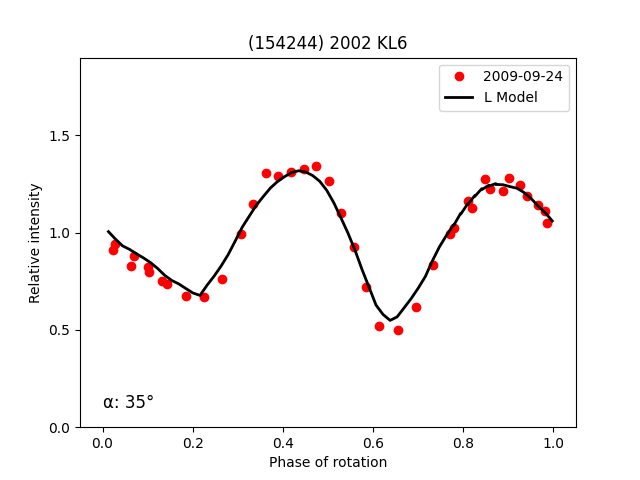}
    \end{subfigure}
    \begin{subfigure}{0.49\textwidth}
        \centering
        \includegraphics[width=\textwidth]{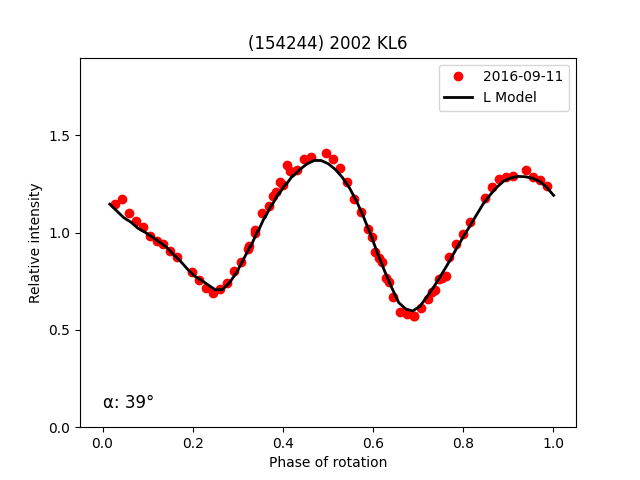}
    \end{subfigure}
    \begin{subfigure}{0.49\textwidth}
        \centering
        \includegraphics[width=\textwidth]{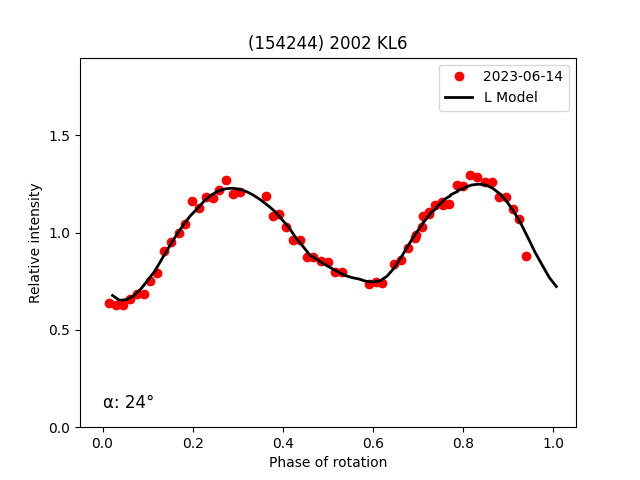}
    \end{subfigure}
    \begin{subfigure}{0.49\textwidth}
        \centering
        \includegraphics[width=\textwidth]{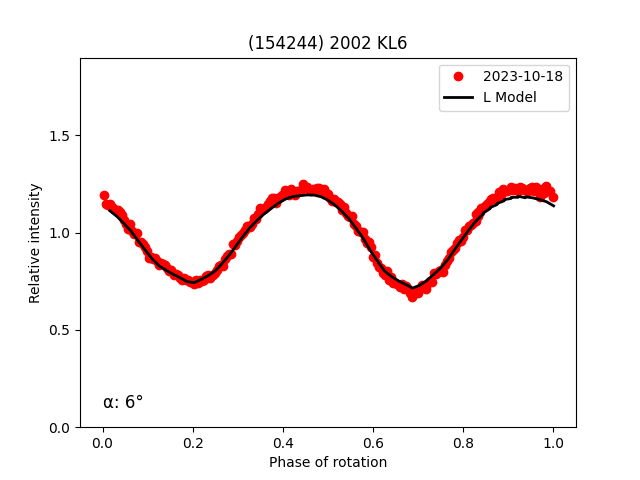}
    \end{subfigure}
    \caption{Fit between three of the ALCDEF light-curves and one (18 October 2023) of the new light-curves from (154244) 2002 KL6 presented in this work and the best-fitting linearly increasing period model (L Model) around ($\lambda = 330^{\circ}$, $\beta = -89^{\circ}$). The data is plotted as red dots for each observation, while the model is plotted as a solid black line. The geometry is described by its solar phase angle $\alpha$.}
    \label{fig:fit_154244_y_330}
\end{figure*}

Since the data set time span is relatively large ($\sim$ 14 years), it was worth an attempt to check if the the asteroid is affected by YORP. As the obtained period ($P=4.610235$) fitted good enough the data, it was used as the initial period to the YORP code. The medium search obtained again two solutions: $P=4.610232$ h corresponding to 18 June 2009, $\lambda = 335^{\circ}$, $\beta = -90^{\circ}$, $\upsilon = -7.48\times10^{-9}$ rad d$^{-2}$,  $\chi_\mathrm{red}^{2}=1.12$ and $P=4.610232$ h corresponding to the same date as the previous model, $\lambda = 155^{\circ}$, $\beta = -90^{\circ}$, $\upsilon = -7.48\times10^{-9}$ rad d$^{-2}$,  $\chi_\mathrm{red}^{2}=1.12$ (see Figure \ref{fig:15422_plot_y} for a graphical representation of the medium pole search). As in the No YORP code medium search, the solutions are 180$^{\circ}$ away from each other. A fine search around each of those solutions was made with the following results: $P=4.610232$ h, $\lambda = 151^{\circ}$, $\beta = -90^{\circ}$, $\upsilon = -7.07\times10^{-9}$ rad d$^{-2}$, $\epsilon \sim 178^{\circ}$, $\chi_\mathrm{red}^{2}=1.09$ and $P=4.610232$ h, $\lambda = 330^{\circ}$, $\beta = -89^{\circ}$, $\upsilon = -6.95\times10^{-9}$ rad d$^{-2}$, $\epsilon \sim 178^{\circ}$, $\chi_\mathrm{red}^{2}=1.08$ (see Figures \ref{fig:154244_155_y_shape_model} and \ref{fig:154244_335_y_shape_model} for shape models of both solutions and Figures \ref{fig:fit_154244_y_150}, \ref{fig:IAC_fit_154244_y_150} and \ref{fig:fit_154244_y_330}, \ref{fig:fit_154244_y_330} for the fit to the data respectively).

Again the uncertainties were calculated in the same way as previously, obtaining: $P=4.610232 \pm 0.000001$ h, $\lambda = 333^{\circ} \pm 18^{\circ}$, $\beta = -89^{\circ} \pm 2^{\circ}$, $\upsilon = (-7.14\pm1.93)\times10^{-9}$ rad d$^{-2}$ and $\epsilon = 177^{\circ} \pm 2^{\circ}$ (iteration around $\lambda=335^{\circ}$, $\beta=-90^{\circ}$, $P=4.610233$ h) and $P=4.610232 \pm 0.000001$ h, $\lambda = 152^{\circ} \pm 15^{\circ}$, $\beta = -90^{\circ} \pm 2^{\circ}$, $\upsilon = (-7.12\pm1.65)\times10^{-9}$ rad d$^{-2}$ and $\epsilon = 177^{\circ} \pm 2^{\circ}$ (iteration around $\lambda=155^{\circ}$, $\beta=-90^{\circ}$, $P=4.610233$ h).

\begin{figure*}
    \centering
    \begin{subfigure}{0.49\textwidth}
        \centering
        \includegraphics[width=\linewidth,keepaspectratio]{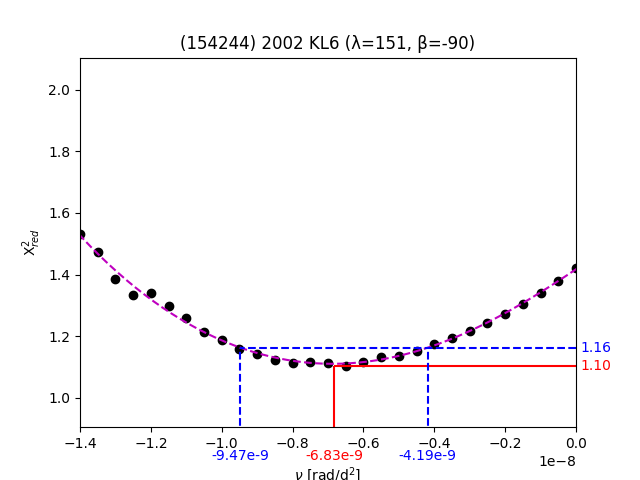}
    \end{subfigure}
    \begin{subfigure}{0.49\textwidth}
        \centering
        \includegraphics[width=\linewidth,keepaspectratio]{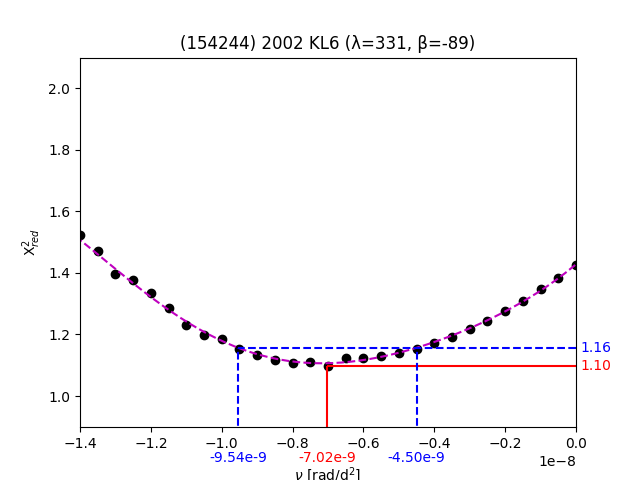}
    \end{subfigure}
    \caption{Change in $\chi_\mathrm{red}^{2}$ for both solutions of (154244) 2002 KL6, with a fixed pole solution at the best fine solutions and varying $\upsilon$ from $-1.4\times10^{-9}$ to 0. Left side: for $\lambda = 151^{\circ}$, $\beta = -90^{\circ}$ the lowest $\chi_\mathrm{red}^{2}$ value is at $\upsilon = -6.83 \times10^{-9}$, with $\chi_\mathrm{red}^{2}=1.10$ (red solid lines). The $3\sigma$ value  corresponds to $\chi_\mathrm{red}^{2}=1.16$ which is reached at $\upsilon = -9.47 \times10^{-9}$ rad d$^{-2}$ and $\upsilon = -4.19 \times10^{-9}$ rad d$^{-2}$(blue dashed lines). Right side: for $\lambda = 330^{\circ}$, $\beta = -89^{\circ}$ the lowest $\chi_\mathrm{red}^{2}$ value is at $\upsilon = -7.02 \times10^{-9}$, with $\chi_\mathrm{red}^{2}=1.10$ (red solid lines). The $3\sigma$ value  corresponds to $\chi_\mathrm{red}^{2}=1.16$ which is reached at $\upsilon = -9.54 \times10^{-9}$ rad d$^{-2}$ and $\upsilon = -4.50 \times10^{-9}$ rad d$^{-2}$(blue dashed lines)}
    \label{fig:154244_yorp_range}
\end{figure*}

As an alternate way of estimating the uncertainty of the YORP effect, the $3\sigma$ method explained in Section \ref{sec:meth} was applied, iterating the values of $\upsilon$ around the best solutions (($P=4.610232$ h, $\lambda = 151^{\circ}$, $\beta = -90^{\circ}$) and ($P=4.610232$ h, $\lambda = 330^{\circ}$, $\beta = -89^{\circ}$)) between $-1.4\times10^{-9}$ and 0 with steps of $0.05\times10^{-9}$. The solutions obtained with this method were $\upsilon = (-6.83\pm2.70)\times10^{-9}$ rad d$^{-2}$ and $\upsilon = (-7.02\pm2.70)\times10^{-9}$ rad d$^{-2}$ respectively (See Figure \ref{fig:154244_yorp_range}) which is in agreement with the obtained solutions.

We present in this work four solutions, two with constant rotation period and the other two with linearly increasing rotation period, all of which fit the data well, but slightly better if the YORP effect is taken into account, but more future observations are needed to confirm these results. One hint of the YORP effect being present is the value of $\epsilon$, in this case of $\epsilon \sim 178^{\circ}$, which is pretty close to the extreme value of 180$^{\circ}$, which is a known consequence of this effect taking place \citep{2013A&A...551A..67H}. If this asteroid is indeed affected by YORP, it would be another addition to the short list of asteroids affected by it. What is even more important is the negative value of $\upsilon$, which implies that unlike all the other asteroids known to be under YORP effect, (154244) 2002 KL6 is being decelerated rather than accelerated. This could be the second occasion an asteroid is reported to be decelerating, as it happened with (25143) Itokawa in \cite{2007A&A...472L...5K}, where it was reported a $\upsilon = (-8.95\pm0.15)\times10^{-8}$ rad d$^{-2}$, but later revised in \cite{2014A&A...562A..48L} where, with a wider temporal span an acceleration of $\upsilon = (3.54\pm0.38)\times10^{-8}$ rad d$^{-2}$ was deemed as the best fitting. Moreover, Itokawa was again studied in \cite{2015MNRAS.450.2104S}, where a positive value of $\upsilon$ in the order of $10^{-7}$ rad d$^{-2}$ was found as the best fitting. The Itokawa case is specially hard to study since that asteroid is a contact binary, which is not the situation for our asteroid.

We also followed the method proposed in \cite{2013MNRAS.430.1376R} where the modulus of the YORP effect can be estimated following the equation $|d\omega/dt| = 1.20^{+1.66}_{-0.86} \times 10^{-2} (a^2\sqrt{1 - e^2}D^2)^{-1}$, where $a$ is the semi major axis in AU, while $e$ is the eccentricity and $D$ is the asteroid's diameter in km. We computed that equation for the smallest and the biggest diameters (0.5 km and 2 km as explained in Section \ref{sec:154244} introduction), being $a=2.307249 AU$, $e=0.548644$, obtaining $\nu=8.1^{+11.2}_{-5.8} \times 10^{-8}$ rad d$^{-2}$ for $D=0.5$ km and $\nu=5.1^{+7.0}_{-3.6} \times 10^{-9}$ rad d$^{-2}$ for $D=2.0$ km. We find that a diameter $>$ 1 km makes our obtained result to be in agreement with this estimation, while a value of $\sim$ 1.7 km makes it the mean estimated value.

To support this claim regarding the negative value, our method to calculate the uncertainties can be useful, since for the 100 computed models created with light-curves with random measures removed the value of $\upsilon$ was consistently negative.

\subsection{(159402) 1999 AP10}\label{sec:159402}

This asteroid also belongs to the Amor group and its diameter has been reported to be 1.20 km \citep{2010AJ....140..770T} and 1.20$\pm$0.29 km \citep{2011AJ....141..109M}, while its spectral class is reported as: Sq \citep{2014Icar..228..217T}, Sw \citep{2019Icar..324...41B} and in the S complex \citep{2021A&A...656A..89H}. 
Radar observations were obtained with Arecibo on October 2009 and with Goldstone on October 2020 but there is no report of the results published published yet\footnote{\url{http://mel.epss.ucla.edu/radar/object/info.php?search=159402}}.

As the other asteroids this one's period was already studied, with the following results: $P=7.908 \pm 0.001$ h \citep{2010MPBu...37...83F}, $P=7.911 \pm 0.001$ h \citep{2018PASJ...70..114H}, $P=7.9219 \pm 0.0003$ h \citep{2021MPBu...48...30W}, $P=7.9186 \pm 0.0004$ h and $P=7.9219 \pm 0.0003$ h \citep{2021MPBu...48..170W} and $P=7.92 \pm 0.01$ h, $P=7.922 \pm 0.004$ h and $P=7.919 \pm 0.005$ h \citep{2023MPBu...50...74S}. As for the shape model, the one presented on this work is the first. 

\begin{figure}
    \centering
    \includegraphics[width=\linewidth,keepaspectratio]{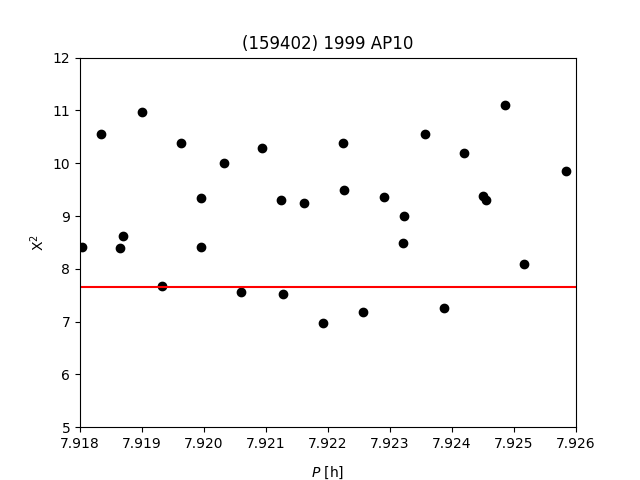}
    \caption{Period search tool output plot for (159402) 1999 AP10. This search was made in a interval from 7.918 h to 7.926 h, with a coefficient $p$ of 0.8. Each obtained period is represented as a black dot, with the red line representing a 10\% threshold from the lowest $\chi^{2}$ obtained. The presence of several values under this threshold is an indicator that more future observations are needed to refine this value.}
    \label{fig:159402_period}
\end{figure}

For finding the best fitting period to our data set of light-curves (46 light-curves obtained from ALCDEF and 3 new presented in this work), which cover a temporal span from 21 September 2009 to 22 January 2021, the period search tool was applied in an interval between 7.918 to 7.926 h, being the best fitting period $P=7.921915$ h (see Figure \ref{fig:159402_period} for a graphical representation of the obtained periods).

\begin{figure}
    \centering
    \includegraphics[width=\linewidth,keepaspectratio]{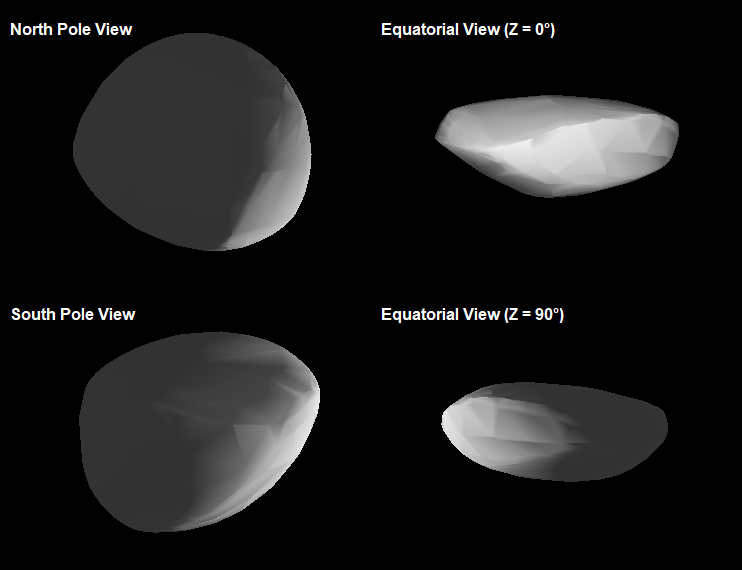}
    \caption{Shape model of (159402) 1999 AP10 obtained with a constant rotation period.. Left top: North Pole View (Y axis = 0$^{\circ}$). Left bottom: South Pole View (Y axis = 180$^{\circ}$). Right top and bottom: Equatorial Views with Z axis rotated 0$^{\circ}$ and 90$^{\circ}$.}
    \label{fig:159402_shape_model}
\end{figure}

\begin{figure*}
    \centering
    \begin{subfigure}{0.49\textwidth}
        \centering
        \includegraphics[width=\textwidth]{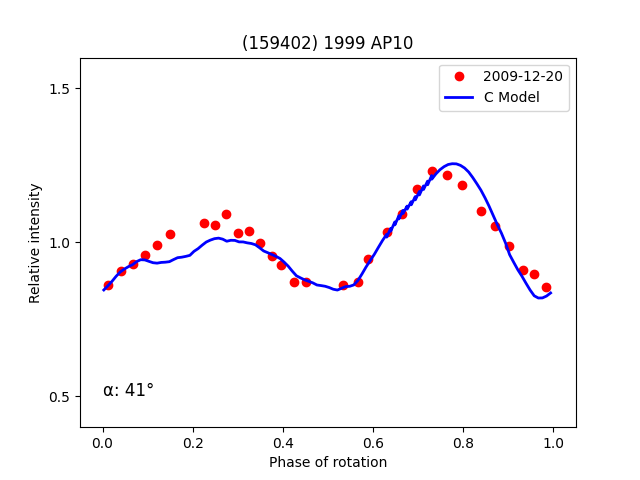}
    \end{subfigure}
    \begin{subfigure}{0.49\textwidth}
        \centering
        \includegraphics[width=\textwidth]{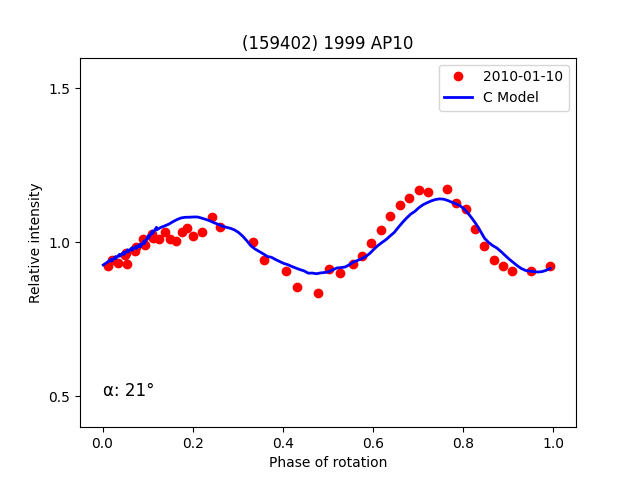}
    \end{subfigure}
    \begin{subfigure}{0.49\textwidth}
        \centering
        \includegraphics[width=\textwidth]{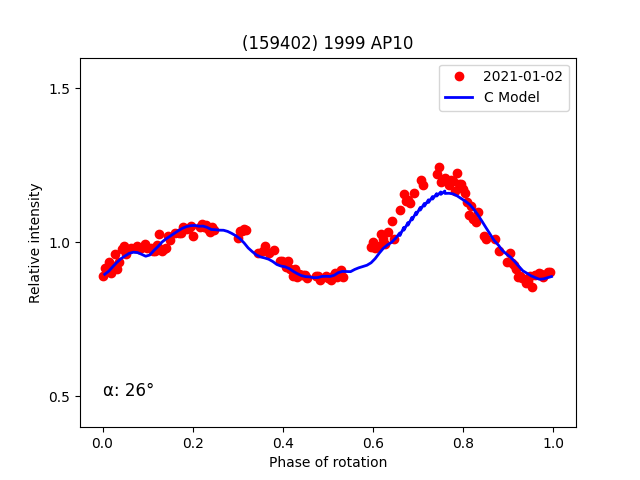}
    \end{subfigure}
    \begin{subfigure}{0.49\textwidth}
        \centering
        \includegraphics[width=\textwidth]{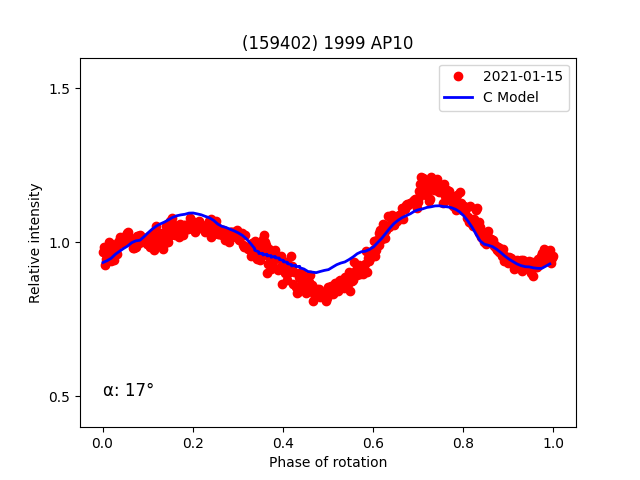}
    \end{subfigure}
    \caption{Fit between three of the ALCDEF database light-curves and one (15 January 2020) of the new light-curves from (159402) 1999 AP10 presented in this paper and the best-fitting constant period model (C Model). The data is plotted as red dots for each observation, while the model is plotted as a solid blue line. The geometry is described by its solar phase angle $\alpha$.}
    \label{fig:fit_159402}
\end{figure*}

Adopting this period, a medium search was conducted, resulting in: $P=7.921919$ h, $\lambda = 50^{\circ}$, $\beta = -60^{\circ}$, $\chi_\mathrm{red}^{2}=1.85$ (see Figure \ref{fig:159402_plot} for a graphical representation of the solutions) as the best fit. The fine search around the best fitting medium search solution was: $P=7.921917$ h, $\lambda = 49^{\circ}$, $\beta = -60^{\circ}$, $\chi_\mathrm{red}^{2}=1.80$ and $\epsilon = 155^{\circ}$, which again implies retrograde rotation. The best fitting computed shape model is shown in Figure \ref{fig:159402_shape_model}, with Figures \ref{fig:fit_159402} and \ref{fig:IAC_fit_159402} showing a graphical representation of the fit between the light-curves and the shape model.

The uncertainties for the solution were obtained randomly removing 25\% of the measures from the main data set ($\sim$ 8000 measures) to create the 100 subsets, repeating the fine for each dataset around the best fitting medium search. The results obtained were: $P=7.921917 \pm 0.000005$ h $\lambda = 49^{\circ} \pm 2^{\circ}$, $\beta = -60^{\circ} \pm 3^{\circ}$ and $\epsilon = 155^{\circ} \pm 2^{\circ}$.
Since the temporal span is relatively large enough ($\sim$ 14 years), as for (154244) 2002 KL6, we conducted a search with the YORP code to see if it could be detected with our data, but the attempt was not conclusive, so with the data used in this work we rule out this possibility for this asteroid.

\section{Conclusions}\label{sec:con}

In this work we present 38 new light-curves obtained with five different telescopes located at Teide Observatory (Tenerife, Spain), along with new the derived shape models and rotation state parameters for NEAs (7335) 1989 JA, (7822) 1991 CS, (154244) 2002 KL6 and (159402) 1999 AP10.

For (7335) 1989 JA, a rotation period of $P=2.590432 \pm 0.000391$ h is found, which is in agreement with previous results. Additionally, a pole solution of $\lambda = 243^{\circ} \pm 17^{\circ}$, $\beta = -61^{\circ} \pm 6^{\circ}$ and $\epsilon = 147^{\circ} \pm 8^{\circ}$ is found to be the best fitting. Additionaly, at least four mutual events of this binary system may have been identified in our dataset.

For (7822) 1991 CS, the period found as the best fitting was $P=2.390157 \pm 0.000002$ h, again, in agreement with previous results. The best fitting pole solution obtained from the data is $\lambda = 242^{\circ} \pm 9^{\circ}$, $\beta = -57^{\circ} \pm 7^{\circ}$ and $\epsilon = 175^{\circ} \pm 7^{\circ}$.

For (159402) 1999 AP10, we found a period of $P=7.921917 \pm 0.000005$ h, in agreement with previously reported results, the best fitting pole solution is: $\lambda = 49^{\circ} \pm 2^{\circ}$, $\beta = -60^{\circ} \pm 3^{\circ}$ and $\epsilon = 155^{\circ} \pm 2^{\circ}$.

For (154244) 2002 KL6, a period of $P=4.610235\pm0.000001$ h is found, with 2 pole solutions yielding the same fit to the data, $\lambda = 334^{\circ} \pm 28^{\circ}$, $\beta = -90^{\circ} \pm 4^{\circ}$, $\epsilon = 176^{\circ} \pm 3^{\circ}$ and $\lambda = 153^{\circ} \pm 21^{\circ}$, $\beta = -90^{\circ} \pm 4^{\circ}$, $\epsilon = 177^{\circ} \pm 3^{\circ}$. Since the time span is large enough, a search for detecting the YORP effect was made, and another 2 solutions were found to fit the data slightly better than the constant period solutions. The initial period obtained taking into account the YORP effect was $P=4.610232\pm0.000001$ h, obtaining: $\lambda = 333^{\circ} \pm 18^{\circ}$, $\beta = -89^{\circ} \pm 2^{\circ}$, $\upsilon = (-7.14\pm1.93)\times10^{-9}$ rad d$^{-2}$, $\epsilon = 177^{\circ} \pm 2^{\circ}$ and $\lambda = 152^{\circ} \pm 15^{\circ}$, $\beta = -90^{\circ} \pm 2^{\circ}$, $\upsilon = (-7.12\pm1.65)\times10^{-9}$ rad d$^{-2}$ and $\epsilon = 177^{\circ} \pm 2^{\circ}$. The YORP detection can not be ruled out since, as previously mentioned, the fit is slightly better, but also the uncertainties are lower than with a constant period. It is worth mentioning that in the 100 models computed to obtain the uncertainties, in all of them, the value of $\upsilon$ was negative. If so, (154244) 2002 KL6 would be the very first detection of YORP decelerating an asteroid.

\section*{Acknowledgements}

We thank Dr. Josef Ďurech for providing us the inversion code that includes the Yarkovsky–O’Keefe–Radzievskii–Paddack (YORP) acceleration and for his advises in using the inversion codes.

The work has been funded by HUNOSA through the collaboration agreement with reference SV-21-HUNOSA-2.
JL, MRA and MS-R acknowledge support from the Agencia Estatal de Investigacion del Ministerio de Ciencia e Innovacion (AEI-MCINN) under grant "Hydrated Minerals and Organic Compounds in Primitive Asteroids" with reference PID2020-120464GB-100. 

This article includes observations made in the Two-meter Twin Telescope (TTT) sited at the Teide Observatory of the IAC, that Light Bridges operates in the Island of Tenerife, Canary Islands (Spain). The 
Observing Time Rights (DTO) used for this research were provided by IAC. This article also includes observations made with the Telescopio IAC80 and TAR2 telescopes operated on the island of Tenerife by the Instituto de Astrof\'{\i}sica de Canarias in the Spanish Observatorio del Teide. This work uses data obtained from the Asteroid Lightcurve Data Exchange Format (ALCDEF) data base, which is supported by funding from NASA grant 80NSSC18K0851.

This work uses software MPO LC Invert from Brian Warner to plot the asteroids shapes as seen in Figures \ref{fig:7335_shape_model}, \ref{fig:7822_shape_model}, \ref{fig:154244_150_ny_shape_model}, \ref{fig:154244_330_ny_shape_model}, \ref{fig:154244_155_y_shape_model}, \ref{fig:154244_335_y_shape_model}, \ref{fig:159402_shape_model}.

\section*{Data Availability}

The data underlying this article will be shared on reasonable request to the corresponding author.



\bibliographystyle{mnras}
\bibliography{bibliography} 




\appendix

\section{Summary of archival light-curves used in this work}

\begin{table*}

\caption{Archival observations from ALCDEF database for (7822) 1991 CS. The table includes the date, the starting and end time (UT) of the observations, the phase angle ($\alpha$), the heliocentric ($r$) and geocentric ($\Delta$) distances, phase angle bisector longitude (PABLon) and latitude (PABLat) of the asteroid at the time of observation. \textbf{References:} WAR16: \protect\cite{2016MPBu...43...66W}; KLI16: \protect\cite{2016MPBu...43..234K}; WAR16b: \protect\cite{2016MPBu...43..240W}; WAR22: \protect\cite{2022MPBu...49...16W}}
\label{tab:7822 archive}
\begin{tabular}{lllrrrrrl}
\hline
 Date        & UT (start)   & UT (end)    &   $\alpha (^\circ)$ &   $r$ (au) &   $\Delta$ (au) &   PABLon (deg) &   PABLat (deg) & Reference   \\
\hline
 2015-Aug-06 & 04:01:56.842 & 0:52:43.392 &               63.01 &     1.1237 &          0.3479 &       308.36  &        49.13 & WAR16           \\
 2015-Aug-08 & 04:14:55.738 & 0:19:17.011 &               63.35 &     1.1184 &          0.3305 &       308.35  &        48.26 & WAR16           \\
 2016-Feb-27 & 03:50:07.584 & 5:33:19.526 &               21.96 &     1.1496 &          0.1742 &       144.78  &        -7.08 & KLI16           \\
 2016-Feb-27 & 05:56:31.344 & 7:40:01.085 &               21.98 &     1.1498 &          0.1745 &       144.75  &        -6.91 & KLI16           \\
 2016-Feb-27 & 07:48:35.856 & 8:49:28.416 &               22.01 &     1.1500 &          0.1747 &       144.73  &        -6.76 & KLI16           \\
 2016-Feb-28 & 02:49:36.106 & 4:27:05.818 &               22.43 &     1.1521 &          0.1774 &       144.51  &        -5.26 & KLI16           \\
 2016-Feb-28 & 05:56:29.098 & 7:34:56.957 &               22.52 &     1.1524 &          0.1779 &       144.47  &        -5.02 & KLI16           \\
 2016-Feb-29 & 04:56:44.880 & 7:11:02.717 &               23.41 &     1.1549 &          0.1818 &       144.25   &        -3.25 & KLI16           \\
 2016-Feb-29 & 07:20:17.578 & 9:34:10.186 &               23.52 &     1.1551 &          0.1823 &       144.23  &        -3.08 & KLI16           \\
 2016-Mar-15 & 02:54:12.326 & 9:45:46.310 &               42.72 &     1.1915 &          0.2959 &       145.46  &        15.97 & WAR16b           \\
 2016-Mar-15 & 09:19:34.003 & 0:19:18.134 &               42.94 &     1.1921 &          0.2986 &       145.54  &        16.19 & WAR16b           \\
 2016-Mar-16 & 03:04:24.557 & 0:21:56.506 &               43.55 &     1.1938 &          0.3057 &       145.75  &        16.81   & WAR16b           \\
 2016-Mar-17 & 03:04:45.206 & 9:49:01.834 &               44.31 &     1.1962 &          0.3155 &       146.05  &        17.61   & WAR16b           \\
 2021-Aug-22 & 09:17:36.758 & 0:55:39.389 &               78.62 &     1.0303 &          0.1508 &        16.86 &        -2.89  & WAR22           \\
 2021-Aug-22 & 11:04:01.459 & 2:12:16.214 &               78.68 &     1.0301 &          0.1506 &        16.94 &        -3.11 & WAR22           \\
 2021-Aug-23 & 09:28:48.518 & 0:02:20.256 &               79.45 &     1.0277 &          0.1481 &        17.90 &        -6.00 & WAR22           \\
 2021-Aug-23 & 10:03:56.333 & 1:37:59.117 &               79.47 &     1.0277 &          0.1481 &        17.93  &        -6.08 & WAR22           \\
 2021-Aug-23 & 11:40:26.429 & 2:12:53.885 &               79.53 &     1.0275 &          0.1479 &        17.99  &        -6.30 & WAR22           \\
\hline
\end{tabular}
\end{table*}

\begin{table*}
\caption{Archival observations from ALCDEF database for (154244) 2002 KL6. The table includes the date, the starting and end time (UT) of the observations, the phase angle ($\alpha$), the heliocentric ($r$) and geocentric ($\Delta$) distances, phase angle bisector longitude (PABLon) and latitude (PABLat) of the asteroid at the time of observation. \textbf{References:} SKI19: \protect\cite{2019MPBu...46..458S}; WAR16: \protect\cite{2016MPBu...43..343W};WAR17: \protect\cite{2017MPBu...44..206W}; WAR23: \protect\cite{2023MPBu...50..304W}; BEN23$^*$.}
\label{tab:154244 archive}
\resizebox{0.95\textwidth}{!}{\begin{tabular}{lllrrrrrl}
\hline
 Date        & UT (start)   & UT (end)    &   $\alpha (^\circ)$ &   $r$ (au) &   $\Delta$ (au) &   PABLon (deg) &   PABLat (deg) & Reference   \\
\hline
 2009-Jun-18 & 07:46:31.267 & 0:59:00.442 &               44.18 &     1.1258 &          0.1618 &       293.14  &        11.86 & SKI19           \\
 2009-Jun-19 & 07:14:08.477 & 1:00:32.803 &               45.50 &     1.1214 &          0.1592 &       294.72  &        12.17 & SKI19           \\
 2009-Jun-22 & 07:15:02.736 & 1:06:26.784 &               49.77 &     1.1084 &          0.1529 &       299.77  &        13.09 & SKI19           \\
 2009-Jun-23 & 06:56:36.384 & 1:03:32.602 &               51.23 &     1.1043 &          0.1514 &       301.48  &        13.37 & SKI19           \\
 2009-Jul-01 & 07:16:32.678 & 0:49:15.686 &               62.96 &     1.0756 &          0.1488 &       315.88  &        14.96 & SKI19           \\
 2009-Jul-13 & 07:57:52.531 & 1:11:35.318 &               74.71 &     1.0482 &          0.1712 &       335.81  &        14.58 & SKI19           \\
 2009-Sep-24 & 05:58:31.958 & 2:13:02.611 &               35.17 &     1.2889 &          0.3788 &        31.23 &         2.70 & SKI19           \\
 2009-Sep-25 & 05:54:10.598 & 1:15:23.328 &               34.15 &     1.2955 &          0.3815 &        31.47 &         2.60 & SKI19           \\
 2009-Sep-26 & 05:45:44.294 & 0:57:45.965 &               33.13 &     1.3021 &          0.3842 &        31.70 &         2.50 & SKI19           \\
 2009-Oct-16 & 04:20:20.083 & 2:12:35.222 &               12.38 &     1.4421 &          0.4607 &        34.65 &         0.75 & SKI19           \\
 2009-Oct-19 & 03:59:28.838 & 2:16:08.890 &                9.40 &     1.4638 &          0.4771 &        34.93 &         0.53 & SKI19           \\
 2009-Oct-20 & 03:59:09.917 & 0:33:29.174 &                8.42 &     1.4711 &          0.4830 &        35.02  &         0.46 & SKI19           \\
 2009-Oct-22 & 05:35:59.626 & 1:57:43.747 &                6.42 &     1.4863 &          0.4957 &        35.21 &         0.32 & SKI19           \\
 2016-Jun-10 & 04:19:43.018 & 0:06:57.427 &               13.78 &     1.2161 &          0.2080 &       253.78  &         7.84 & WAR17           \\
 2016-Jun-12 & 04:12:02.160 & 0:03:44.669 &               15.48 &     1.2044 &          0.1974 &       254.72   &         8.40 & WAR17           \\
 2016-Jun-13 & 04:10:00.941 & 0:01:55.027 &               16.36 &     1.1986 &          0.1922 &       255.21  &         8.69 & WAR17           \\
 2016-Jun-14 & 04:23:17.981 & 0:01:11.654 &               17.26 &     1.1928 &          0.1870 &       255.71  &         9.00 & WAR17           \\
 2016-Jun-14 & 05:42:15.466 & 0:34:31.037 &               17.31 &     1.1925 &          0.1868 &       255.74  &         9.02 & WAR17           \\
 2016-Jun-15 & 03:46:43.421 & 0:23:16.166 &               18.14 &     1.1873 &          0.1821 &       256.21  &         9.31 & WAR17           \\
 2016-Jun-15 & 04:14:45.197 & 9:58:55.574 &               18.16 &     1.1872 &          0.1820 &       256.22  &         9.32  & WAR17           \\
 2016-Jun-15 & 19:56:03.754 & 0:24:33.206 &               18.75 &     1.1835 &          0.1788 &       256.58  &         9.53 & WAR17           \\
 2016-Jun-16 & 04:11:24.835 & 9:58:15.917 &               19.08 &     1.1816 &          0.1771 &       256.75  &         9.65 & WAR17           \\
 2016-Jun-17 & 04:11:48.422 & 9:32:10.090 &               20.02 &     1.1761 &          0.1723 &       257.29  &         9.99 & WAR17           \\
 2016-Jun-20 & 04:40:35.213 & 9:21:01.354 &               22.90 &     1.1600 &          0.1582 &       259.03  &        11.12 & WAR17           \\
 2016-Jun-21 & 04:14:31.286 & 6:36:26.266 &               23.86 &     1.1550 &          0.1537 &       259.64  &        11.51 & WAR17           \\
 2016-Jun-22 & 05:50:44.794 & 9:37:40.483 &               24.90 &     1.1495 &          0.1490 &       260.32  &        11.96 & WAR17           \\
 2016-Jun-22 & 20:59:16.973 & 0:35:53.347 &               25.51 &     1.1464 &          0.1462 &       260.75  &        12.24  & WAR17           \\
 2016-Jun-23 & 21:20:15.130 & 0:06:28.454 &               26.52 &     1.1414 &          0.1419 &       261.43  &        12.70 & WAR17           \\
 2016-Jun-23 & 21:14:35.750 & 1:30:39.398 &               26.52 &     1.1414 &          0.1419 &       261.43   &        12.70 & WAR17           \\
 2016-Jun-26 & 20:53:26.362 & 2:39:12.614 &               29.53 &     1.1273 &          0.1295 &       263.62  &        14.18 & WAR17           \\
 2016-Jun-27 & 04:13:32.275 & 9:45:25.402 &               29.86 &     1.1259 &          0.1283 &       263.85  &        14.34 & WAR17           \\
 2016-Jul-11 & 23:12:49.968 & 3:31:34.896 &               46.13 &     1.0707 &          0.0804 &       281.86  &        25.19 & WAR16           \\
 2016-Jul-14 & 00:18:31.968 & 0:31:04.512 &               48.71 &     1.0652 &          0.0762 &       285.99   &        27.02 & WAR16           \\
 2016-Sep-09 & 05:54:59.414 & 2:12:25.546 &               41.03 &     1.1493 &          0.1998 &        13.82 &         7.44 & WAR17           \\
 2016-Sep-10 & 05:52:00.739 & 2:27:04.061 &               39.91 &     1.1544 &          0.2033 &        14.24 &         7.18 & WAR17           \\
 2016-Sep-11 & 05:49:48.288 & 2:27:19.526 &               38.79 &     1.1595 &          0.2069 &        14.64 &         6.93 & WAR17           \\
 2016-Oct-14 & 03:55:23.722 & 1:43:21.821 &                3.27 &     1.3657 &          0.3691 &        23.42 &         1.49  & WAR17           \\
 2016-Oct-18 & 03:30:25.632 & 1:12:25.171 &                1.44 &     1.3938 &          0.3977 &        24.27 &         1.10 & WAR17           \\
 2016-Oct-19 & 03:22:42.010 & 1:20:27.542 &                1.95 &     1.4009 &          0.4052 &        24.48 &         1.01 & WAR17           \\
 2023-Jun-13 & 04:31:47.568 & 9:13:35.357 &               22.47 &     1.2199 &          0.2251 &       247.28  &         7.40 & WAR23           \\
 2023-Jun-14 & 04:10:41.462 & 9:07:37.315 &               23.53 &     1.2140 &          0.2205 &       247.61  &         7.63 & WAR23           \\
 2023-Jun-15 & 04:10:29.021 & 9:06:30.096 &               24.61 &     1.2081 &          0.2160 &       247.96  &         7.87 & WAR23           \\
 2023-Jul-31 & 19:41:40.531 & 2:24:30.211 &               63.55 &     1.0427 &          0.0660 &       298.23  &        33.89 & BEN23           \\
 2023-Jul-31 & 22:28:24.355 & 1:08:54.413 &               63.57 &     1.0427 &          0.0658 &       298.59  &        33.96 & BEN23           \\
 2023-Aug-01 & 01:11:53.952 & 2:14:37.363 &               63.58 &     1.0426 &          0.0657 &       298.93  &        34.01 & BEN23           \\
 2023-Aug-02 & 20:34:03.245 & 3:15:55.296 &               63.43 &     1.0419 &          0.0643 &       304.85  &        34.64 & BEN23           \\
 2023-Aug-02 & 23:20:29.530 & 2:01:24.730 &               63.44 &     1.0419 &          0.0642 &       305.23  &        34.67 & BEN23           \\
 2023-Aug-03 & 19:24:38.938 & 2:04:38.323 &               63.29 &     1.0418 &          0.0638 &       308.07  &        34.77 & BEN23           \\
 2023-Aug-03 & 22:10:45.178 & 0:51:07.805 &               63.28 &     1.0418 &          0.0638 &       308.46  &        34.78 & BEN23           \\
 2023-Aug-04 & 00:55:17.328 & 1:54:43.488 &               63.27 &     1.0418 &          0.0638 &       308.83  &        34.79  & BEN23           \\
 2023-Aug-04 & 19:18:47.376 & 1:38:54.614 &               63.08 &     1.0418 &          0.0636 &       311.47  &        34.75 & BEN23           \\
 2023-Aug-04 & 21:45:37.670 & 3:54:06.192 &               63.08 &     1.0418 &          0.0636 &       311.81  &        34.75 & BEN23           \\
 2023-Aug-05 & 00:32:54.154 & 1:19:45.005 &               63.06 &     1.0418 &          0.0636 &       312.19  &        34.74 & BEN23           \\
\hline
\end{tabular}}\\
\flushleft\footnotesize{\quad \quad \quad $^*$ Data available on ALCDEF with no manuscript associated.}
\end{table*}

\begin{table*}

\caption{Archival observations from ALCDEF database for (159402) 1999 AP10. The table includes the date, the starting and end time (UT) of the observations, the phase angle ($\alpha$), the heliocentric ($r$) and geocentric ($\Delta$) distances, phase angle bisector longitude (PABLon) and latitude (PABLat) of the asteroid at the time of observation. \textbf{References:} SKI23: \protect\cite{2023MPBu...50...74S}; WAR21: \protect\cite{2021MPBu...48..170W}; ASE20$^*$ and WAR21$^*$.}
\label{tab:159402 archive}
\captionsetup{singlelinecheck=false}
\begin{tabular}{lllrrrrrl}
\hline
 Date        & UT (start)   & UT (end)    &   $\alpha (^\circ)$ &   $r$ (au) &   $\Delta$ (au) &   PABLon (deg) &   PABLat (deg) & Reference   \\
\hline
 2009-Sep-21 & 05:47:15.446 & 8:21:51.696 &               20.15 &     1.1578 &          0.1655 &       344.99  &        -1.36 & SKI23           \\
 2009-Sep-22 & 03:30:04.637 & 8:32:34.166 &               20.99 &     1.1525 &          0.1610 &       345.37   &        -0.92 & SKI23           \\
 2009-Sep-23 & 03:05:16.051 & 8:14:53.606 &               21.93 &     1.1467 &          0.1562 &       345.80  &        -0.42 & SKI23           \\
 2009-Oct-10 & 02:41:03.840 & 5:38:15.965 &               42.78 &     1.0632 &          0.0905 &       356.74  &        15.11 & SKI23           \\
 2009-Oct-11 & 03:05:09.139 & 5:36:37.210 &               44.44 &     1.0593 &          0.0880 &       357.83   &        16.61 & SKI23           \\
 2009-Oct-12 & 03:19:59.232 & 6:27:10.368 &               46.14 &     1.0555 &          0.0856 &       359.00  &        18.17 & SKI23           \\
 2009-Dec-18 & 05:56:20.285 & 3:04:33.744 &               42.95 &     1.1520 &          0.2496 &       107.78  &        23.51 & SKI23           \\
 2009-Dec-20 & 05:32:28.205 & 3:08:29.875 &               40.84 &     1.1638 &          0.2568 &       108.66  &        22.88 & SKI23           \\
 2010-Jan-10 & 03:42:35.107 & 2:12:57.341 &               20.88 &     1.3040 &          0.3516 &       115.35  &        16.84 & SKI23           \\
 2020-Aug-24 & 06:53:04.790 & 0:38:33.994 &               11.30 &     1.3504 &          0.3485 &       337.35  &        -7.36 & WAR21           \\
 2020-Aug-25 & 06:39:30.038 & 0:13:39.878 &               10.85 &     1.3432 &          0.3405 &       337.69  &        -7.25 & WAR21           \\
 2020-Aug-26 & 06:43:30.835 & 9:15:31.565 &               10.41 &     1.3359 &          0.3326 &       338.04  &        -7.14 & WAR21           \\
 2020-Aug-27 & 06:47:14.352 & 0:07:50.822 &                9.98 &     1.3286 &          0.3248 &       338.39  &        -7.02 & WAR21           \\
 2020-Aug-28 & 07:10:01.373 & 0:29:17.837 &                9.56 &     1.3213 &          0.3170 &       338.74   &        -6.89 & WAR21           \\
 2020-Aug-29 & 08:12:21.888 & 0:26:05.942 &                9.15 &     1.3138 &          0.3091 &       339.11  &        -6.75 & WAR21           \\
 2020-Aug-30 & 06:15:26.640 & 0:21:31.622 &                8.82 &     1.3072 &          0.3022 &       339.43  &        -6.63 & WAR21           \\
 2020-Aug-31 & 06:11:02.861 & 0:13:18.365 &                8.49 &     1.3001 &          0.2949 &       339.79  &        -6.48 & WAR21           \\
 2020-Sep-05 & 06:13:05.290 & 9:38:00.787 &                7.59 &     1.2653 &          0.2600 &       341.60  &        -5.61 & WAR21           \\
 2020-Sep-06 & 05:33:39.485 & 9:54:13.219 &                7.61 &     1.2586 &          0.2535 &       341.96  &        -5.41 & WAR21           \\
 2020-Sep-07 & 05:30:19.987 & 9:52:44.918 &                7.70 &     1.2518 &          0.2469 &       342.34  &        -5.19 & WAR21           \\
 2020-Sep-09 & 05:11:10.435 & 9:54:15.466 &                8.12 &     1.2385 &          0.2342 &       343.10  &        -4.73 & WAR21           \\
 2020-Sep-13 & 06:12:07.142 & 9:21:20.880 &                9.81 &     1.2121 &          0.2097 &       344.70  &        -3.61 & WAR21           \\
 2020-Sep-14 & 04:38:54.384 & 9:35:50.150 &               10.33 &     1.2062 &          0.2042 &       345.09  &        -3.32 & WAR21           \\
 2020-Sep-15 & 04:53:54.154 & 9:41:36.787 &               10.95 &     1.1998 &          0.1985 &       345.51  &        -2.98 & WAR21           \\
 2020-Sep-16 & 05:48:14.976 & 9:39:53.194 &               11.62 &     1.1933 &          0.1927 &       345.95  &        -2.62 & WAR21           \\
 2020-Sep-17 & 04:15:18.461 & 9:34:45.091 &               12.27 &     1.1876 &          0.1875 &       346.35  &        -2.27 & WAR21           \\
 2020-Oct-04 & 20:32:34.685 & 1:30:10.339 &               29.49 &     1.0928 &          0.1082 &       356.57  &         9.36 & ASE20           \\
 2020-Oct-05 & 18:55:17.846 & 0:48:29.837 &               30.72 &     1.0887 &          0.1051 &       357.36  &        10.36  & ASE20           \\
 2020-Oct-06 & 00:46:24.067 & 2:06:46.051 &               31.08 &     1.0876 &          0.1043 &       357.56  &        10.63 & ASE20           \\
 2020-Oct-07 & 00:31:03.216 & 2:37:14.966 &               32.45 &     1.0834 &          0.1012 &       358.45  &        11.76 & ASE20           \\
 2020-Oct-07 & 03:16:28.070 & 5:16:03.331 &               32.63 &     1.0829 &          0.1009 &       358.56  &        11.90 & ASE20           \\
 2020-Oct-08 & 19:30:47.434 & 1:03:50.515 &               35.07 &     1.0760 &          0.0961 &         0.23 &        13.95 & ASE20           \\
 2020-Oct-12 & 03:30:09.302 & 5:47:34.627 &               40.54 &     1.0635 &          0.0883 &         4.13 &        18.51 & ASE20           \\
 2020-Oct-14 & 23:50:50.496 & 4:50:36.211 &               45.67 &     1.0539 &          0.0836 &         8.31 &        22.76 & ASE20           \\
 2020-Oct-16 & 16:59:47.357 & 9:31:30.720 &               48.85 &     1.0487 &          0.0819 &        11.28 &        25.36   & ASE20           \\
 2021-Jan-01 & 06:31:41.405 & 3:49:19.056 &               27.02 &     1.2478 &          0.3082 &       111.37  &        19.03 & WAR21           \\
 2021-Jan-02 & 04:13:23.030 & 2:02:16.512 &               26.22 &     1.2539 &          0.3125 &       111.65  &        18.78   & WAR21           \\
 2021-Jan-04 & 04:02:25.786 & 6:57:57.168 &               24.50 &     1.2675 &          0.3223 &       112.24  &        18.22 & WAR21           \\
\hline
\end{tabular}\\
\flushleft\footnotesize{\quad \quad \quad \quad \quad \quad $^*$ Data available on ALCDEF with no manuscript associated.}
\end{table*}

\section{Statistical Plot of Pole solutions}\label{sec:poleplot}

\begin{figure*}
    \centering
    \includegraphics[width=\textwidth]{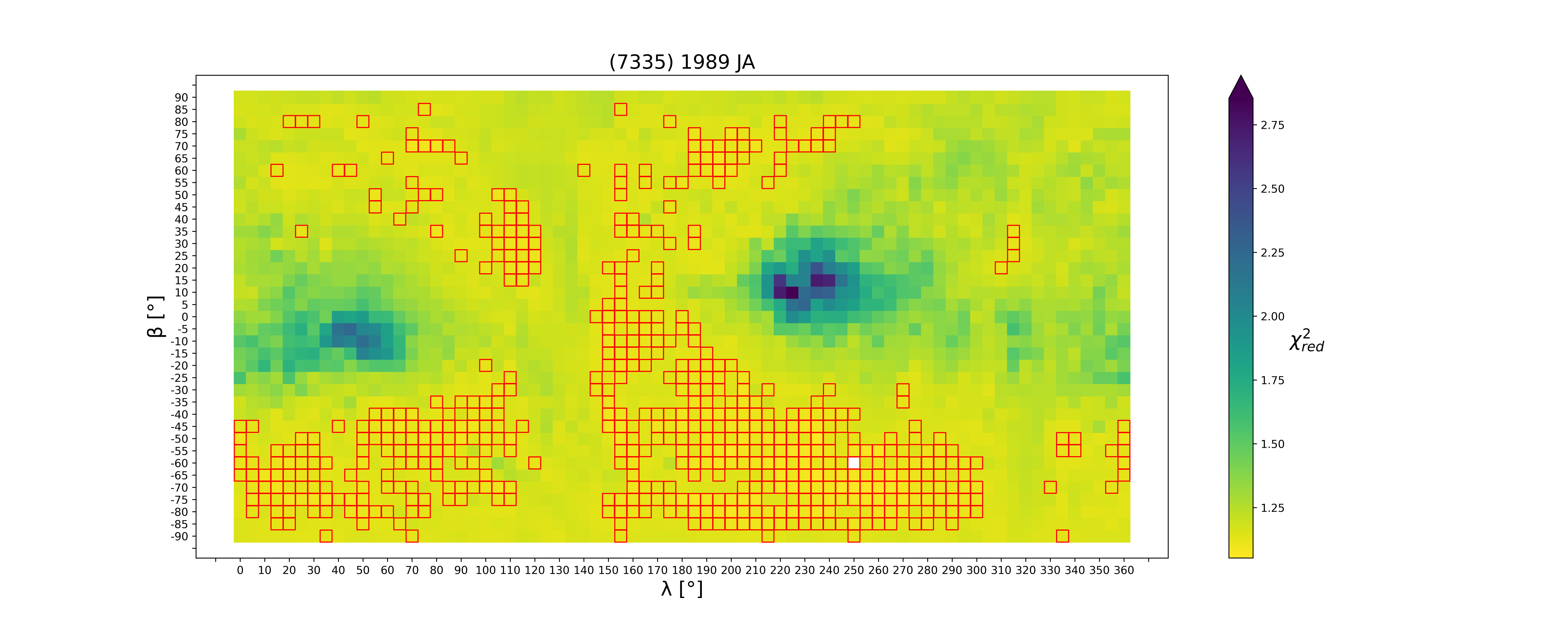}
    \caption{Pole solutions statistical quality for (7335) 1989 JA applying the constant period code. The solutions are shaded by its $\chi_\mathrm{red}^{2}$ value, with the best solution obtained represented as a white square ($\lambda=250 ^{\circ}, \beta=-60 ^{\circ}$) with $\chi_\mathrm{red}^{2}=1.05$ (normalized for the 3312 data points). The solutions highlighted with a red border are within a margin of 6.9\% (3$\sigma$) of the best obtained solution.}
    \label{fig:7335_plot}
\end{figure*}

\begin{figure*}
    \centering
    \includegraphics[width=\textwidth]{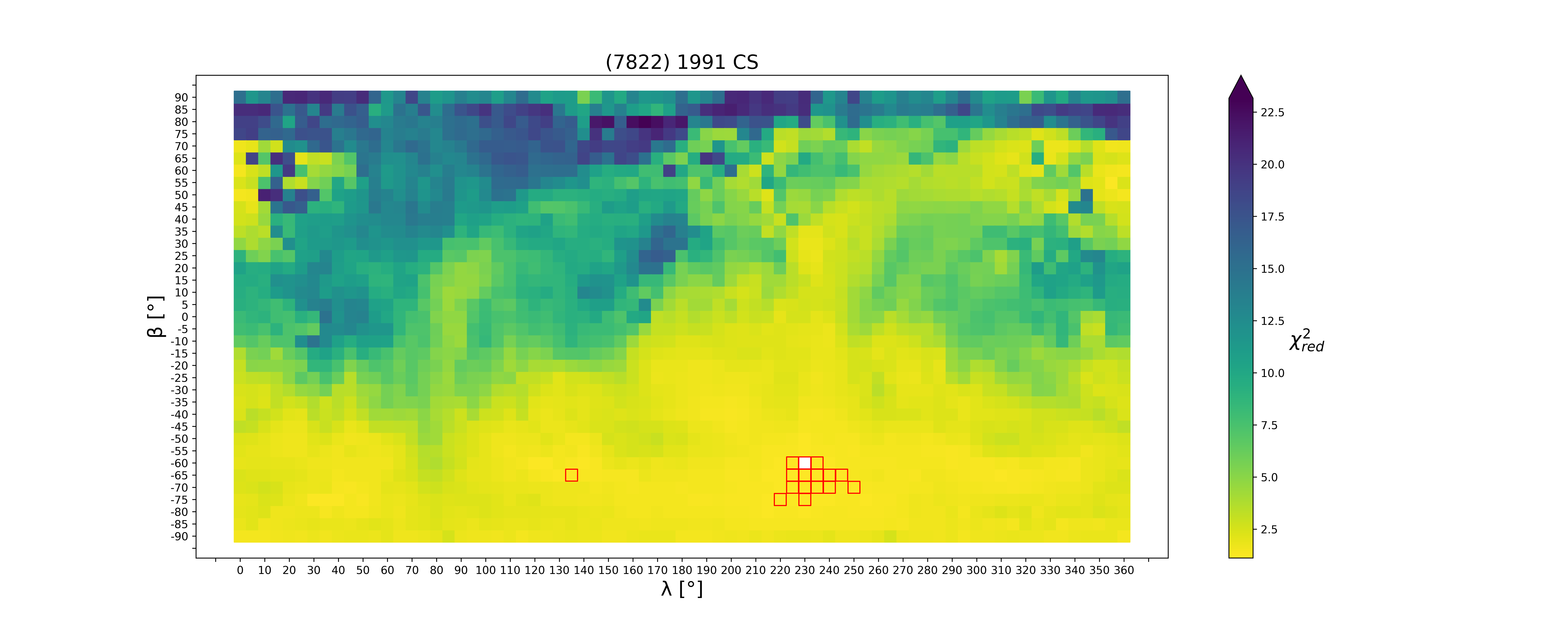}
    \caption{Pole solutions statistical quality for (7822) 1991 CS applying the constant period code. The solutions are shaded by its $\chi_\mathrm{red}^{2}$ value, with the best solution obtained represented as a white square ($\lambda=230 ^{\circ}, \beta=-60 ^{\circ}$) with $\chi_\mathrm{red}^{2}=1.12$ (normalized for the 3035 data points). The solutions highlighted with a red border are within a margin of 7.3\% (3$\sigma$) of the best obtained solution.}
    \label{fig:7822_plot}
\end{figure*}

\begin{figure*}
    \centering
    \includegraphics[width=\textwidth]{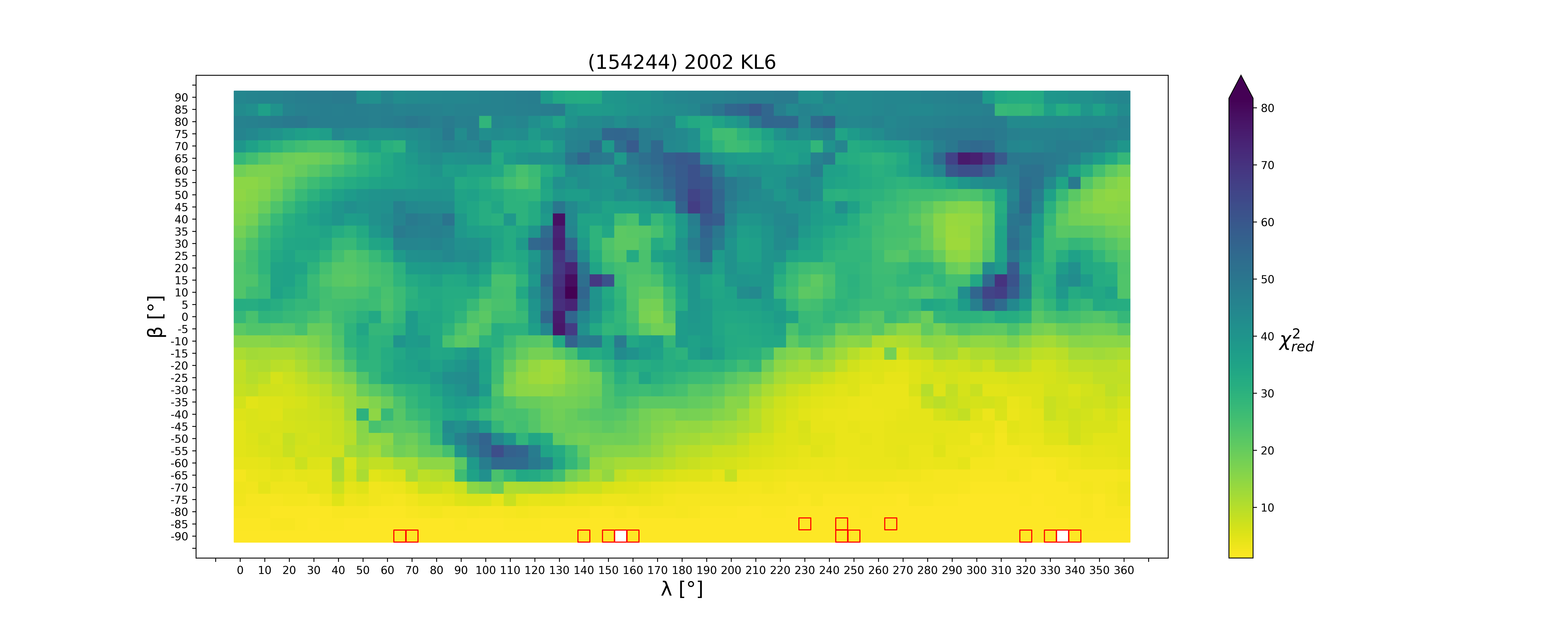}
    \caption{Pole solutions statistical quality for (154244) 2002 KL6 applying the constant period code. The solutions are shaded by its $\chi_\mathrm{red}^{2}$ value, with the best solutions obtained represented as white squares ($\lambda=330 ^{\circ}, \beta=-90 ^{\circ}$ and $\lambda=150 ^{\circ}, \beta=-90 ^{\circ}$) with $\chi_\mathrm{red}^{2}=1.18$ (normalized for the 5022 data points). The solutions highlighted with a red border are within a margin of 5.7\% (3$\sigma$) of the two best obtained solutions.}
    \label{fig:15422_plot_ny}
\end{figure*}

\begin{figure*}
    \centering
    \includegraphics[width=\textwidth]{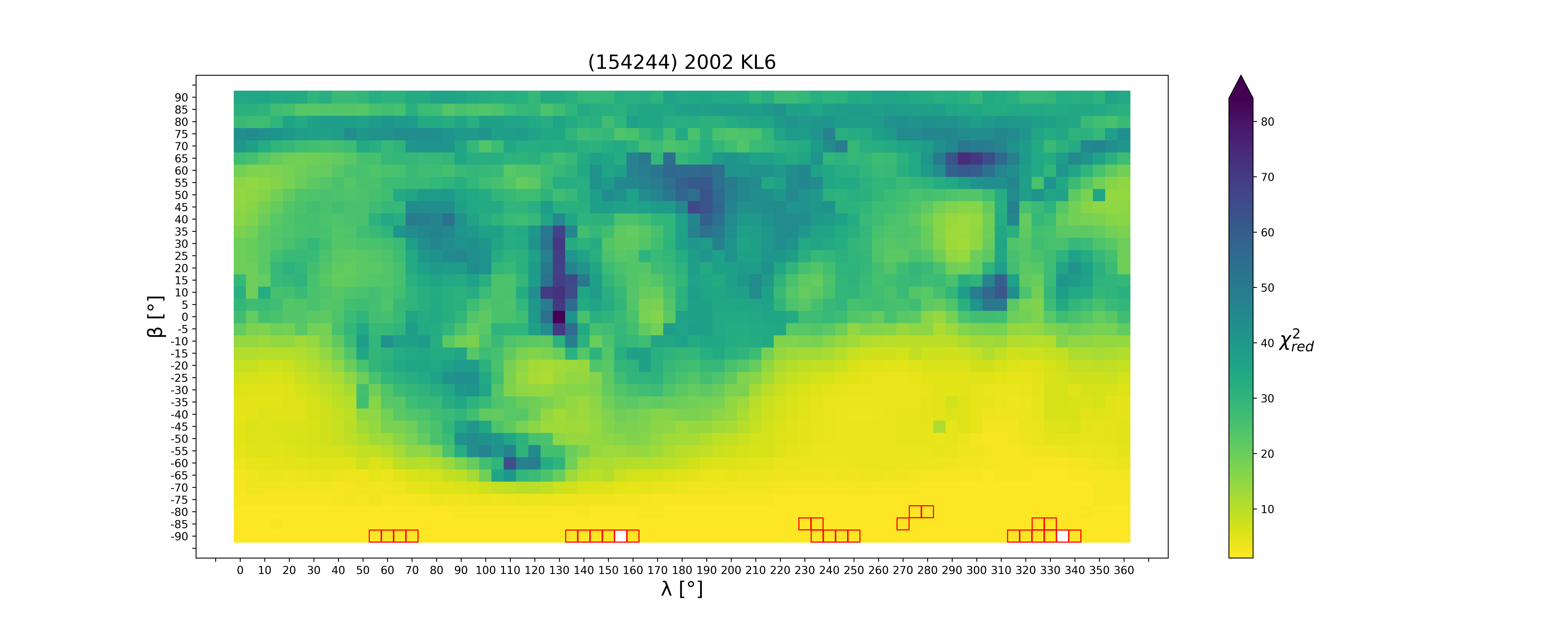}
    \caption{Pole solutions statistical quality for (154244) 2002 KL6 applying the linearly increasing period code. The solutions are shaded by its $\chi_\mathrm{red}^{2}$ value, with the best solutions obtained represented as white squares ($\lambda=335 ^{\circ}, \beta=-90 ^{\circ}$ and $\lambda=155 ^{\circ}, \beta=-90 ^{\circ}$) with $\chi_\mathrm{red}^{2}=1.15$ (normalized for the 5022 data points). The solutions highlighted with a red border are within a margin of 5.7\% (3$\sigma$) of the two best obtained solutions.}
    \label{fig:15422_plot_y}
\end{figure*}

\begin{figure*}
    \centering
    \includegraphics[width=\textwidth]{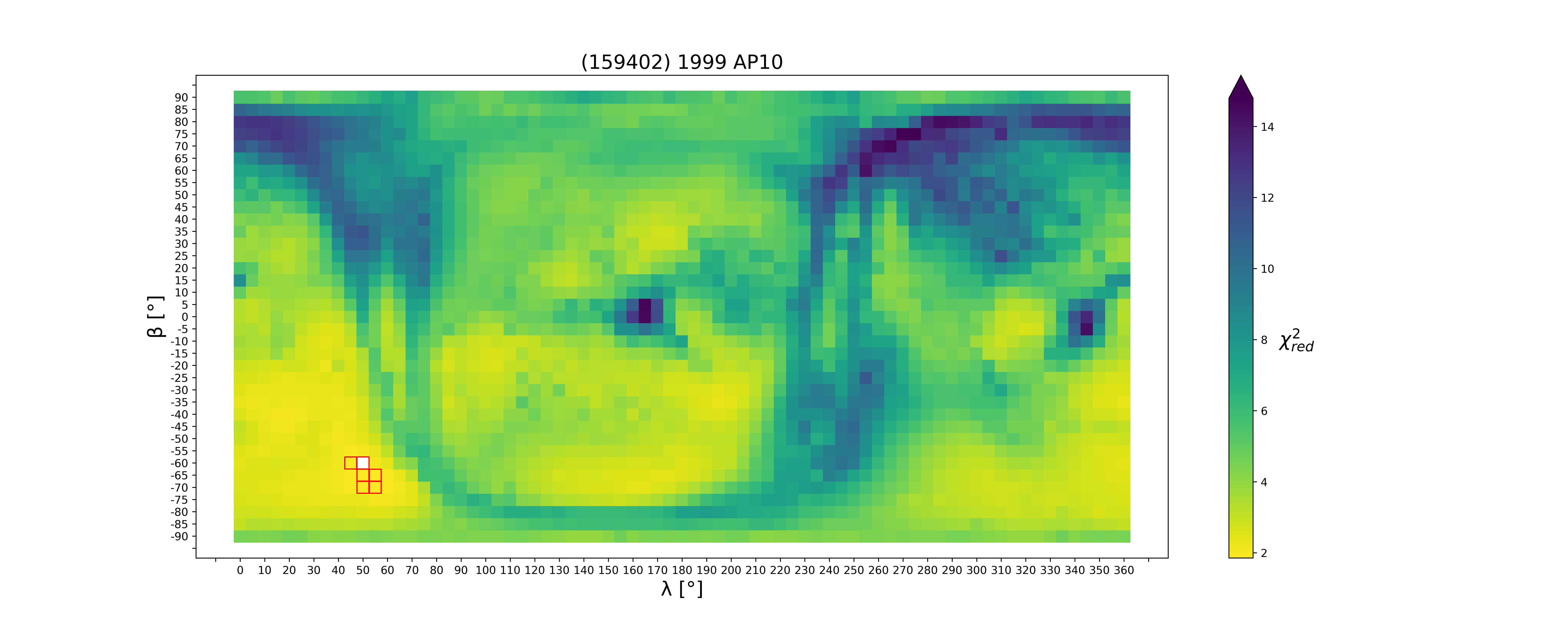}
    \caption{Pole solutions statistical quality for (159402) 1999 AP10 applying the constant period code. The solutions are shaded by its $\chi_\mathrm{red}^{2}$ value, with the best solutions obtained represented as white squares ($\lambda=50 ^{\circ}, \beta=-60 ^{\circ}$ with $\chi_\mathrm{red}^{2}=1.85$ (normalized for the 8049 data points). The solutions highlighted with a red border are within a margin of 4.5\% (3$\sigma$) of the two best obtained solutions.}
    \label{fig:159402_plot}
\end{figure*}

\section{Model and data fit}\label{sec:modelfit}

\begin{figure*}
    \centering
    \begin{subfigure}{0.3\textwidth}
        \centering
        \includegraphics[width=\textwidth]{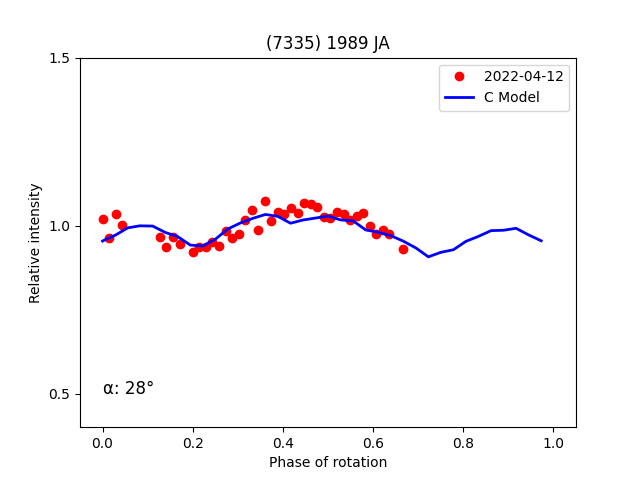}
    \end{subfigure}
    \begin{subfigure}{0.3\textwidth}
        \centering
        \includegraphics[width=\textwidth]{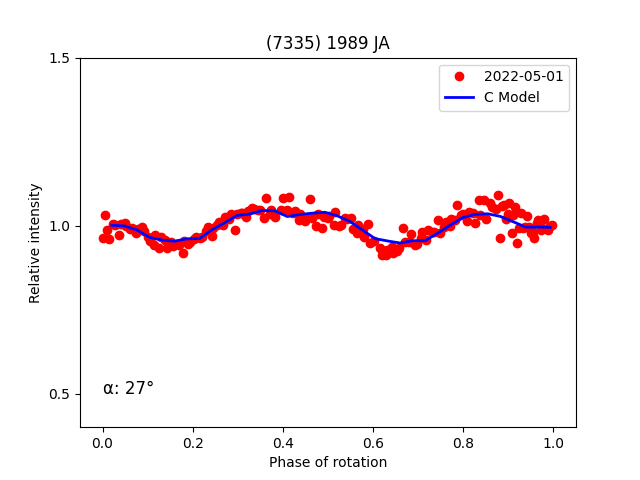}
    \end{subfigure}
    \begin{subfigure}{0.3\textwidth}
        \centering
        \includegraphics[width=\textwidth]{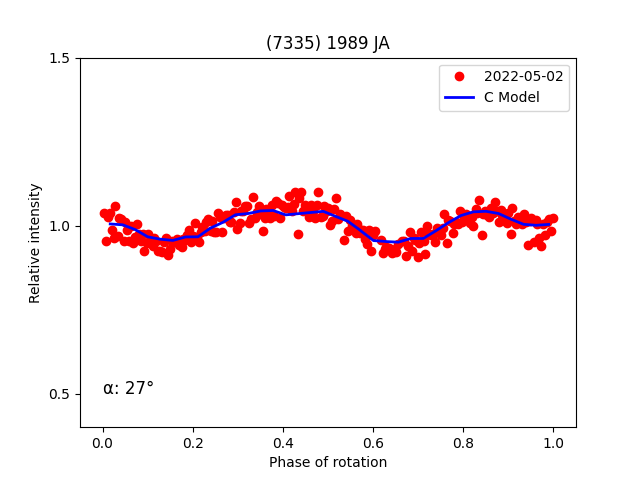}
    \end{subfigure}
    \begin{subfigure}{0.3\textwidth}
        \centering
        \includegraphics[width=\textwidth]{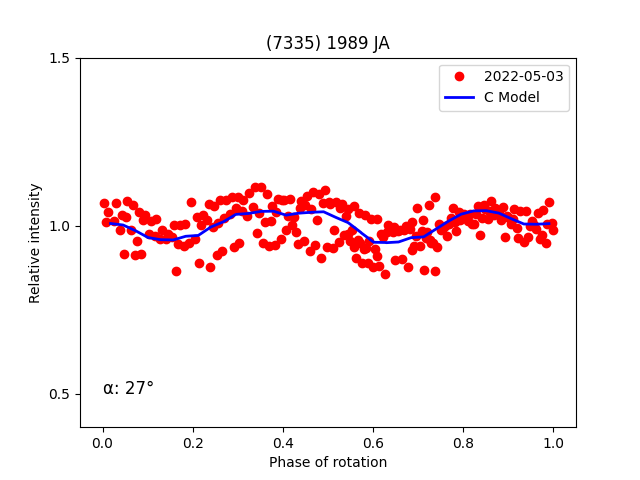}
    \end{subfigure}
    \begin{subfigure}{0.3\textwidth}
        \centering
        \includegraphics[width=\textwidth]{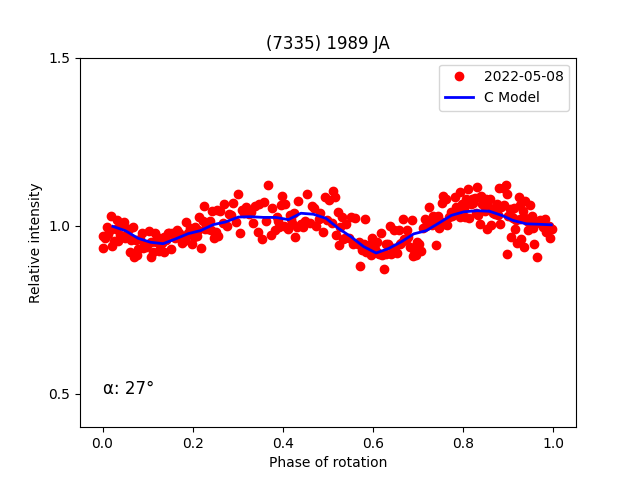}
    \end{subfigure}
        \begin{subfigure}{0.3\textwidth}
        \centering
        \includegraphics[width=\textwidth]{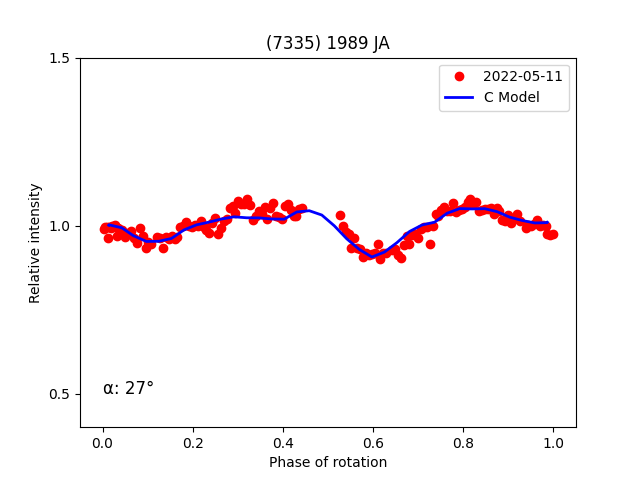}
    \end{subfigure}
    \begin{subfigure}{0.3\textwidth}
        \centering
        \includegraphics[width=\textwidth]{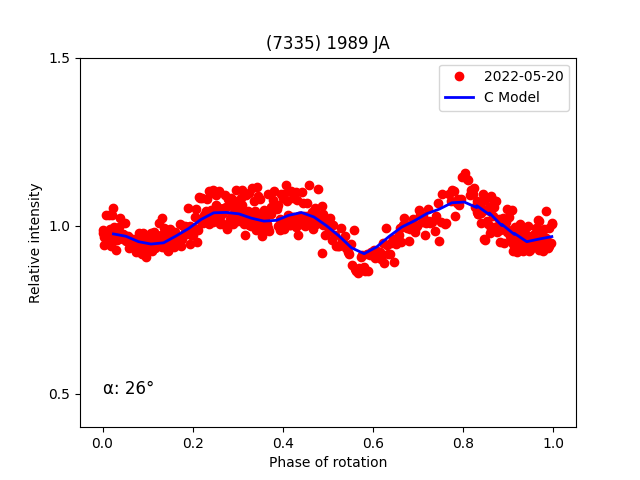}
    \end{subfigure}
    \begin{subfigure}{0.3\textwidth}
        \centering
        \includegraphics[width=\textwidth]{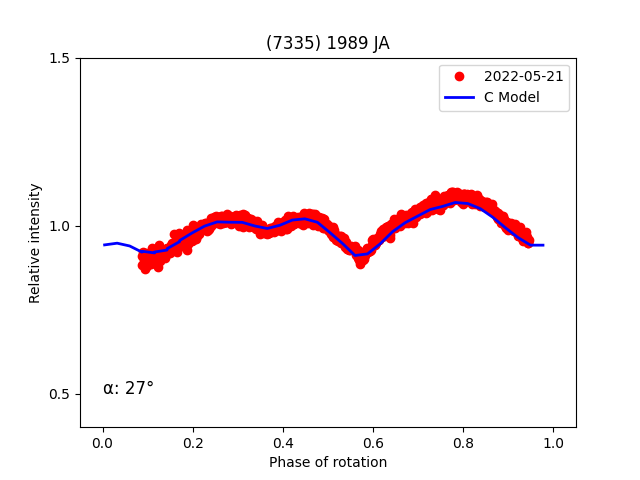}
    \end{subfigure}
    \caption{Fit between the rest (8 light-curves) of the new light-curves from (7335) 1989 JA presented in this work and the best-fitting constant period model (C Model). The data is plotted as red dots for each observation, while the model is plotted as a solid blue line. The geometry is described by its solar phase angle $\alpha$.}
    \label{fig:IAC_fit_7335_2}
\end{figure*}

\begin{figure*}
    \centering
    \begin{subfigure}{0.49\textwidth}
        \centering
        \includegraphics[width=\textwidth]{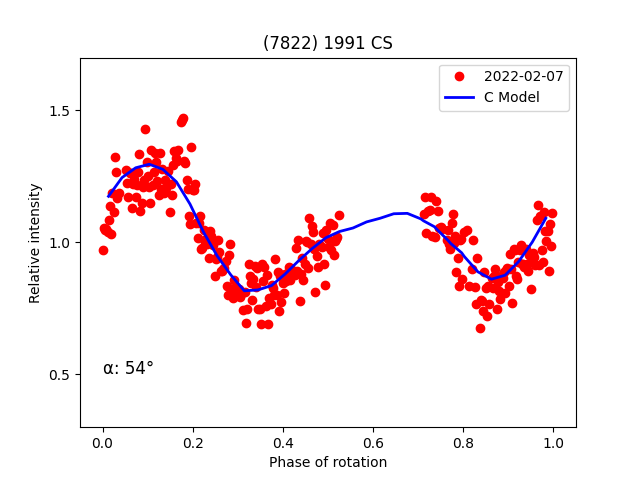}
    \end{subfigure}
    \begin{subfigure}{0.49\textwidth}
        \centering
        \includegraphics[width=\textwidth]{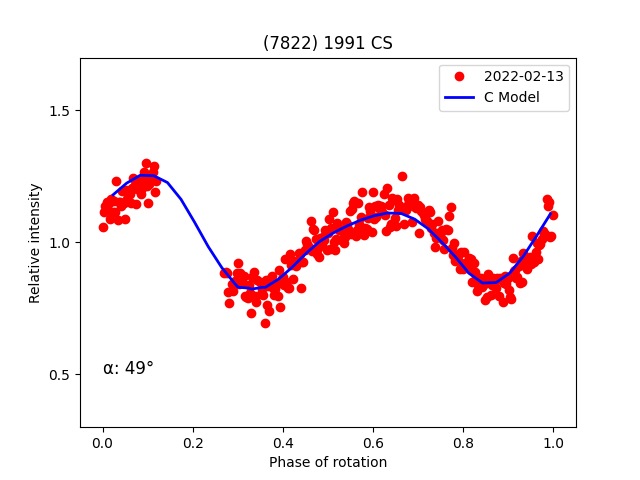}
    \end{subfigure}
    \begin{subfigure}{0.49\textwidth}
        \centering
        \includegraphics[width=\textwidth]{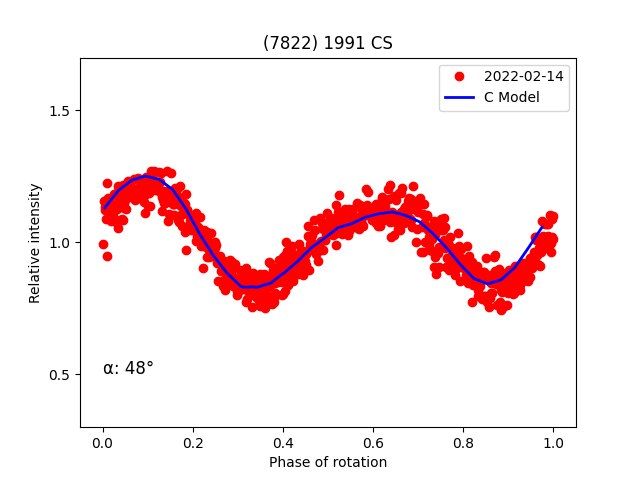}
    \end{subfigure}
    \begin{subfigure}{0.49\textwidth}
        \centering
        \includegraphics[width=\textwidth]{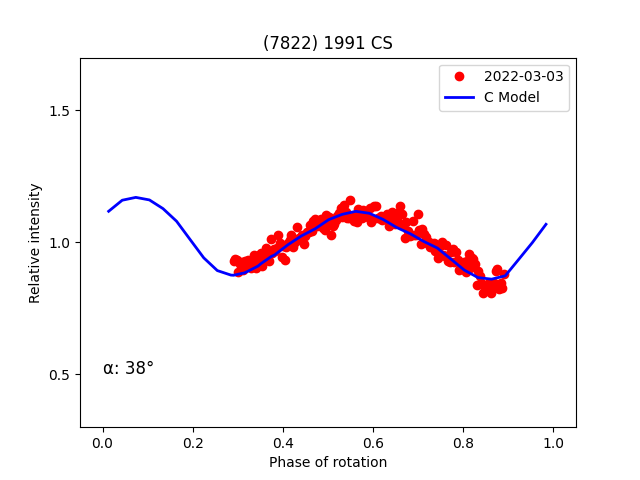}
    \end{subfigure}
    \caption{Fit between the rest (4 light-curves) of new light-curves from (7822) 1991 CS presented in this work and the best-fitting constant period model (C Model). The data is plotted as red dots for each observation, while the model is plotted as a solid blue line. The geometry is described by its solar phase angle $\alpha$.}
    \label{fig:IAC_fit_7822}
\end{figure*}

\begin{figure*}
    \centering
    \begin{subfigure}{0.3\textwidth}
        \centering
        \includegraphics[width=\textwidth]{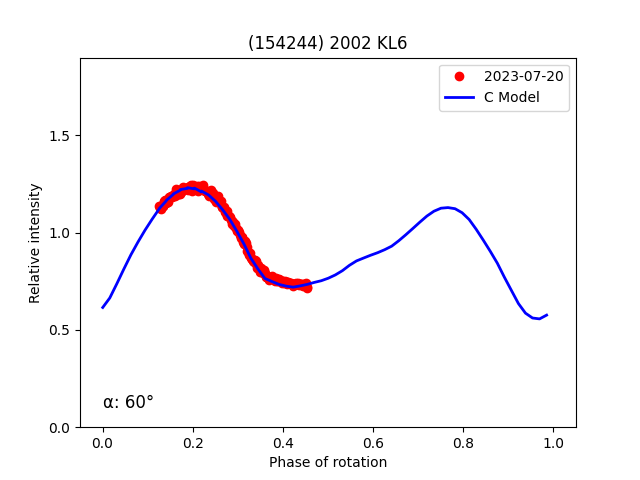}
    \end{subfigure}
    \begin{subfigure}{0.3\textwidth}
        \centering
        \includegraphics[width=\textwidth]{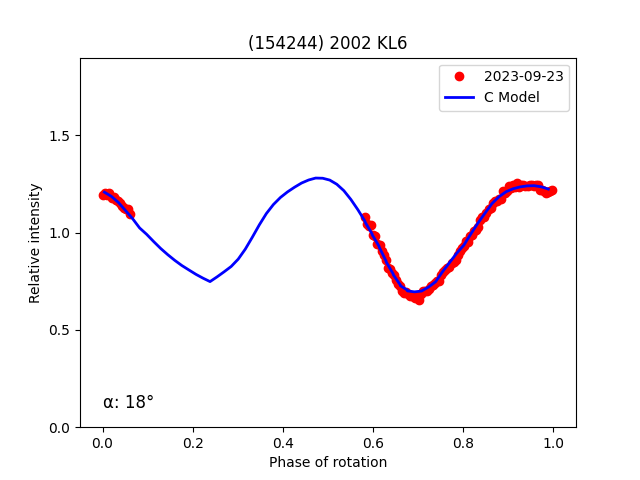}
    \end{subfigure}
    \begin{subfigure}{0.3\textwidth}
        \centering
        \includegraphics[width=\textwidth]{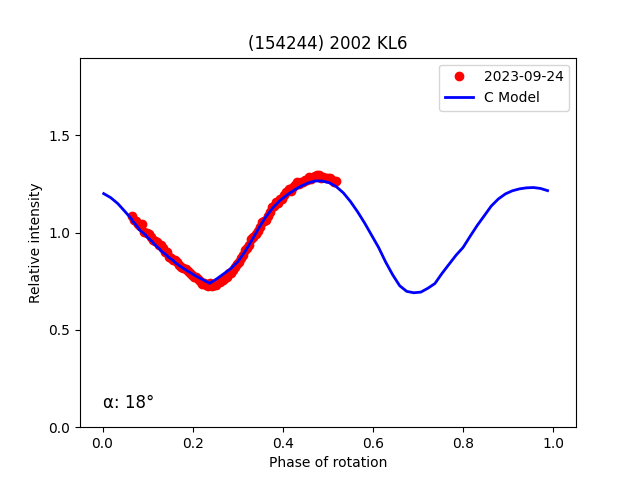}
    \end{subfigure}
    \begin{subfigure}{0.3\textwidth}
        \centering
        \includegraphics[width=\textwidth]{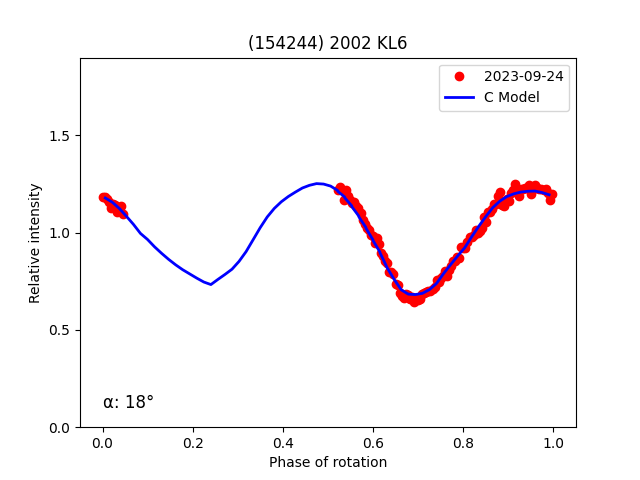}
    \end{subfigure}
    \begin{subfigure}{0.3\textwidth}
        \centering
        \includegraphics[width=\textwidth]{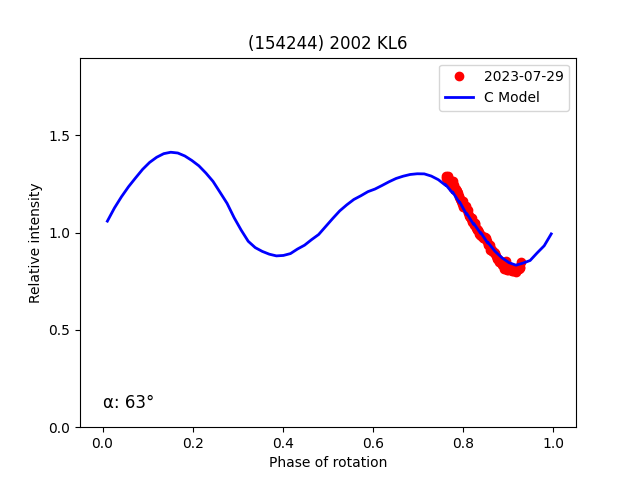}
    \end{subfigure}
    \begin{subfigure}{0.3\textwidth}
        \centering
        \includegraphics[width=\textwidth]{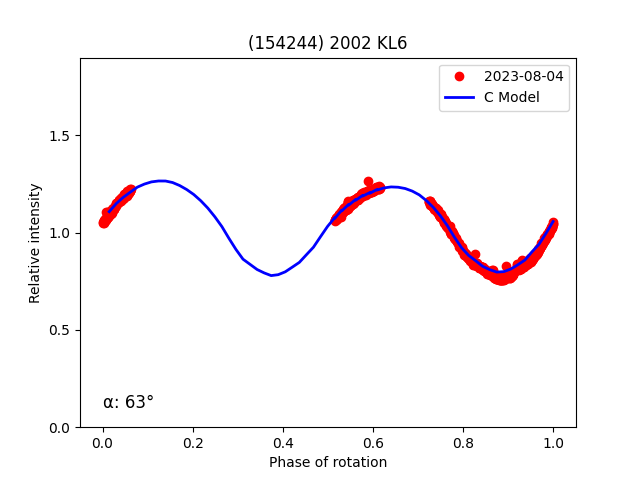}
    \end{subfigure}
    \begin{subfigure}{0.3\textwidth}
        \centering
        \includegraphics[width=\textwidth]{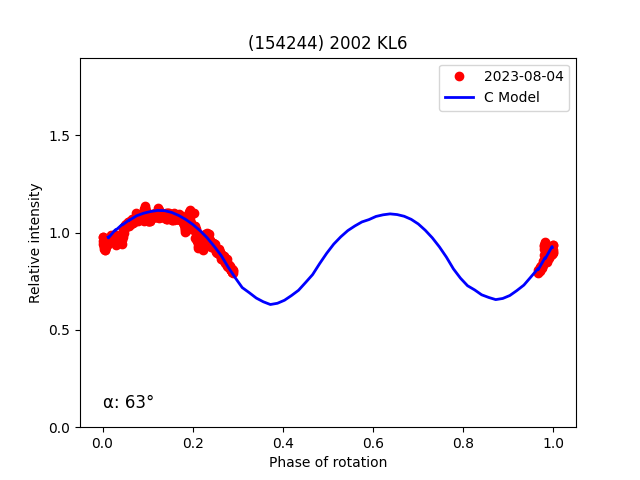}
    \end{subfigure}
    \begin{subfigure}{0.3\textwidth}
        \centering
        \includegraphics[width=\textwidth]{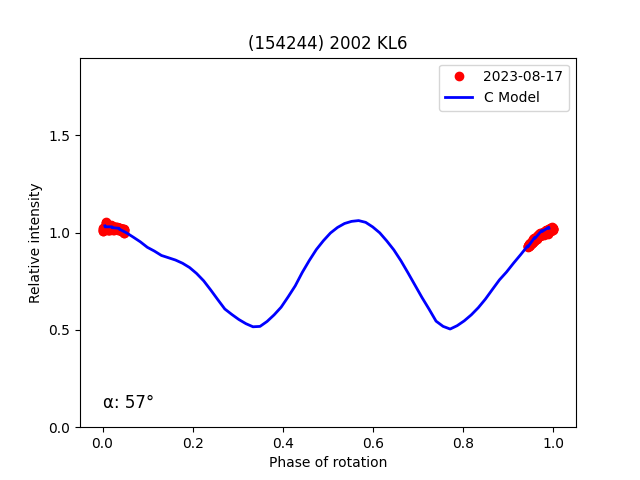}
    \end{subfigure}
    \begin{subfigure}{0.3\textwidth}
        \centering
        \includegraphics[width=\textwidth]{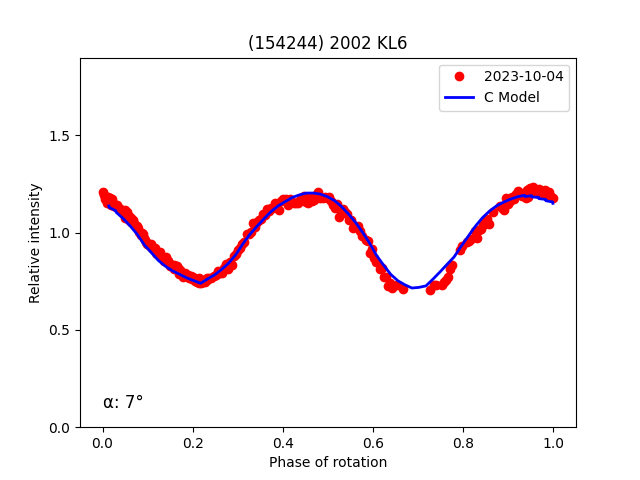}
    \end{subfigure}
    \begin{subfigure}{0.3\textwidth}
        \centering
        \includegraphics[width=\textwidth]{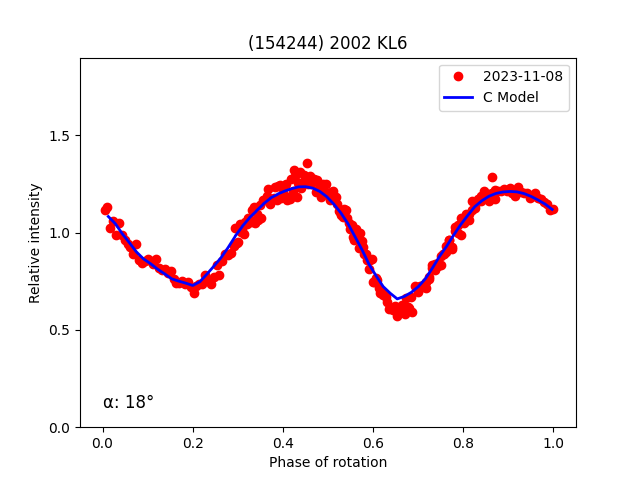}
    \end{subfigure}
    \begin{subfigure}{0.3\textwidth}
        \centering
        \includegraphics[width=\textwidth]{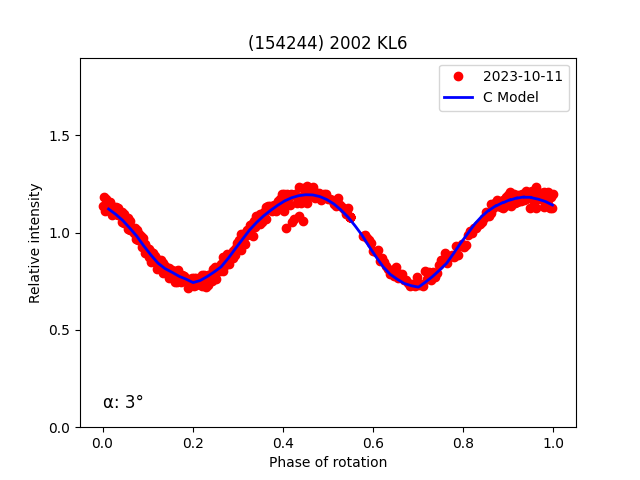}
    \end{subfigure}
    \caption{Fit between a selection (11 light-curves) of the new light-curves from (154244) 2002 KL6 presented in this work and the best-fitting constant period model (C Model) around ($\lambda = 149^{\circ}$, $\beta = -90^{\circ}$). The data is plotted as red dots for each observation, while the model is plotted as a solid blue line. The geometry is described by its solar phase angle $\alpha$.}
    \label{fig:IAC_fit_154244_ny_155}
\end{figure*}

\begin{figure*}
    \centering
    \begin{subfigure}{0.3\textwidth}
        \centering
        \includegraphics[width=\textwidth]{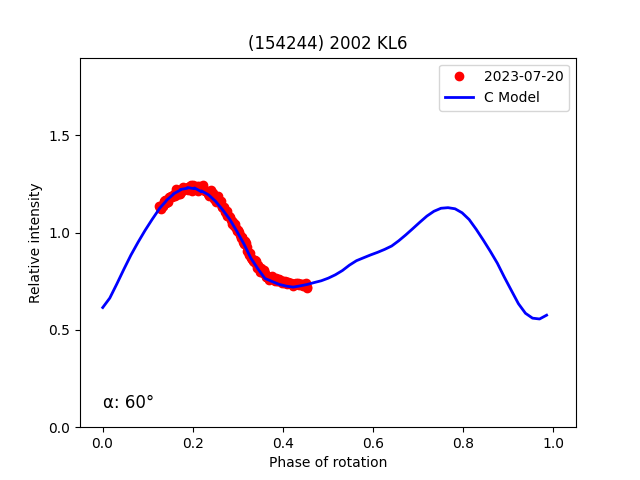}
    \end{subfigure}
    \begin{subfigure}{0.3\textwidth}
        \centering
        \includegraphics[width=\textwidth]{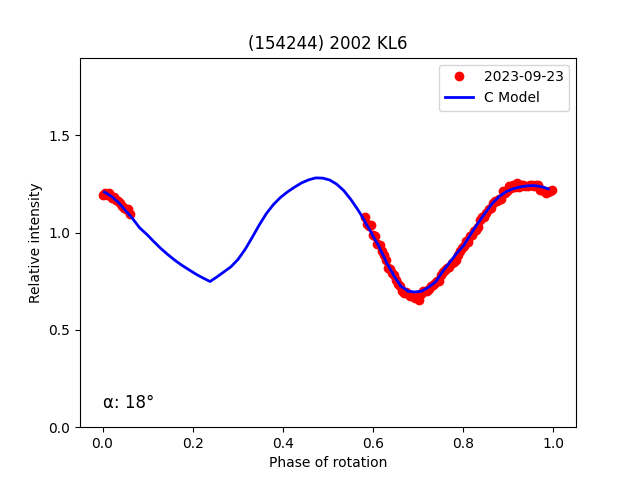}
    \end{subfigure}
    \begin{subfigure}{0.3\textwidth}
        \centering
        \includegraphics[width=\textwidth]{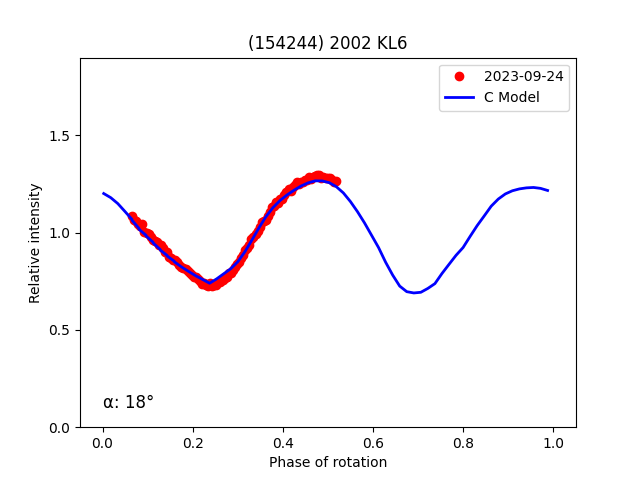}
    \end{subfigure}
    \begin{subfigure}{0.3\textwidth}
        \centering
        \includegraphics[width=\textwidth]{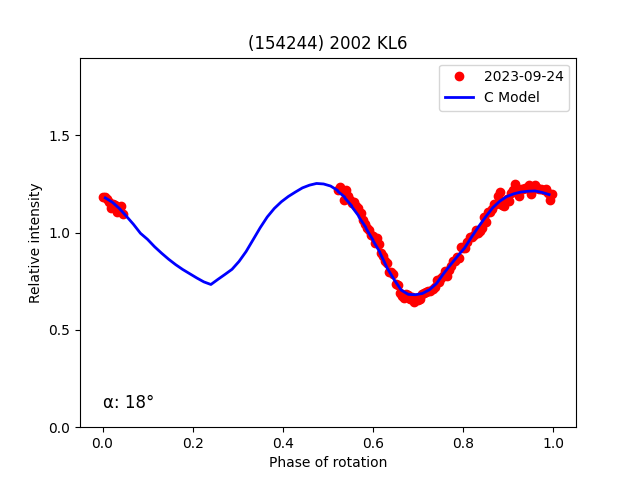}
    \end{subfigure}
    \begin{subfigure}{0.3\textwidth}
        \centering
        \includegraphics[width=\textwidth]{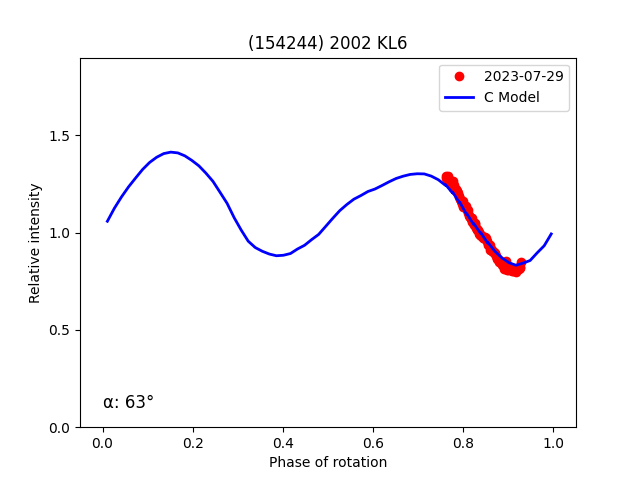}
    \end{subfigure}
    \begin{subfigure}{0.3\textwidth}
        \centering
        \includegraphics[width=\textwidth]{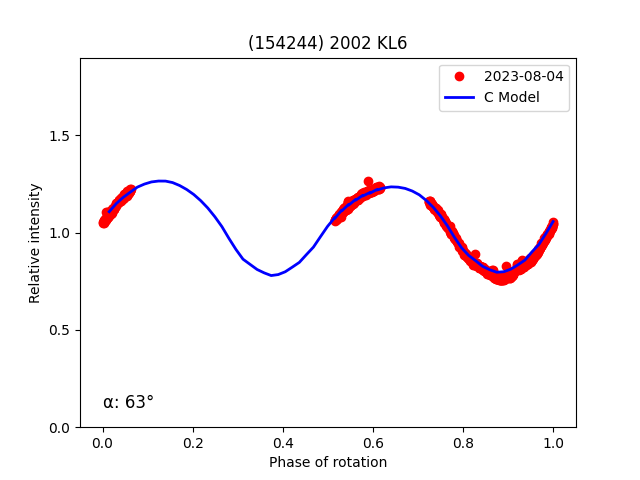}
    \end{subfigure}
    \begin{subfigure}{0.3\textwidth}
        \centering
        \includegraphics[width=\textwidth]{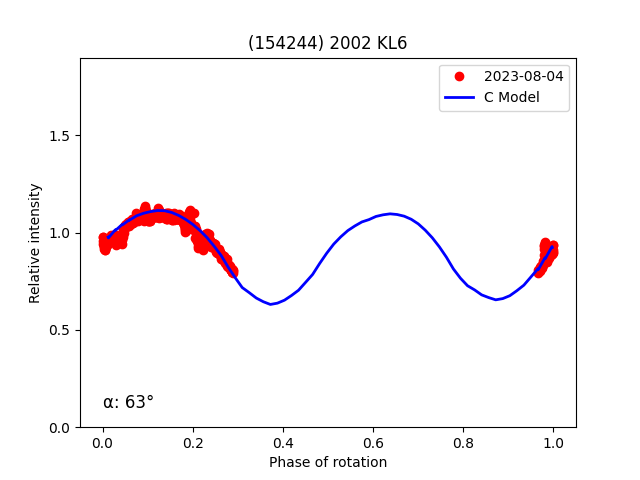}
    \end{subfigure}
    \begin{subfigure}{0.3\textwidth}
        \centering
        \includegraphics[width=\textwidth]{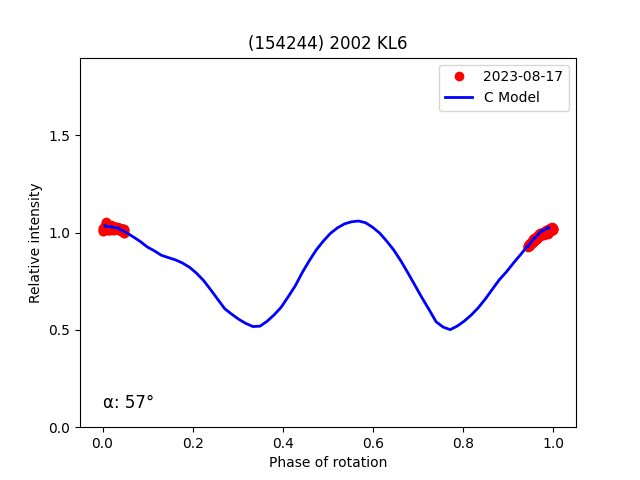}
    \end{subfigure}
    \begin{subfigure}{0.3\textwidth}
        \centering
        \includegraphics[width=\textwidth]{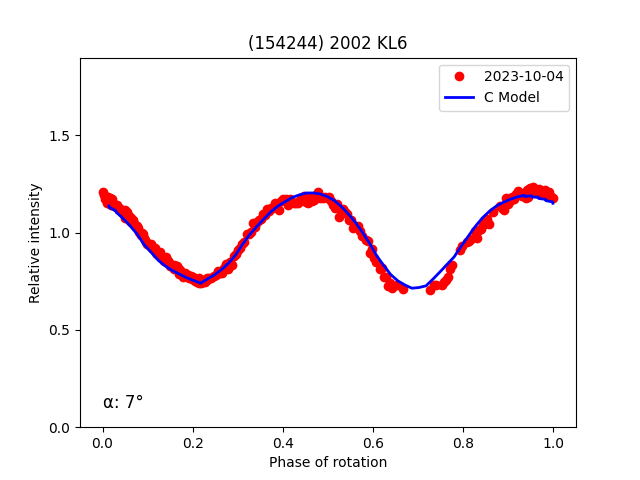}
    \end{subfigure}
    \begin{subfigure}{0.3\textwidth}
        \centering
        \includegraphics[width=\textwidth]{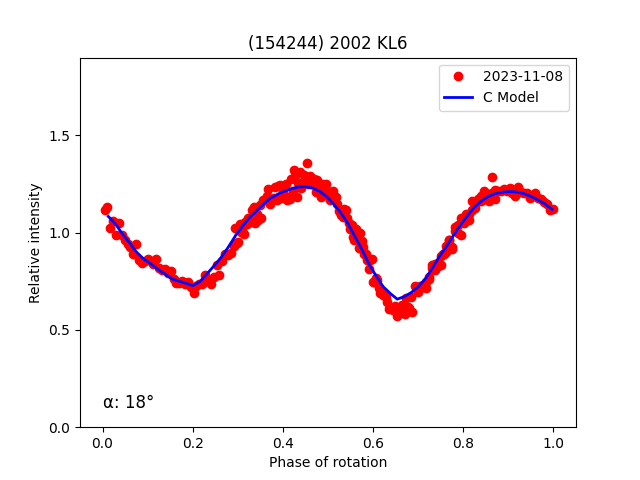}
    \end{subfigure}
    \begin{subfigure}{0.3\textwidth}
        \centering
        \includegraphics[width=\textwidth]{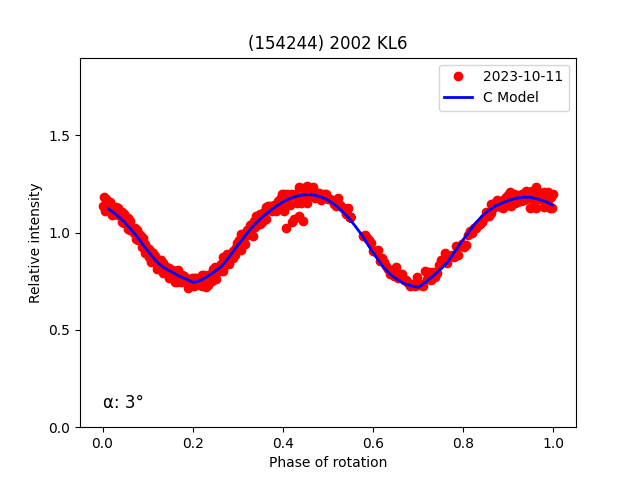}
    \end{subfigure}
    \caption{Fit between a selection (11 light-curves) the new light-curves from (154244) 2002 KL6 presented in this work and the best-fitting constant period model (C Model) around ($\lambda = 326^{\circ}$, $\beta = -88^{\circ}$). The data is plotted as red dots for each observation, while the model is plotted as a solid blue line. The geometry is described by its solar phase angle $\alpha$.}
    \label{fig:IAC_fit_154244_ny_335}
\end{figure*}

\begin{figure*}
    \centering
    \begin{subfigure}{0.3\textwidth}
        \centering
        \includegraphics[width=\textwidth]{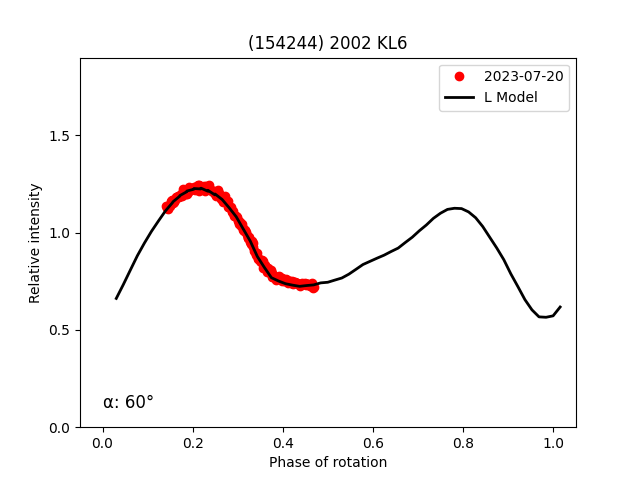}
    \end{subfigure}
    \begin{subfigure}{0.3\textwidth}
        \centering
        \includegraphics[width=\textwidth]{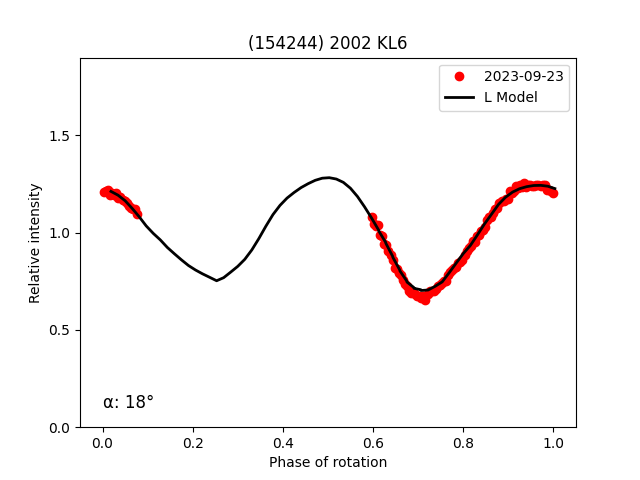}
    \end{subfigure}
    \begin{subfigure}{0.3\textwidth}
        \centering
        \includegraphics[width=\textwidth]{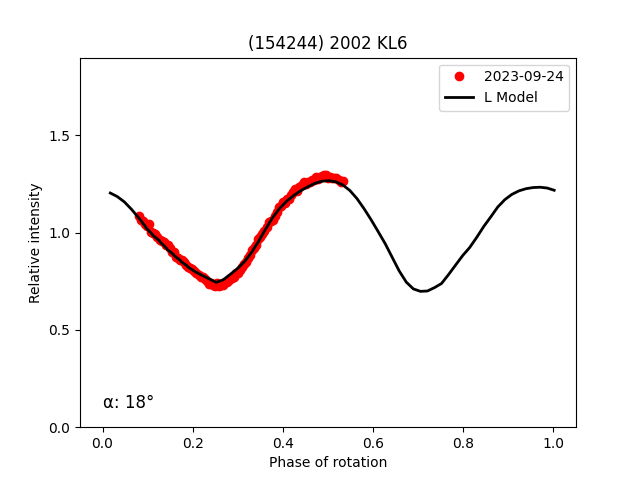}
    \end{subfigure}
    \begin{subfigure}{0.3\textwidth}
        \centering
        \includegraphics[width=\textwidth]{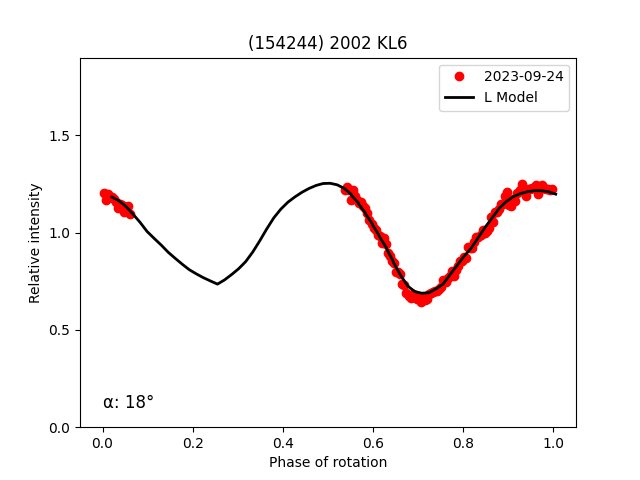}
    \end{subfigure}
    \begin{subfigure}{0.3\textwidth}
        \centering
        \includegraphics[width=\textwidth]{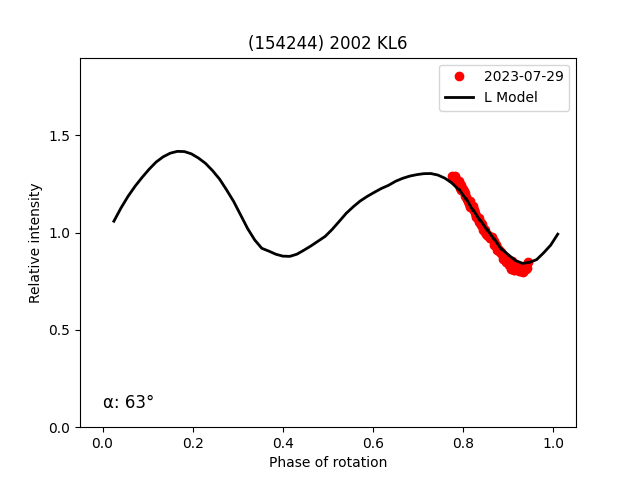}
    \end{subfigure}
    \begin{subfigure}{0.3\textwidth}
        \centering
        \includegraphics[width=\textwidth]{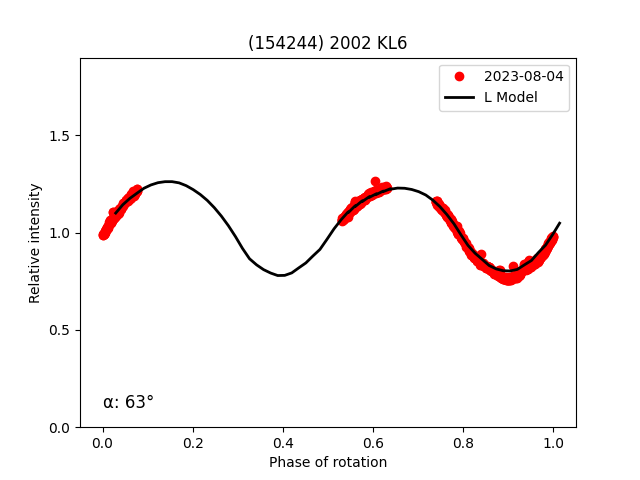}
    \end{subfigure}
    \begin{subfigure}{0.3\textwidth}
        \centering
        \includegraphics[width=\textwidth]{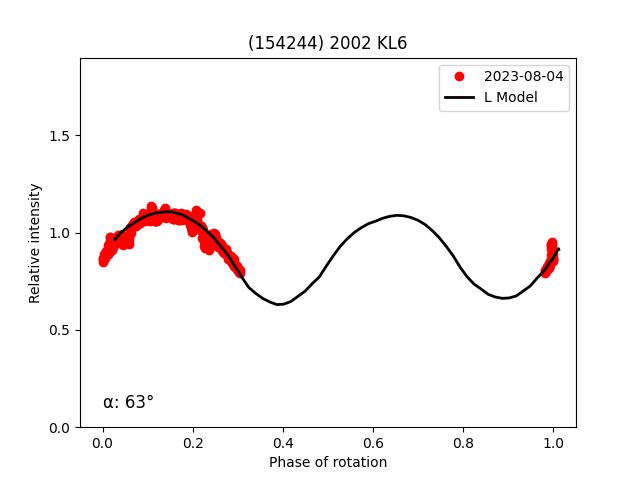}
    \end{subfigure}
    \begin{subfigure}{0.3\textwidth}
        \centering
        \includegraphics[width=\textwidth]{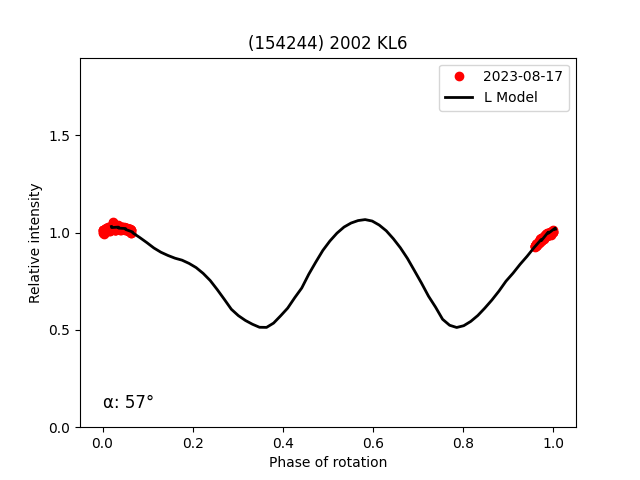}
    \end{subfigure}
    \begin{subfigure}{0.3\textwidth}
        \centering
        \includegraphics[width=\textwidth]{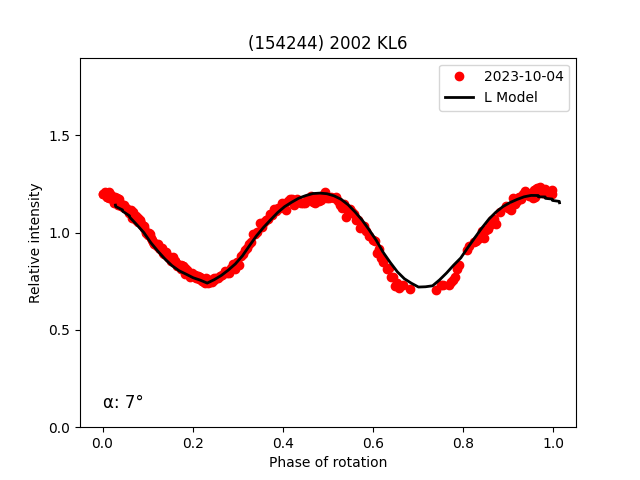}
    \end{subfigure}
    \begin{subfigure}{0.3\textwidth}
        \centering
        \includegraphics[width=\textwidth]{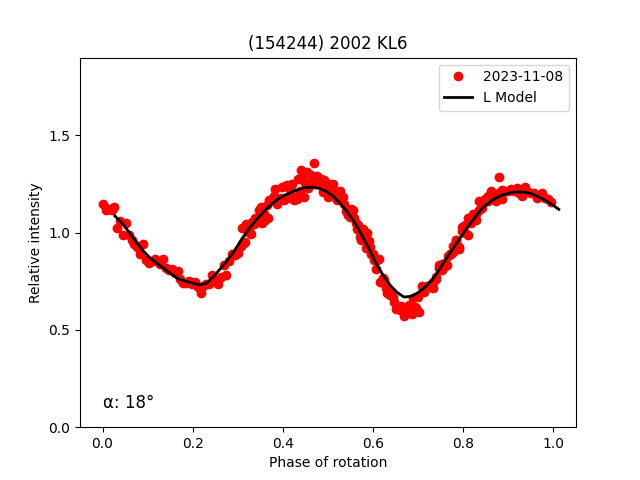}
    \end{subfigure}
    \begin{subfigure}{0.3\textwidth}
        \centering
        \includegraphics[width=\textwidth]{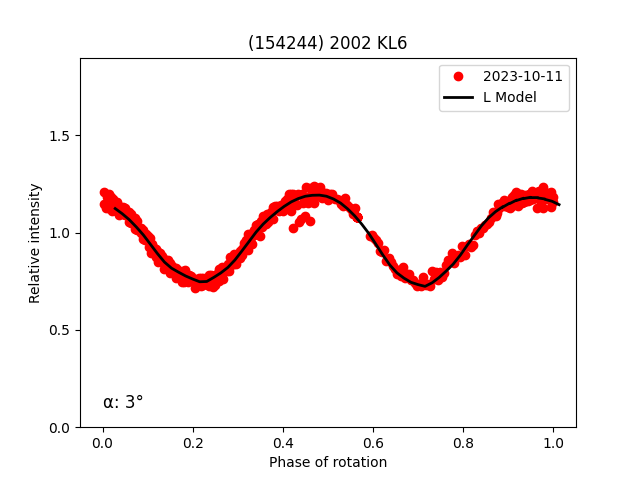}
    \end{subfigure}
    \caption{Fit between a selection (11 light-curves) the new light-curves from (154244) 2002 KL6 presented in this work and the best-fitting linearly increasing period model (L Model) around ($\lambda = 151^{\circ}$, $\beta = -90^{\circ}$). The data is plotted as red dots for each observation, while the model is plotted as a solid black line. The geometry is described by its solar phase angle $\alpha$.}
    \label{fig:IAC_fit_154244_y_150}
\end{figure*}

\begin{figure*}
    \centering
    \begin{subfigure}{0.3\textwidth}
        \centering
        \includegraphics[width=\textwidth]{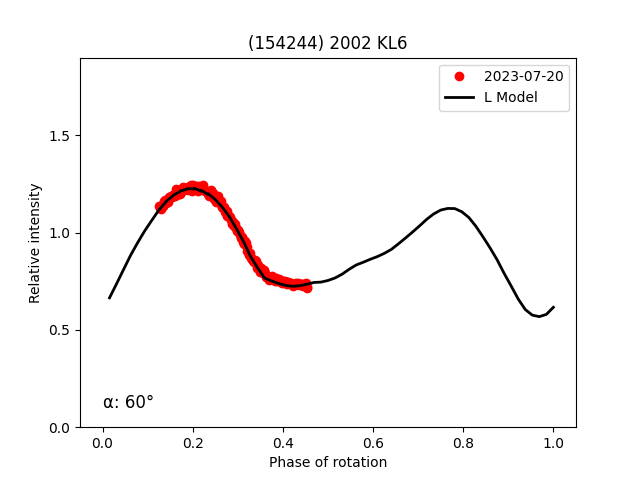}
    \end{subfigure}
    \begin{subfigure}{0.3\textwidth}
        \centering
        \includegraphics[width=\textwidth]{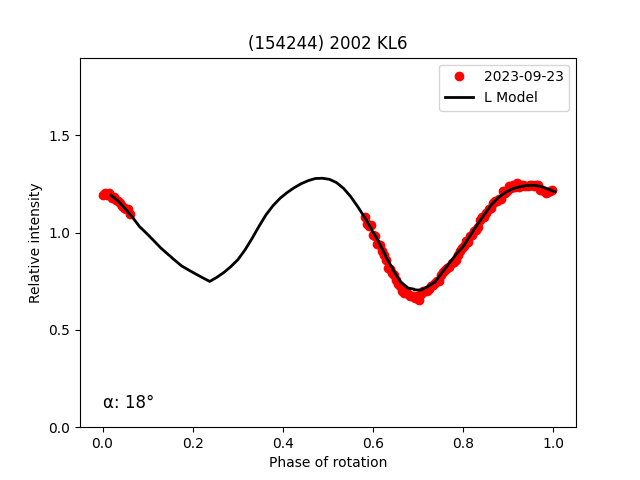}
    \end{subfigure}
    \begin{subfigure}{0.3\textwidth}
        \centering
        \includegraphics[width=\textwidth]{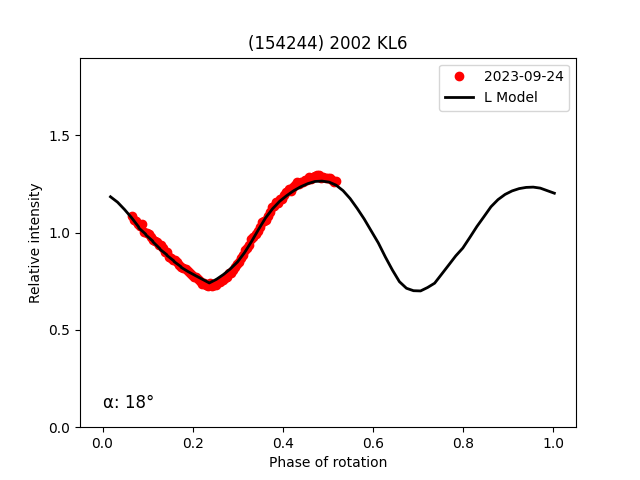}
    \end{subfigure}
    \begin{subfigure}{0.3\textwidth}
        \centering
        \includegraphics[width=\textwidth]{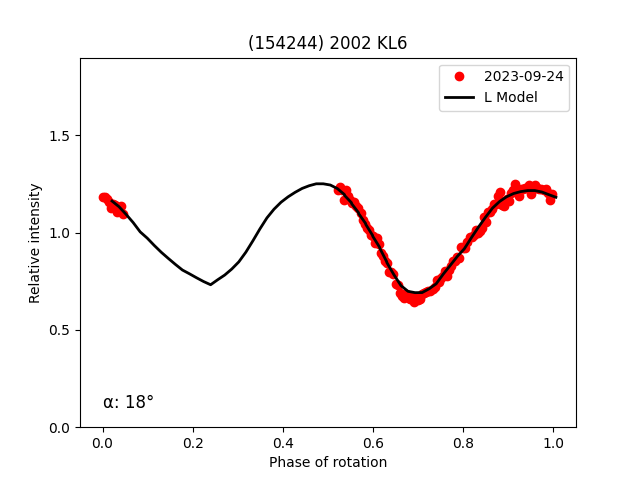}
    \end{subfigure}
    \begin{subfigure}{0.3\textwidth}
        \centering
        \includegraphics[width=\textwidth]{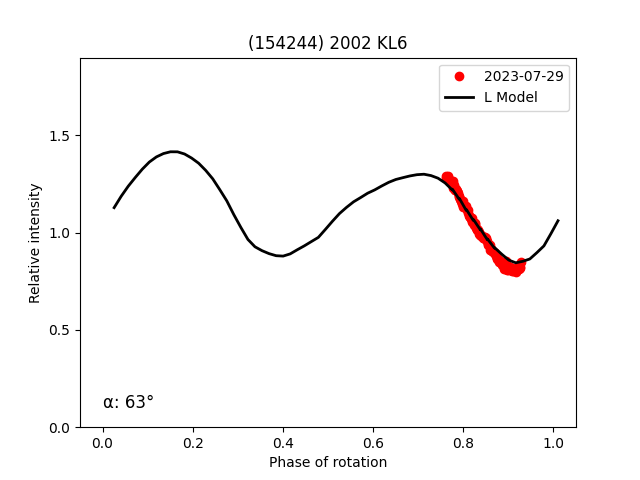}
    \end{subfigure}
    \begin{subfigure}{0.3\textwidth}
        \centering
        \includegraphics[width=\textwidth]{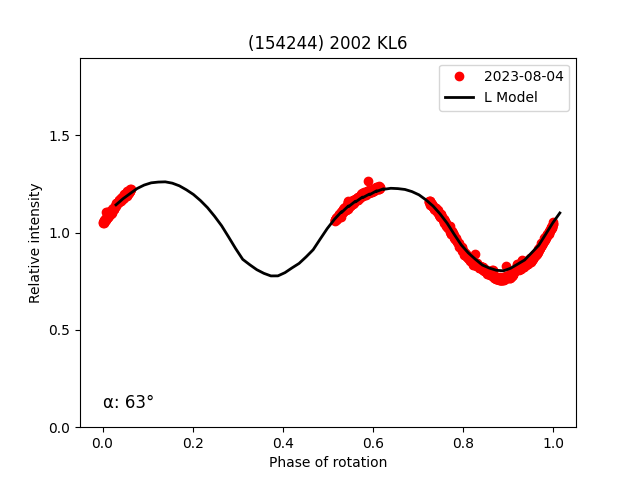}
    \end{subfigure}
    \begin{subfigure}{0.3\textwidth}
        \centering
        \includegraphics[width=\textwidth]{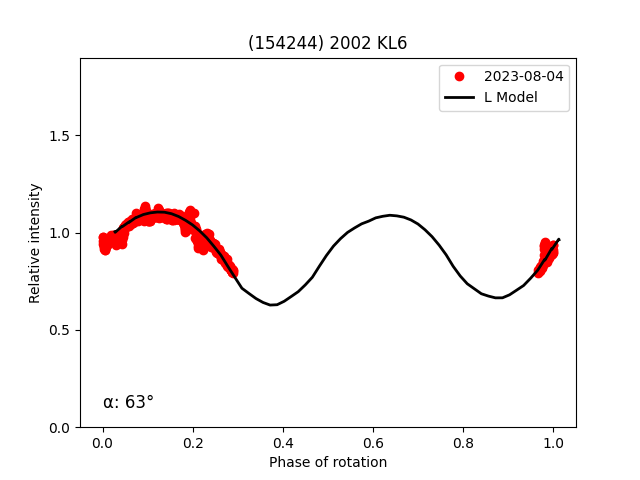}
    \end{subfigure}
    \begin{subfigure}{0.3\textwidth}
        \centering
        \includegraphics[width=\textwidth]{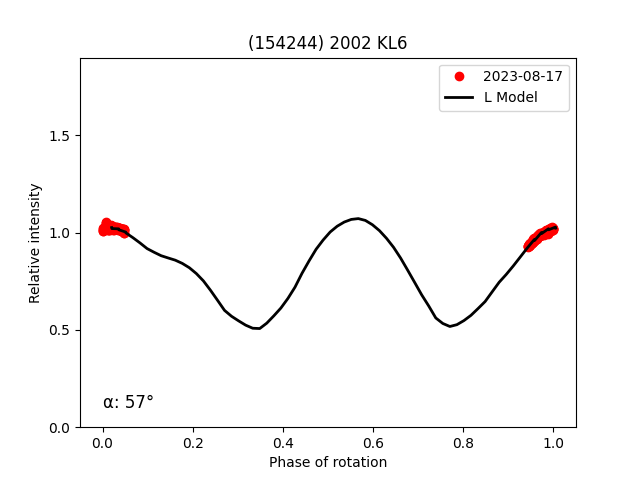}
    \end{subfigure}
    \begin{subfigure}{0.3\textwidth}
        \centering
        \includegraphics[width=\textwidth]{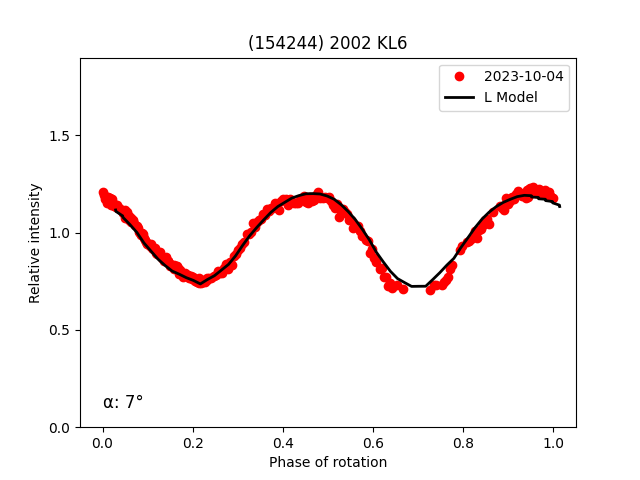}
    \end{subfigure}
    \begin{subfigure}{0.3\textwidth}
        \centering
        \includegraphics[width=\textwidth]{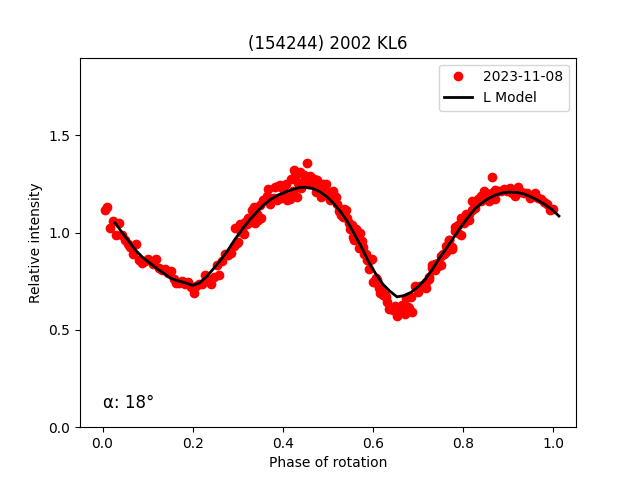}
    \end{subfigure}
    \begin{subfigure}{0.3\textwidth}
        \centering
        \includegraphics[width=\textwidth]{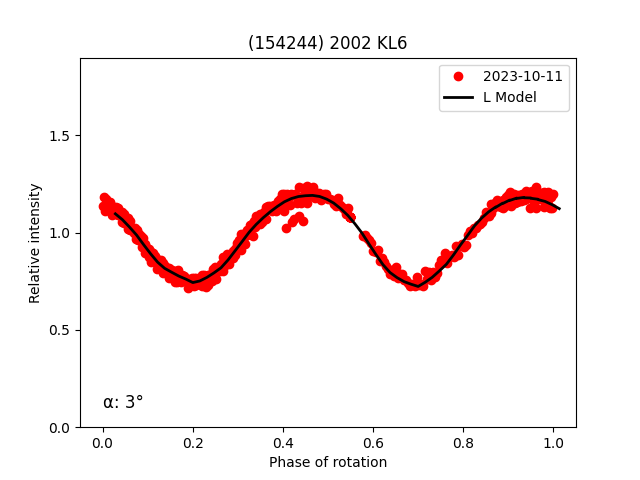}
    \end{subfigure}
    \caption{Fit between a selection (11 light-curves) of the new light-curves from (154244) 2002 KL6 presented in this work and the best-fitting linearly increasing period model (L Model) around ($\lambda = 330^{\circ}$, $\beta = -89^{\circ}$). The data is plotted as red dots for each observation, while the model is plotted as a solid black line. The geometry is described by its solar phase angle $\alpha$.}
    \label{fig:IAC_fit_154244_y_330}
\end{figure*}

\begin{figure*}
    \centering
    \begin{subfigure}{0.3\textwidth}
        \centering
        \includegraphics[width=\textwidth]{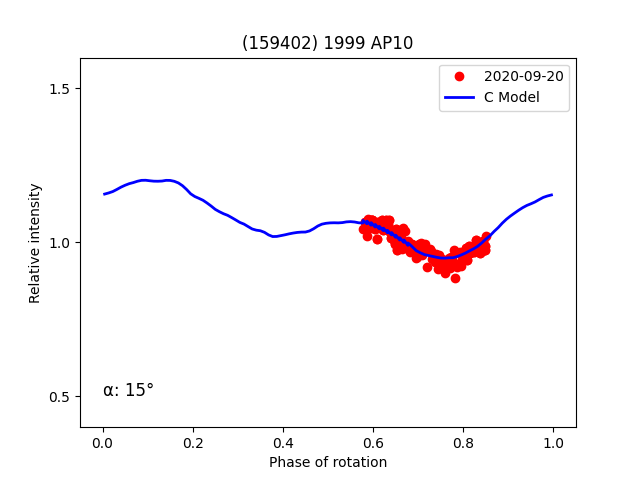}
    \end{subfigure}
    \begin{subfigure}{0.3\textwidth}
        \centering
        \includegraphics[width=\textwidth]{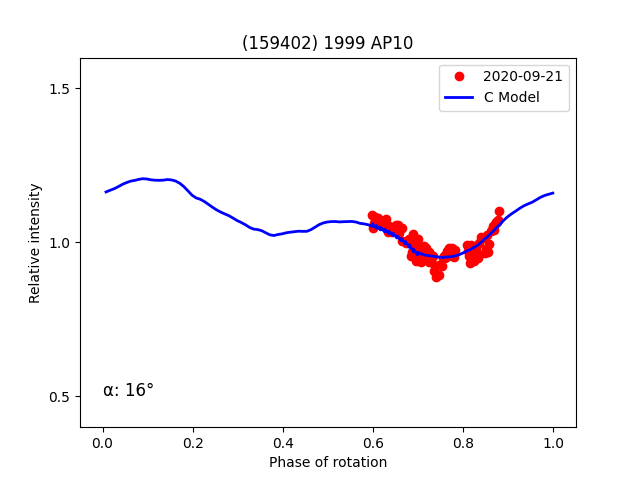}
    \end{subfigure}
    \begin{subfigure}{0.3\textwidth}
        \centering
        \includegraphics[width=\textwidth]{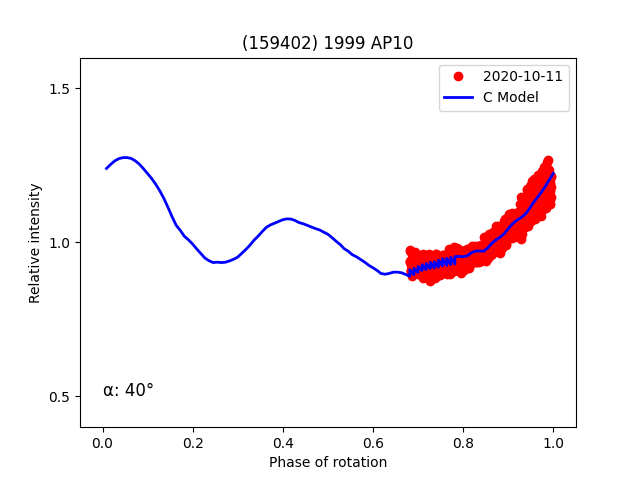}
    \end{subfigure}
    \begin{subfigure}{0.3\textwidth}
        \centering
        \includegraphics[width=\textwidth]{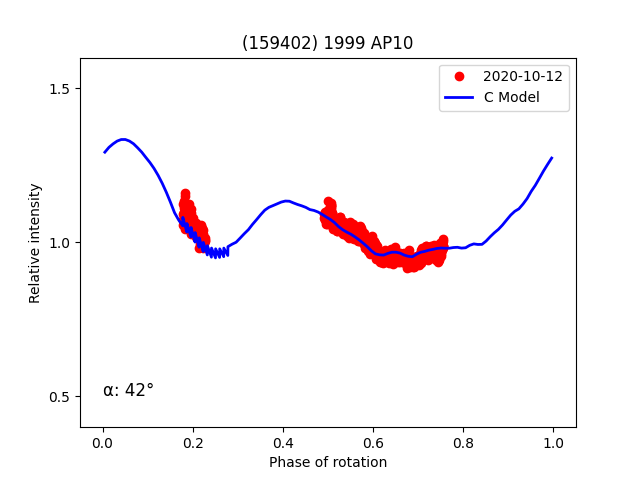}
    \end{subfigure}
    \begin{subfigure}{0.3\textwidth}
        \centering
        \includegraphics[width=\textwidth]{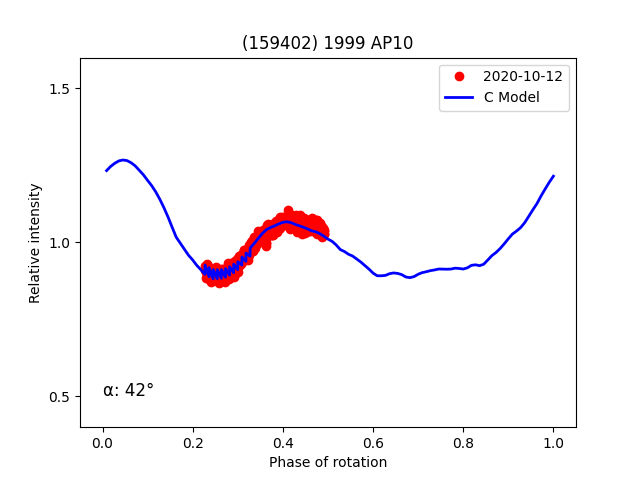}
    \end{subfigure}
    \begin{subfigure}{0.3\textwidth}
        \centering
        \includegraphics[width=\textwidth]{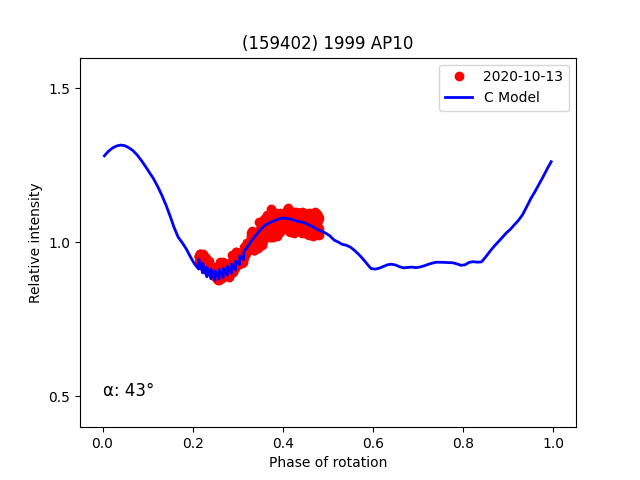}
    \end{subfigure}
    \begin{subfigure}{0.3\textwidth}
        \centering
        \includegraphics[width=\textwidth]{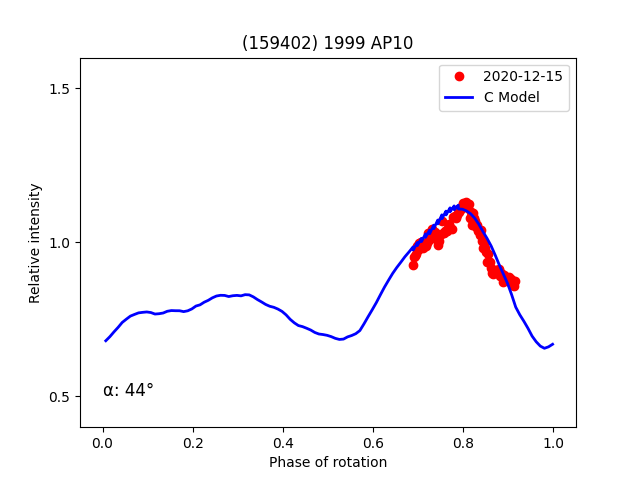}
    \end{subfigure}
    \begin{subfigure}{0.3\textwidth}
        \centering
        \includegraphics[width=\textwidth]{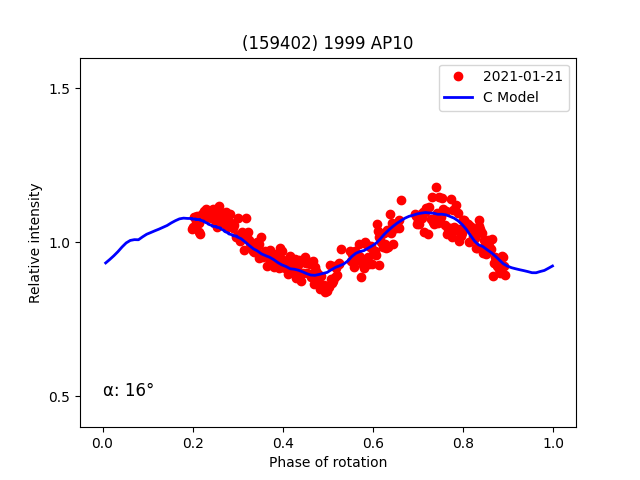}
    \end{subfigure}
    \begin{subfigure}{0.3\textwidth}
        \centering
        \includegraphics[width=\textwidth]{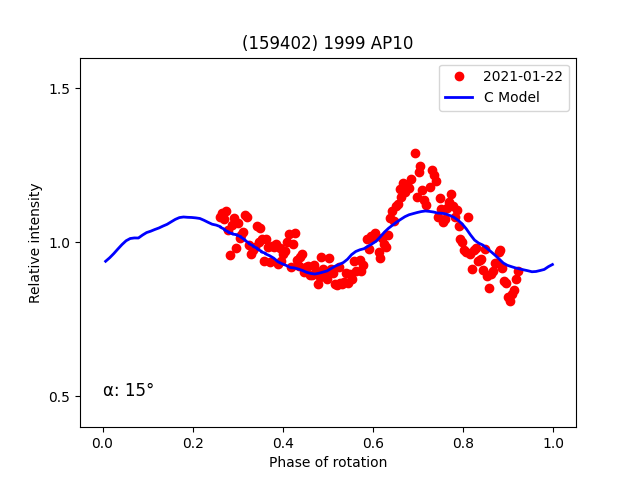}
    \end{subfigure}
    \caption{Fit between a selection (10 light-curves) of new light-curves presented in this work from (159402) 1999 AP10 and the best-fitting constant period model (C Model). The data is plotted as red dots for each observation, while the model is plotted as a solid blue line. The geometry is described by its solar phase angle $\alpha$.}
    \label{fig:IAC_fit_159402}
\end{figure*}

\section{7335 satellite detection}\label{sec:7335_sat}

\begin{figure*}
    \centering
    \begin{subfigure}{0.3\textwidth}
        \centering
        \includegraphics[width=\textwidth]{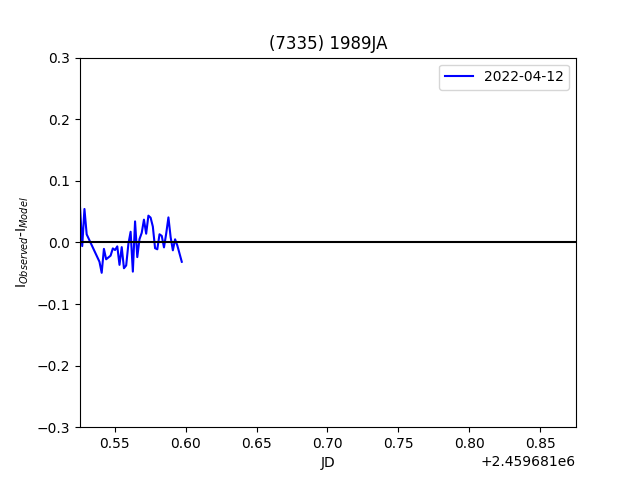}
    \end{subfigure}
    \begin{subfigure}{0.3\textwidth}
        \centering
        \includegraphics[width=\textwidth]{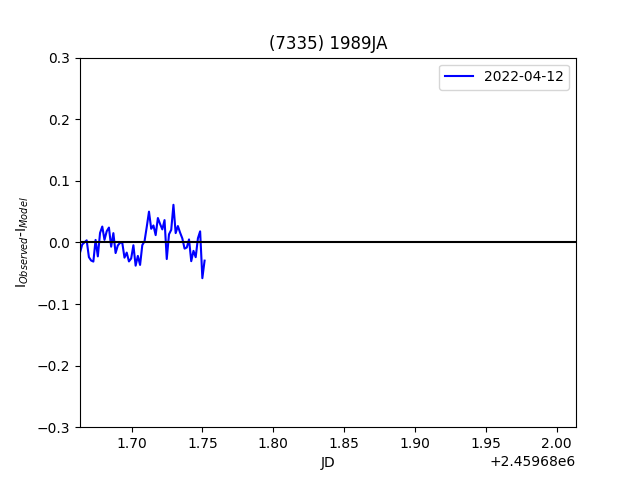}
    \end{subfigure}
    \begin{subfigure}{0.3\textwidth}
        \centering
        \includegraphics[width=\textwidth]{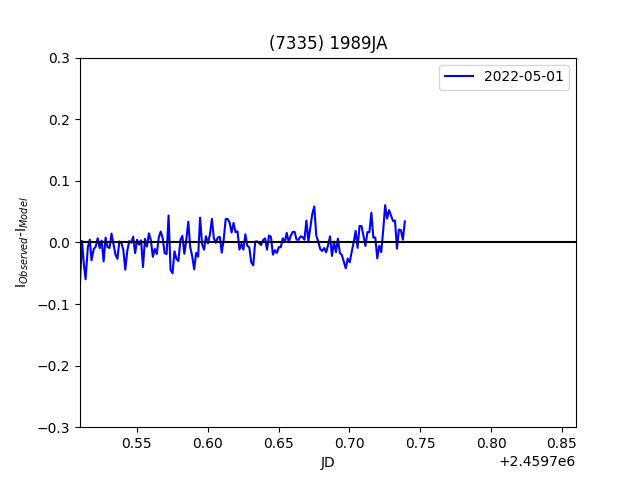}
    \end{subfigure}
    \begin{subfigure}{0.3\textwidth}
        \centering
        \includegraphics[width=\textwidth]{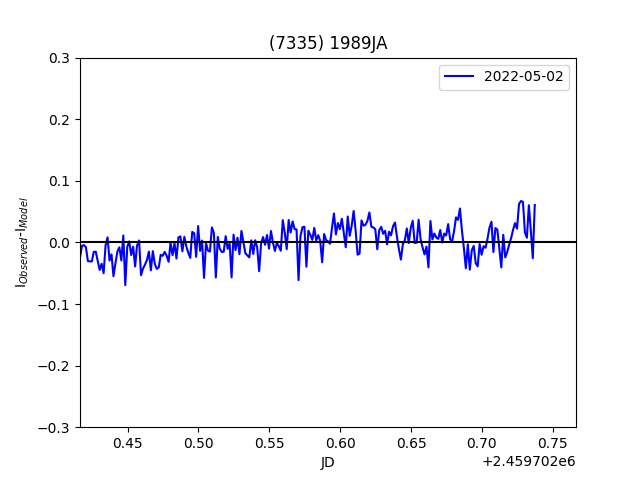}
    \end{subfigure}
    \begin{subfigure}{0.3\textwidth}
        \centering
        \includegraphics[width=\textwidth]{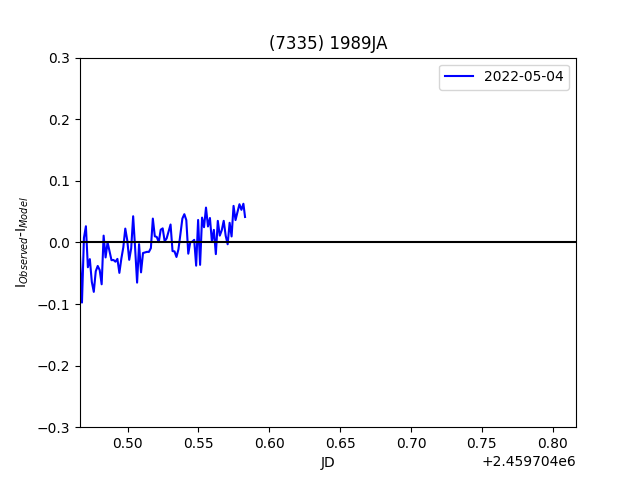}
    \end{subfigure}
    \begin{subfigure}{0.3\textwidth}
        \centering
        \includegraphics[width=\textwidth]{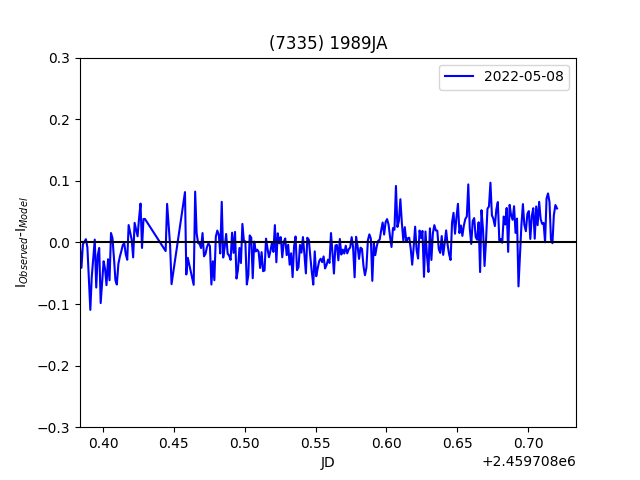}
    \end{subfigure}
    \begin{subfigure}{0.3\textwidth}
        \centering
        \includegraphics[width=\textwidth]{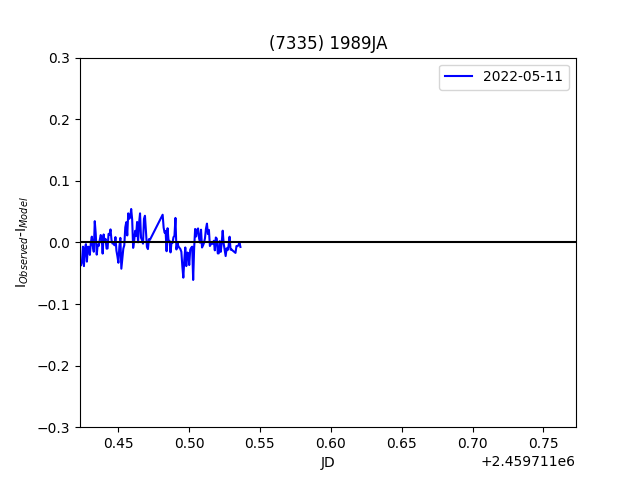}
    \end{subfigure}
    \begin{subfigure}{0.3\textwidth}
        \centering
        \includegraphics[width=\textwidth]{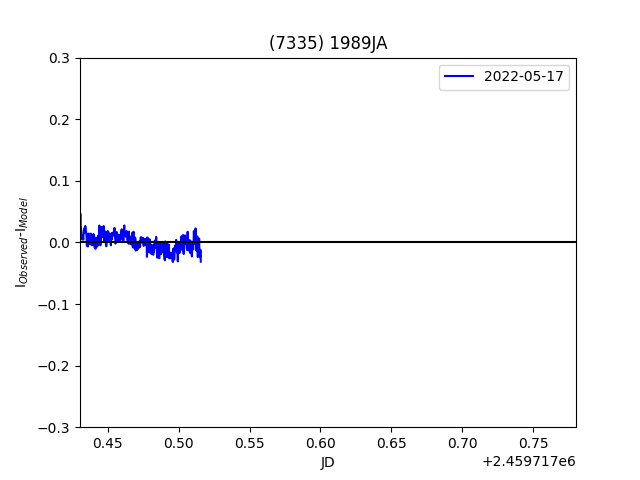}
    \end{subfigure}
    \begin{subfigure}{0.3\textwidth}
        \centering
        \includegraphics[width=\textwidth]{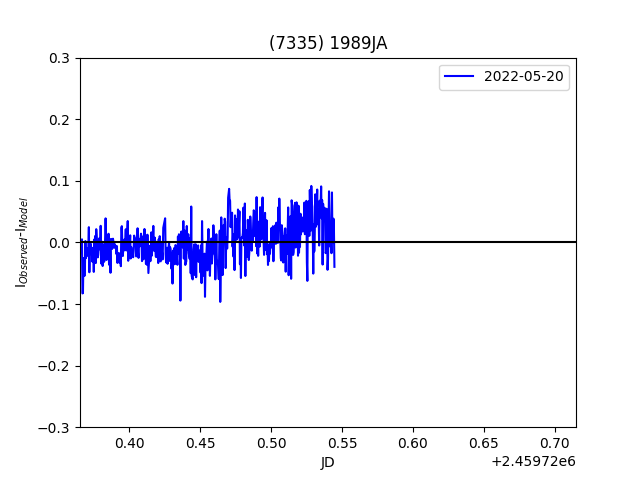}
    \end{subfigure}
    \begin{subfigure}{0.3\textwidth}
        \centering
        \includegraphics[width=\textwidth]{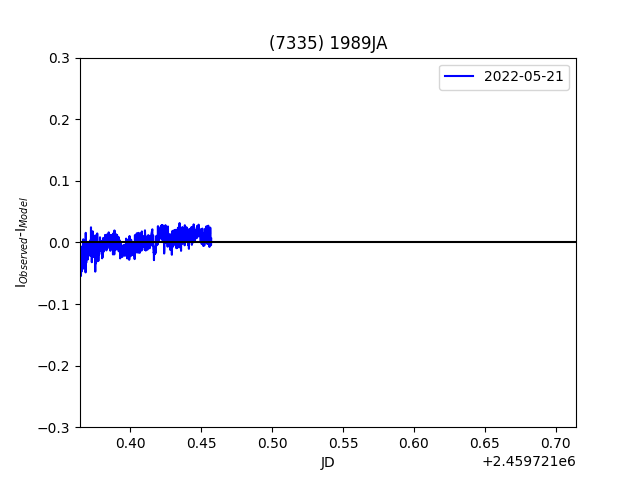}
    \end{subfigure}
    \caption{Graphical representation of the intensity of the data minus the intensity of the model against the Julian date (JD) for the rest of the light-curves.}
    \label{fig:7335_sat_all}
\end{figure*}


\bsp	
\label{lastpage}
\end{document}